\documentstyle[12pt,epsf]{article}
\newcommand{\beq}{\begin{equation}}   
\newcommand{\eeq}{\end{equation}}
\newcommand{\ra}{\rightarrow}
\newcommand{\gsim}{\lower.7ex\hbox{$
\;\stackrel{\textstyle>}{\sim}\;$}}
\newcommand{\lsim}{\lower.7ex\hbox{$
\;\stackrel{\textstyle<}{\sim}\;$}}

\begin{document}

\begin{titlepage}

\begin{center}
\Large{\bf  Theoretical Physics Institute\\
University of Minnesota}
\end{center}

\bigskip
\begin{flushright}
TPI-MINN-98/01-T\\
UMN-TH-1622-98\\
January  1998
\end{flushright}

\bigskip
\begin{center} \Large
{\bf  Snapshots of Hadrons }

\vspace{0.3cm}
or the Story of How the Vacuum Medium Determines the Properties
of the Classical Mesons Which Are Produced, Live and Die in the QCD 
Vacuum
\end{center}

\vskip 12pt

\begin{center} {\Large
M. Shifman}

\vspace{.4cm}

{\normalsize {\it  Theoretical Physics Institute, Univ. of Minnesota,
Minneapolis, MN 55455}}

\end{center}

\vspace{7cm}

\noindent Lecture  given at
 the 1997 Yukawa International Seminar 
{\em Non-Perturbative QCD -- Structure of the QCD Vacuum},
Kyoto, December 2 -- 12,  1997.

\end{titlepage}

\tableofcontents

\newpage

%\addtocounter{footnote}{-2}
%\renewcommand{\theequation}{1.\arabic{equation}}
%\setcounter{equation}{0}

\section{QCD Sum Rules: Twenty Years After}

I will discuss a method of treating the nonperturbative 
dynamics of QCD  which  was created almost twenty years 
ago \cite{SVZ} 
in an attempt  to understand a variety of  properties and 
behavior patterns 
in the hadronic family in terms of several basic 
parameters of the vacuum state.  The method goes under the name  
{\em
QCD sum rules} -- rather   awkward, for many reasons. 
First and foremost,  it does 
not emphasize the 
essence of the method. 
Second, in Quantum Chromodynamics there exist many other
sum rules,  having nothing to do with those suggested in Ref. 
\cite{SVZ}.  Finally, some authors add further  confusion by using 
{\em ad hoc} names, e.g. the Laplace sum rules, spectral sum rules, 
and so on, which are even foggier and are not generally accepted. 

It would be more  accurate to say  ``the method 
of expansion of the correlation functions in the vacuum condensates 
with  the subsequent 
matching via the dispersion relations". This is evidently  far too long 
a string to put  into
circulation. 
Therefore, for clarity I will refer to the 
Shifman-Vainshtein-Zakharov (SVZ) sum rules. 
Sometimes, I will resort to   abbreviations such as ``the 
condensate expansion". 

Twenty years ago, next to nothing was known about 
nonperturbative aspects of QCD. The condensate expansion was the 
first quantitative 
approach which proved to be successful in dozens of problems. Since 
then, many things changed. 
Various new ideas and models were suggested  concerning the 
peculiar
 infrared behavior in Quantum Chromodynamics. Lattice QCD grew 
into a powerful 
computational scheme which  promises, with time, to produce the 
most accurate results,  if not for the whole set of the   hadronic 
parameters, at least, for a significant part.  

It seems  timely to survey the ideas and technology constituting the 
core of the SVZ sum rules from the modern perspective, when the 
method
became  just one among several  theoretical components in a modern 
highly 
competitive environment. An exhaustive review of a wealth of 
``classical", old  elements of the method and 
applications   was given 
in Ref.  \cite{RV}. 
There is hardly any need in an abbreviated 
version of such a report.  New applications which were worked out in 
the last decade or so definitely do deserve a detailed discussion. As 
far as 
I know,  no comprehensive coverage of the topic exists in the 
literature. Unfortunately, in these lectures I will not be able to 
provide such a coverage, which thus remains a task for the future
\footnote{Work on systematically reviewing a variety
of developments that took place since the mid-1980's and numerous 
new 
applications is under way (a private communication from B.L. Ioffe).
A survey devoted to the relation between the sum rule and lattice 
results is being written by A. Khodjamirian.}.  
Instead,  I will focus on those 
qualitative aspects  where understanding became deeper. This is 
the first goal. Secondly, 
selected new applications will be considered to the extent that they
illustrate  the  theoretical ideas of the last decade. And last but 
not  least,  
I will
 try to outline an ecological niche which belongs to the SVZ
method today. As a matter of fact, over the years,  slow but steady 
advances  were taking place in our knowledge  of the hadronic 
world. Some old and largely forgotten  predictions of the SVZ 
sum rules were confirmed recently by other investigations based on 
totally different principles. 
These predictions are extremely 
nontrivial. For instance, about 15 years ago, it was discovered 
\cite{NSVZ}  that not all 
hadrons are alike;
there are remarkable distinctions between them, especially in the
glueball sector.  The fact that not all hadrons are alike is now 
becoming more and more evident from the lattice results as well. 
Other examples of this type are known too. 
By confronting  them with alternative  sources of 
information, 
such as lattices, we get  a much fuller  picture of the QCD vacuum.
This process might be very beneficial to both sides. Unfortunately, at 
present the lattice and analytic QCD communities are largely 
disconnected, and rarely talk to each other. My task is to show that
borrowing from each other makes everybody richer! 

\section{QCD Vacuum and  Basics of the SVZ Method}

\subsection{General ideas}

The color dynamics described by QCD is very peculiar. If we have 
two 
probe color charges,  their interaction approaches 
the Coulomb law at  short distances, with a weak coupling constant.  
The Coulomb 
interaction is due to the  one gluon exchange (Fig. 1). At larger 
distances, the 
gluon starts branching (Fig. 2), which leads to a 
remarkable 
phenomenon known as antiscreening, or asymptotic freedom 
\cite{AF}.  In 
normal theories, like QED, the virtual cloud screens the bare charge
making the charge seen at larger distances smaller than the bare one.
This situation is perfectly transparent intuitively.

In QCD, instead of screening, the branching processes result in a 
totally counter-intuitive behavior --  the antiscreening. Those of you 
who would like to know the physical origin of antiscreening
are referred to the very pedagogical review \cite{6avt}, Sect. 1.3. 
If the 
distances 
are not too large one can apply perturbation theory to quantitatively 
describe the gluon branchings, and derive the famous formula of 
asymptotic freedom
\begin{equation}
\alpha_s = {\rm Const} / \ln\frac{r_0}{r}
\label{asyco}
\end{equation}
where $r_0$ is a dynamically generated scale parameter of QCD,
$r_0\sim 1$ fm.  At small  separations the effective coupling 
constant 
dies off logarithmically.

Usually one considers the running coupling constant in the 
momentum space. Then in the leading (one-loop) approximation
corresponding to Eq. (\ref{asyco}),
\beq
\alpha_s(\mu^2)= \frac{4\pi}{b\ln (\mu^2/\Lambda^2)}\, ,\,\,\, 
b=\frac{11}{3}N_c - \frac{2}{3} N_f \, ,
\label{asymo}
\eeq
where $N_c$ is the number of colors and $N_f$ is the number of 
flavors and $b$
is the first coefficient in the Gell-Mann-Low function.  
In the low-energy domain we will be interested in
(below the threshold of the charm production) the number of active 
flavors
is $N_f=3$, and, hence, $b=9$. 

In perturbation theory
the effective coupling is calculated order by order.  The result is 
known up to three loops, and for $N_c=N_f=3$ can be conveniently 
written as \cite{3loop}
\begin{equation}
a(\mu^2)=\frac{1}{\ln \frac{\mu^2}{\Lambda^2}}\, - 0.79
\, \frac{\ln \ln \frac{\mu^2}{\Lambda^2}}{\ln^2 
\frac{\mu^2}{\Lambda^2}}
+  \frac{(0.79)^2}{\ln^3 \frac{\mu^2}{\Lambda^2}} \left[(\ln \ln
\frac{\mu^2}{\Lambda^2})^2 - \ln \ln \frac{\mu^2}{\Lambda^2} + 
0.415 \right] + 
{\cal O}\left(\frac{1}{\ln^4 \frac{\mu^2}{\Lambda^2}}\right)\, ,
\label{deflam}
\end{equation}
where we define 
\beq
a(\mu^2)\equiv\frac{b}{4} \frac{\alpha_s(\mu^2)}{\pi}\, ,  
\eeq
and 
$\Lambda$ is the scale parameter of QCD introduced  in a standard
way \cite{3loop}. In the third and higher loops
the law of running of $\alpha_s$ becomes scheme-dependent. 
The third  term in 
Eq. (\ref{deflam}) refers to the so called modified minimal 
subtraction ($\overline{\mbox{MS}}$)
scheme. More exactly, since Eq.  (\ref{deflam}) describes  running  
with three 
flavors, the parameter $\Lambda$ 
is actually $\Lambda^{(3)}_{\overline{\mbox{MS}}}$. 
Below we will deal exclusively with 
$\Lambda^{(3)}_{\overline{\mbox{MS}}}$; therefore, not to make the 
notation too clumsy, we will suppress the sub(super)scripts.

Let us leave  a while the issue of the effective coupling constant, with 
the intention of returning 
to it later. The only lesson one should remember at this stage is that
the effects caused by perturbative gluon exchanges are logarithmic.

Equations (\ref{asyco}) or (\ref{asymo})
imply that the effective  color 
interaction becomes stronger as the separation between the probe 
color charges increases. Being remarkable by itself, this phenomenon 
carries the seeds of another, even more remarkable property of QCD.
When the distance becomes larger than some number times
$\Lambda^{-1}$, and exceeds a critical one, the 
branchings of gluons become so 
intensive (Fig. 3) that it makes no sense to speak about individual 
gluons.

\begin{figure}
  \epsfxsize=4cm
  \epsfysize = 4.2cm
  \centerline{\epsfbox{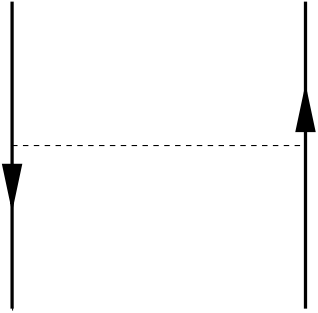}}
  \caption{ Heavy color charges (solid lines)  interacting through the 
one-gluon 
exchange (dashed line).}    
\end{figure}

\begin{figure}
  \epsfxsize=5.5cm
  \epsfysize = 4.2cm
  \centerline{\epsfbox{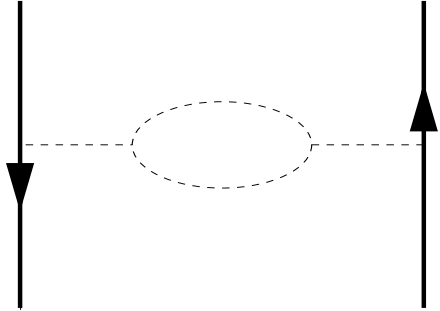}}
  \caption{ Gluon branching responsible for the logarithmic running 
of the effective 
gauge coupling constant in perturbation theory (shown is a sample 
one-loop graph).}    
\end{figure}

\begin{figure}
  \epsfxsize=10cm
  \epsfysize = 4.2cm
  \centerline{\epsfbox{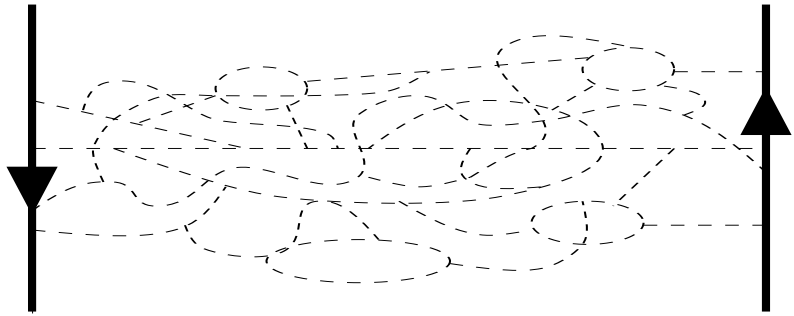}}
  \caption{When the separation between the probe color charges 
exceeds  critical  the nonlinearity becomes so strong 
that it makes no sense to speak about individual gluons. The color 
fields between the probe charges form a flux tube with transverse 
dimensions $\sim \Lambda^{-1}$.}    
\end{figure}

Rather, we should phrase our consideration in terms of the 
chromoelectric and chromomagnetic fields. In ``normal" theories, like 
QED, the field induced by the probe charges separated by a large 
distance, is dispersed all over
 space. In QCD it is conjectured that a specific organization of the 
QCD vacuum makes such a dispersed configuration energetically 
inexpedient \cite{Hooft}. Rather, the chromoelectric field between the 
probe 
charges squeezes itself into a sausage-like configuration. The 
situation is reminiscent of  the Meissner effect in superconductivity. 
As is well-known, the superconducting media do not tolerate the 
magnetic field. If one imposes, as an external boundary condition, a 
certain flux of the magnetic field through such a medium, the 
magnetic field will be squeezed into a thin tube carrying all the 
magnetic flux; the superconducting phase is destroyed inside the 
tube, and there is no magnetic field outside. 

Superconductivity is 
caused by condensation of the Cooper pairs --  pairs of  electric 
charges. A phenomenologically acceptable picture of the infrared 
dynamics in QCD requires chromoelectric flux tubes, not magnetic 
ones. If something like condensation of ``chromomagnetic monopoles" 
took place in the 
QCD vacuum then this would naturally explain nonproliferation of the 
chromoelectric field from the external probe color charges in the 
entire space -- a dual Meissner effect would force the field to form 
flux tubes in the QCD vacuum, in this way ensuring  color 
confinement \cite{Hooft}. 

Nobody ever succeeded in  proving   
that the dual 
Meissner effect does indeed take place in QCD. There are
reasons to believe, however, that it does. First, some evidence came 
from the numerical study in the lattice QCD (see e.g. \cite{ADG}). 
Quite recently, a 
breakthrough \cite{SW} was achieved in understanding of the 
infrared 
dynamics in a theory which might be considered a  relative of 
QCD, the so called $N=2$ supersymmetric (SUSY) Yang-Mills theory, 
in which it was  
analytically 
shown that the monopoles do condense in the strong coupling regime,
thus providing a basis for the dual Meissner mechanism.

Even if, as a result of a breakthrough  sometimes in the future,
it becomes clear
how a similar mechanism might develop in QCD, 
we have  a long way to go from a qualitative picture of the 
phenomenon to 
 a quantitative approach allowing one to exactly  {\em calculate} 
all the variety of the hadronic properties from the first principles. 
It may be  that such string-based exact approach will 
never emerge. While the issue of the exact solution is still under 
investigation, let us see what analytic QCD can offer as an 
approximate solution. 

The basic idea lying behind  the SVZ method is quite 
transparent. The quarks comprising the 
low-lying
hadronic states, e.g. classical mesons or baryons, are not that far from 
each other, on average. The distance between them is of order 
$\Lambda^{-1}$. Under the circumstances, 
 the string-like chromoelectric flux tubes, connecting 
well-separated probe color charges,  hardly have a chance to be 
developed. Moreover, the valence quark pair injected in the vacuum, 
in a sense, 
perturbs it only slightly. Then we 
do not need the full machinery of the QCD strings, whatever it 
might mean, to approximately describe the properties
of the low-lying states. Their basic parameters depend on how the 
valence quarks of which they are built interact with  typical {\em 
vacuum} field fluctuations.

It is established that the QCD vacuum is characterized by 
various condensates \cite{SVZ}.   Half-dozen of them are known:
the gluon condensate $G^2_{\mu\nu}$, the quark condensate
$\bar q q$, the mixed condensate $\bar q \sigma G q$, and so on. 
The
task is  to
 determine the 
regularities and parameters of  the classical mesons and baryons 
from a few 
simple  condensates. 

 Of course,
without invoking the entire infinite set of condensates one can hope 
to capture only  gross features of the vacuum medium, at best. 
Correspondingly, any calculation of the hadronic parameters 
 of this type is admittedly  approximate. Since  
hadronic 
physics is deprived of  small parameters, in the vast 
majority of problems high accuracy is by far not the most desired 
requirement to  calculation, however. Does it make any difference if, 
say, the proton magnetic moment is predicted theoretically to two or 
to three digits? I do not think so. Rather, it is high reliability of 
predictions in a {\em wide range of problems where the answer is 
not known a priori} and   theoretical 
control over qualitative aspects  that are  the primary goals of the 
theory of 
hadrons. Below I will try to demonstrate that the SVZ method is 
both a
reliable and controllable approach which has enjoyed enormous 
success 
in dozens and dozens of instances. There are a few cases where it 
fails \cite{NSVZ}, but we do understand why.  And 
the very fact of 
the failure of the standard strategy teaches us a lot. This happens for 
specific, 
``nonclassical" hadrons, with a very strong coupling to the vacuum 
fluctuations,  which are very different from say the $\rho$ meson or 
nucleon. So, we can reverse the argument and convert the failure 
into 
success by saying that the method predicts that {\em not all hadrons 
are 
alike}. 

\subsection{Getting started/Playing with  toy models}

The basic microscopic degrees of freedom of Quantum 
Chromodynamics are the quarks and gluons. Their interaction is 
described by the Lagrangian
\beq
{\cal L} = -\frac{1}{4} G^a_{\mu\nu}G^a_{\mu\nu}
+\sum_f \bar q i\not\!\!{D}q
\label{qcdlag}
\eeq
where $f$ is the flavor index, and the color index of the quark fields 
$q$ is suppressed. For simplicity we will assume that the quark mass 
terms vanish. In other words, we will work in the chiral limit.
In this limit the pion is massless, $m_\pi^2 = 0$. This is known
to be quite a good approximation to the real world.

Neither quarks nor gluons are asymptotic states. Experimentally
observed are hadrons -- color-singlet bound states. In order to study 
the properties of the classical hadrons it is convenient to start from 
the empty space -- the vacuum -- inject there a quark-antiquark 
pair, and then follow the evolution of the valence quarks injected in 
the
vacuum medium. The injection is achieved by external currents.
The most popular are the vector and axial currents. Their popularity 
is due to the fact that they actually exist in nature: virtual photons 
and $W$ bosons couple to the  vector and axial quark currents.
Therefore, they are experimentally accessible in the $e^+e^-$ 
annihilation into hadrons or hadronic $\tau$ decays.   

Thus, the objects we will work with are the correlation functions of 
the quark currents. More concretely, let us consider the
vector current with the isotopic spin $I=1$,
\begin{equation}
J_\mu = \frac{\bar u\gamma_\mu u -\bar d\gamma_\mu 
d}{\sqrt{2}}\, ,\,\,\, \mbox{or}\,\,\, J_\mu =\bar u\gamma_\mu d
.
\label{currents}
\end{equation}
The first current shows up in the $e^+e^-$ annihilation
while the second is relevant to the $\tau$ decays. The two-point 
function $\Pi_{\mu\nu}$ is defined as
\beq
\Pi_{\mu\nu} = i \int {\rm e}^{iqx} d^4 x \langle 0|T\{ J_\mu (x) 
J^\dagger_\nu 
(0) \} |0\rangle\, .
\label{Tproduct}
\eeq
where $q$ is the total momentum of the quark-antiquark pair
injected in the vacuum.
Due to the current conservation $\Pi_{\mu\nu}$ is transversal and, 
hence,
\beq
 \Pi_{\mu\nu}
= (q_\mu q_\nu -q^2g_{\mu\nu})\Pi 
(q^2)\, .
\eeq
For historical reasons $\Pi (q^2)$ is often called the {\em polarization 
operator};  we will use this nomenclature in what follows. 
It is easy to count that the function $\Pi (q^2)$ is dimensionless.
The imaginary part of $\Pi (q^2)$ at positive values
of $q^2$ (i.e. above the physical threshold of the hadron production)
is called {\em the spectral density},
\beq
\rho (s) =\frac{12\pi}{N_c}\, \mbox{Im}\,\Pi (s)\, , \,\,\,  s\equiv 
q^2\, .
\label{rhoim}
\eeq
Up to  normalization 
it coincides with the cross section of $e^+e^-$ annihilation into 
hadrons (measured in the units $\sigma (e^+e^-\rightarrow
\mu^+\mu^-)$) or
the $\tau$ decay distribution function.  The numerical factor in the 
definition of $\rho (s)$ in Eq. (\ref{rhoim}) is introduced for 
convenience, as will become apparent shortly.  In the real world 
$N_c=3$ but we will keep this factor explicit for a while, to keep 
track of the $N_c$ dependence. 

The spectral density carries full information about the spectrum and 
widths of 
hadrons with  given quantum numbers.
Every QCD 
practitioner dreams of the exact calculation of the spectral density. 
Later on we will see how the SVZ sum rules constrain 
the parameters of the lowest-lying state, the $\rho$ meson in the 
case at hand. But  first, prior to submerging into technical aspects
of the SVZ approach, let us make an 
educated guess of how the spectral density might look, in gross 
features,  to get an idea of what is expected for 
$\Pi (q^2)$.

To begin with, consider the limit $N_c\rightarrow\infty$.
As is well-known \cite{THW},  in this limit all hadrons are 
infinitely narrow, since all decay widths are suppressed by powers of 
$1/N_c$. Correspondingly,  $\rho (s)$ is a sum of delta functions.

Moreover, there are good reasons to believe that these delta 
functions must be  approximately equidistant, at least 
asymptotically,
for highly excited states.  A string-like picture of color confinement  
naturally leads to (approximately) linear Regge trajectories,
and, hence, $m_n^2 \propto n$ at large $n$ where $m_n$ is the mass 
of the $n$-th $\rho$-meson excitation. 
In the world of
purely linear Regge trajectories there are an infinite number of 
daughter 
trajectories associated with 
each Regge trajectory. The daughter   trajectories are parallel to the 
parent trajectory and are shifted
by integers. Thus, in  the old Veneziano model (a review
is given e.g. in 
\cite{revven})
$m_n^2=m_\rho^2+n/{\alpha '} $ ($\alpha '$ is the slope of the 
Regge trajectory). In some of the later versions, which are more
``QCD-friendly",  $m_n^2=m_\rho^2+2n/{\alpha '} $, see e.g.
\cite{DKS}.  Since for now the focus is on  a qualitative 
picture, the distinction between these two scenarios is not  essential 
for our illustrative purposes.  For definiteness, let us accept that the 
distance between two consecutive excitations contributing to $\rho 
(s)$
is $2/\alpha ' \approx  2$ GeV$^2$. 

Assembling all these elements together one obtains a sketch 
 of the spectral density (Fig. 4),
\beq
\rho (s) =3\,\sum_{n=0}^\infty \delta (s-1-3n) \, .
\label{delmod}
\eeq
Here $m_\rho^2$ is set equal to unity; all dimensional quantities are 
measured in these units, for instance,
$2/\alpha ' \approx  3$.  The couplings of all 
mesons
to the current $J_\mu$ are chosen to be equal. This choice is 
represented  by the overall numerical factor 3 in the sum 
(\ref{delmod}).
The  explaination for this  will become clear momentarily.

\begin{figure}
  \epsfxsize=11cm
  \centerline{\epsfbox{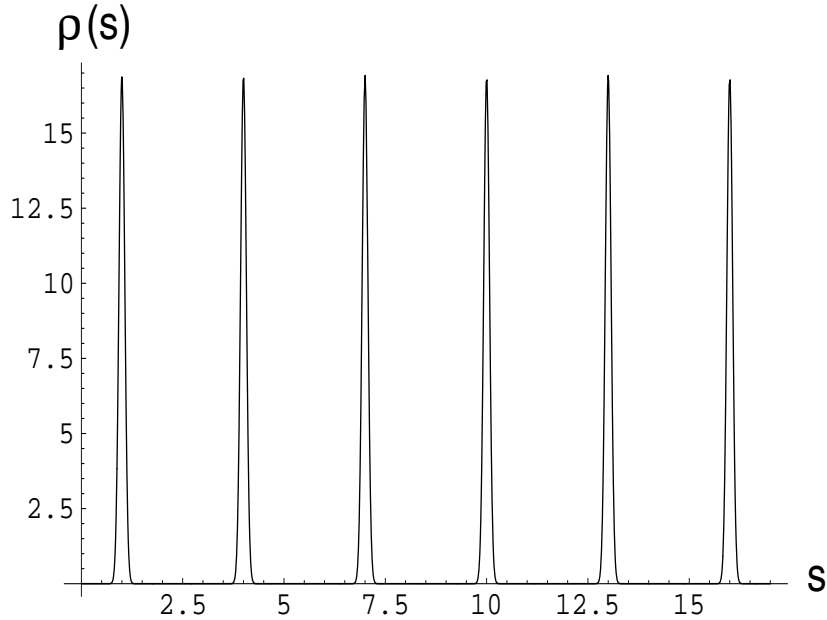}}
  \caption{The spectral density in the toy model. For clarity I gave 
a tiny width to the $\delta$ functions in  Eq. (\ref{delmod}).}    
\end{figure}

Now, given the imaginary part (\ref{delmod}), it is not difficult to 
reconstruct the polarization operator $\Pi$ itself,
\beq
\Pi (Q^2) = -\frac{N_c}{12\pi^2}\, \psi (z) +\,  \mbox{Const.}\, ,
\label{pipsi}
\eeq
$$
Q^2 \equiv - q^2\, , \,\,\, z = \frac{Q^2 +1}{3}\, ,
$$
where $\psi$ is the logarithmic derivative of the $\Gamma$ function,
$$
\psi (z) = - C +\sum_{k=0}^\infty\left[ \frac{1}{k+1} - 
\frac{1}{z+k}\right]\, .
$$
The constant  on the right-hand side of
Eq. (\ref{pipsi}) is irrelevant, since it can be always eliminated 
by an appropriate subtraction. It will be omitted hereafter.

Although our desired target is the spectral density on the physical 
cut, i.e. at positive values of $q^2$, all theoretical calculations in 
 Quantum Chromodynamics 
are carried out off the physical cut, say, at negative values of $q^2$
(positive $Q^2$). The reason is obvious: the QCD Lagrangian 
(\ref{qcdlag})
is formulated in terms of quarks 
and gluons,
not hadrons. In the absence of the final solution of QCD  we can deal 
only with the quark and gluon fields. Working  in the Euclidean 
domain, off the physical cuts,  we can calculate in terms of quarks 
and gluons. This is a common feature of the SVZ  sum rules
and  lattice strategies. 

At positive values of $Q^2$ an asymptotic representation exists for
the
$\psi $ function,
\beq
\psi (z) = \ln z -\frac{1}{2z} - \sum_{n=1}^\infty
\frac{B_{2n}}{2n}\, z^{-2n}\, ,
\label{asypsi}
\eeq
where 
$B_{2n}$ stand for the Bernoulli numbers,
\beq
B_{2n} = (-1)^{n-1}\frac{2(2n)!}{(2\pi)^{2n}}\zeta (2n)\, ;
\label{bernou}
\eeq
here $\zeta$ is the Riemann function. (In some textbooks
$(-1)^{n+1}B_{2n}$ is called the $n$-th Bernoulli number 
and is denoted by $B_n$.)

 Equations (\ref{pipsi}) and  (\ref{asypsi}) at large positive $Q^2$ 
imply that in the leading approximation 
\beq
\Pi (Q^2) \rightarrow -\frac{N_c}{12\pi^2}\ln Q^2\, .
\label{lqpi}
\eeq
This logarithmic formula for the polarization operator
exactly matches what we expect from perturbation theory.
Indeed, at large $Q^2$, in the deep Euclidean domain,  the points of 
injection and annihilation
of the quark pair in the vacuum  are separated by a small space-time 
interval. The quarks have no time to interact with the vacuum 
medium. 
They propagate as free objects (Fig. 5). For free quarks, obviously,
\beq
\Pi_{\mu\nu }(q) =i\int e^{iqx} d^4 x \, \mbox{Tr}\, \{
\gamma_\mu S_0(x,0)\gamma_\nu S_0(0,x)\}\, ,
\label{fqp}
\eeq
where $S_0(x,y)$ is the free-quark Green function
describing propagation from the point $y$ to the point $x$,
\beq
S_0(x,y) =\frac{1}{2\pi^2}\frac{\not\!\! \Delta }{(\Delta^2)^2}\, , \,\,\, 
\Delta\equiv x-y\, .
\eeq
It is  trivial  to obtain
\beq
\mbox{Tr}\, \{
\gamma_\mu S_0(x,0)\gamma_\nu S_0(0,x)\}
=-\frac{N_c}{\pi^4}\, \frac{2x_\mu x_\nu -x^2 g_{\mu\nu}}{x^8}\, .
\label{exer}
\eeq
Note that the expression on the right-hand side is automatically 
transversal. Substituting Eq. (\ref{exer}) in Eq. (\ref{fqp}) and doing 
the Fourier transformation,
we arrive at Eq. (\ref{lqpi}). 

\begin{figure}
  \epsfxsize=5cm
  \centerline{\epsfbox{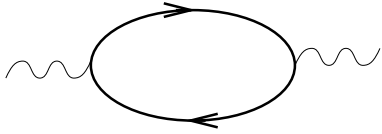}}
  \caption{The two-point function of the vector currents in the 
free-quark approximation. The external current injecting the 
quark-antiquark pair in the vacuum (and then annihilating it) is 
denoted by wavy lines.}    
\end{figure}

Thus, the leading
asymptotic behavior of the polarization operator stemming from the 
free-quark graph of Fig. 5 is purely logarithmic. An infinite comb
of infinitely narrow resonances, taken with one and the same 
residue,
see Eq.  (\ref{delmod}), yields the same logarithm
in the deep Euclidean domain.  This explains why all resonance 
coupling constants are set equal in Eq. (\ref{delmod}).
I hasten to add that  the spectral density  in the real world is much 
more contrived.
The toy spectral density (\ref{delmod})
is supposed to be a  caricature, only roughly reminding 
the actual one. Nevertheless,  the exercise with  the toy spectral 
density
is  instructive in several respects. 

It is clear that the leading logarithmic term (\ref{lqpi}) carries
absolutely no information on the mass of the lowest-lying state or on 
the spacing between the resonances: it has no built-in dimensional 
parameter. This information is encoded in the power corrections. 

Examining the 
$1/Q^2$ expansion in Eq. (\ref{pipsi}), \footnote{In the case at hand 
we deal with pure powers of $1/Q^2$.  Our toy  model  is 
too rude to reproduce the logarithmic corrections 
and logarithmic anomalous dimensions of the condensates typical of 
QCD.} we 
conclude that the high-order  terms have factorially divergent 
coefficients. 
The factorial divergence at large $n$ is due to the factorial growth
of the Bernoulli numbers $B_{2n}$ in Eq. (\ref{asypsi}). In QCD the 
expansion in powers of $1/Q^2$ is nothing but  the condensate 
expansion. The example considered teaches us that the power  
expansion of $\Pi (Q^2)$ has  zero radius of convergence. By itself, 
this is no disaster, one can work with asymptotic series as long
as the expansion parameter is small. In our case the expansion 
parameter is $\tau\equiv 1/Q^2$, and we are interested in the 
expansion in the vicinity of $Q^2 \sim 1$.  (Remember that
we put  $m_\rho^2 = 1$.  If $Q^2$ is substantially larger than 1,
one gets very little  information on the $\rho$ meson
from $\Pi (Q^2)$.)

In the ideal world we would calculate $\Pi (Q^2)$, and, hence, the 
spectral density proportional to Im$\Pi (s)$,  exactly. In reality we 
must   settle for $\Pi (Q^2)$, calculated in the Euclidean domain 
(positive $Q^2$), in the form of a {\em truncated} series. 
The dispersion relation provides a bridge between the
Euclidean calculation and the spectral density,
\beq
\Pi (Q^2) = \frac{1}{\pi}\, \int ds \frac{\mbox{Im}\, \Pi (s)}{s+Q^2} 
+ \mbox{Const.}
\label{disrel}
\eeq 
This explains why the method is referred to as the sum rules. 

The first five terms in the power expansion of the polarization 
operator (\ref{pipsi})
are
\beq
\Pi (Q^2) = -\frac{N_c}{12\pi^2}\left[\ln Q^2 
-\frac{\tau}{2} +\frac{\tau^2}{4}+\frac{\tau^3}{3} -
\frac{13}{40}\,\tau^4 - \tau^5 + ... \right]\,  ,\,\,\,\, \tau \equiv 
1/Q^2\, .
\label{exppiqu}
\eeq
From the structure of the series it is seen that at $\tau \sim 1$ the 
polarization operator $\Pi (Q^2)$
is determined by its expansion with very poor accuracy, up to a 
factor of 2, at best. This accuracy is obviously insufficient
to allow one to estimate, say, $m_\rho^2$ with a reasonable 
precision.

One can drastically improve the accuracy by considering the 
Borel-transform of $\Pi (Q^2)$ \cite{SVZ}. The Borel transformation 
can be defined in various ways. For the functions obeying dispersion 
relations, the most convenient definition is through the following 
limiting procedure:
\beq
\hat{\cal B} = \lim \frac{1}{(n-1)!}(Q^2)^n \left( 
-\frac{d}{dQ^2}\right)^n\, , \,\,\, Q^2\rightarrow\infty , \,\,\, n 
\rightarrow\infty , \,\,\, \frac{Q^2}{n}\equiv M^2 \,\,\, 
\mbox{fixed}\, . 
\label{borelt}
\eeq
$M^2$ is called the {\em Borel parameter}.
It is not difficult to show that applying $\hat{\cal B}$
to $(1/Q^2)^n$ we get
\beq
\hat{\cal B} \left(\frac{1}{Q^2} \right)^n =
\frac{1}{(n-1)!}\left(\frac{1}{M^2} \right)^n\, ,
\eeq
which entails, in turn
\beq
\hat{\cal B} \left(\frac{1}{s+Q^2} \right) = \frac{1}{M^2}\, e^{-
s/M^2}\, .
\eeq
Another trivial but useful relation we will need is
\beq
\hat{\cal B} ( \ln Q^2 ) = -1\, . 
\eeq
The Borel-transformed dispersion relation (\ref{disrel})
takes the form
\beq
\tilde{\Pi} (M^2) \equiv \{ \hat{\cal B} \Pi (Q^2)  \} =
\frac{1}{\pi M^2}\, \int ds  \mbox{Im}\, \Pi (s)\, e^{-
s/M^2}\, .
\label{btdr}
\eeq

The advantages one gains in dealing with the Borel-transformed
sum rules are obvious. First, we improve, factorially, the convergence 
of the power series. Thus, in the toy model we are playing with --
an infinite comb of infinitely narrow peaks presented in  Eq. 
(\ref{delmod}) -- 
the $1/Q^2$ expansion for $\Pi (Q^2) $ is asymptotic, while the
$1/M^2$ expansion of $\tilde{\Pi} (M^2)$ is {\em  convergent}!  
Indeed,
\beq
\tilde{\Pi} (M^2) = \frac{N_c}{4\pi^2}\, \frac{1}{M^2}\,
\frac{e^{-1/M^2}}{1-e^{-3/M^2}}\, ,
\label{rfbpm}
\eeq
and the radius of convergence of the $1/M^2$ expansion is 
\beq
|M_*^2| =\frac{3}{2\pi}\, .
\label{radcon}
\eeq
The factor $2\pi$ in the denominator ensures that we can go down  
to a remarkably  low value of $M^2$; the point $M^2 \sim 1$ is well 
inside 
the convergence radius. Let us examine the first five terms of the 
truncated series,
\beq
\tilde{\Pi} (M^2) = \frac{N_c}{12\pi^2}\, \left[  
1 +\frac{\tau}{2}  -\frac{\tau^2}{4}-\frac{\tau^3}{6}
+\frac{13 \tau^4}{240} + \frac{\tau^5}{24} + ... \right]\,  ,\,\,\,\, \tau 
\equiv 
1/M^2\, .
\label{exbpo}
\eeq
It is seen that in the vicinity of $\tau\sim 1$ our ``theoretical 
prediction" is expected to carry an error of order $\sim 2\%$. Such 
accuracy is already  good enough to get the mass and residue  of the 
lowest-lying state with a reasonable ``resolution". 

On the phenomenological side, proceeding to the Borel-transformed 
dispersion relations we automatically kill possible subtraction 
constants. What is even more important, the exponential weight 
function in Eq. (\ref{btdr}) makes the integral over the imaginary 
part well-convergent.  Thanks to the  improved convergence
the relative role of the $\rho$-meson contribution
is strongly enhanced, as we will prove shortly by quantitative 
estimates.  

To see how it all works let us take a closer look at $\tilde{\Pi} (M^2)$;
to avoid cumbersome numerical factors we will change the overall 
normalization and  will deal with
\beq
I(M^2) = \frac{12\pi^2}{N_c}\tilde{\Pi} (M^2) = \frac{1}{M^2}\, \int \, 
ds e^{-s/M^2} \rho (s)\, .
\label{psim}
\eeq
In the toy model under consideration everything is known 
explicitly:
the exact expression for $I(M^2)$,  its power expansion, and the 
$\rho$-meson contribution to $I(M^2)$. The corresponding plots are 
displayed in Fig. 6. The exact curve has a typical shape: 
at small $M^2$ it is exponentially suppressed, approaching zero as 
$\exp (-m_\rho^2/M^2)$; at large $M^2$ it is flattens off and slowly 
approaches its asymptotic value, which is equal to unity
thanks to a smart choice of the normalization factor in Eq. 
(\ref{psim}). This unity is in one-to-one correspondence with the 
free quark result (\ref{lqpi}). It is convenient to normalize the sum 
rules to the free-quark result at asymptotically large $M^2$, and we 
will 
always do that. The exact curve has a steep (left) shoulder, and a 
shallow (right) one, with a maximum at $M^2$ slightly above 
$m_\rho^2$. Remember this pattern since  we will encounter with a
very similar picture  more than once in real QCD. 

The $\rho$ meson mass and residue  follow immediately
from consideration of $I(M^2)$  in the small $M^2$ domain (the left 
shoulder).
Indeed, in this domain all higher states are exponentially suppressed
(for instance, the contribution of the first excitation relative to that
of $\rho$ is $\sim \exp (-3/M^2)$.  Therefore, $I(M^2)$ is fully  
saturated by $\rho$ in the limit
$M^2 \rightarrow 0$. Fitting the left shoulder by
$C_1 \exp (-C_2/M^2) $ we would find $C_{1,2}$, the  $\rho$-meson 
residue   and mass, respectively.

 Alas... In the real world 
$I(M^2)$ is not known at small $M^2$. All that is  known is the 
large-$M^2$
asymptotics, in the form of the expansion
in the coupling constant $\alpha_s (M^2)$, and a (truncated) 
condensate expansion. Therefore, let us pretend that in our toy  
model we 
have the same: a truncated power series (\ref{exbpo}). The 
corresponding curve is shown in Fig. 6 (where I have actually 
included also the $O(\tau^6)$ term). The expansion is convergent 
only to the right from $M^2 \approx 0.5$. If we want the truncated 
series, with a few first terms included, to accurately represent
the theoretical curve, we have to stop at $M^2 \approx 0.7$
(arrow A in Fig. 6). This is the left boundary of what the sum-rule 
practitioners usually call {\em the window} or {\em working 
window}. 

The right boundary of the window (arrow B) is provided by the 
requirement that the first and higher excitations show up at the level 
not exceeding, say, 20 to 30 \%. In this way we ensure sufficient 
sensitivity to the $\rho$-meson parameters. 

The larger the $M^2$, the weaker  the $\rho$-meson dominance
will be. 
The requirement of the $\rho$-meson dominance forces us to move 
to smaller values of $M^2$,  while keeping control over the power 
expansion suggests that we move to larger values of $M^2$. 
Thus, these two requirements are self-contradictory. {\em A priori}, 
in any given problem, it is not evident  that the AB window exists at 
all. If it does, we can obviously use this fact to  approximately
calculate 
 the mass and the residue of the lowest-lying state.
The reasons why the window exists in the problems of the classical 
mesons and baryons
will be discussed later. 

\begin{figure}
  \epsfxsize=14cm
  \centerline{\epsfbox{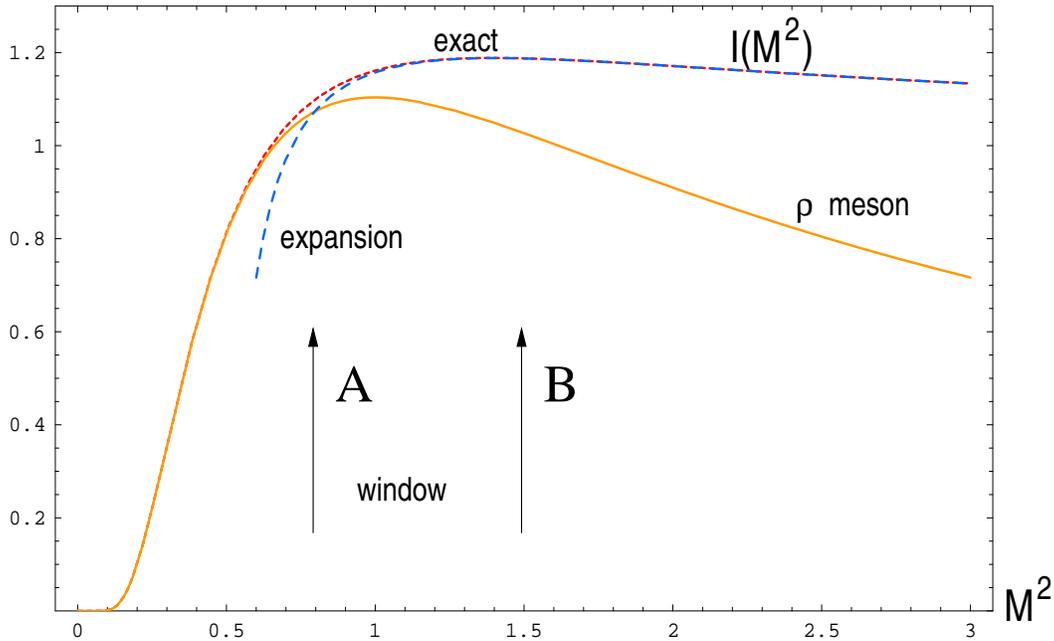}}
  \caption{ The sum rule for the $\rho$ meson in the toy model.}    
\end{figure}

To get an idea of the accuracy one can expect
from the sum rule  calculations of the $\rho$-meson parameters
we must  investigate how sensitive our results are to various 
assumptions regarding the spectral density. The spectral density
presented in Fig. 4 is not fully realistic, for many reasons. 
If  minor details affected the results too strongly, 
the SVZ method would be useless. The right edge of the window
(arrow B) is chosen in such a way as to minimize the impact of the
(unknown) details of the spectral density above the $\rho$-meson 
peak. Yet, it is important to obtain a quantitative measure of the
corresponding uncertainty.

The most obvious unrealistic feature of the spectral density
(\ref{delmod}) is the infinitely narrow width of all resonances from 
the comb.  It is true that in the multicolor QCD the width-to-mass
ratio of all hadrons built from quarks is proportional to
$1/N_c$ \cite{THW} and, thus, vanishes at $N_c\ra\infty$.
If $N_c$ is fixed, however, and we are interested in the widths
of the excited states, there is another large parameter, the excitation 
number, which compensates for the effect of large $N_c$. Moving to 
the right along the $s$ axis in Fig. 4, very soon we find ourselves in 
the situation with the overlapping resonances --  the $\delta$ 
functions in the comb representing the spectral density are  smeared 
and the spectral density becomes perfectly smooth. Let us discuss 
the consequences of such  {\em dynamical smearing}.

First of all, we must establish the dependence of the width 
$\Gamma_n$
on the excitation number $n$. If one believes that a string-like 
picture of color confinement develops in QCD, an asymptotic 
estimate of 
$\Gamma_n$ at large $n$ becomes a simple exercise. 
Unlike the $\rho$ meson and other low-lying classical states,
for highly excited states the string-like picture of the chromoelectric 
tubes is expected to be fully relevant. 

When a highly excited meson state is created by a local source, it can
 be considered, quasiclassically, as a pair of (almost free) 
ultrarelativistic
quarks; each of them with energy $m_n/2$. These quarks are 
created 
at the origin, and then fly back-to-back, creating behind them a flux 
tube
of the chromoelectric field. The length of the tube $L\sim 
m_n/\Lambda^2$
where $\Lambda^2$ represents  the string tension. The decay 
probability is 
determined, to order $1/N_c$, 
 by the probability of producing an extra quark-antiquark pair. Since 
the pair creation 
can happen anywhere inside the flux tube, it is natural to expect that
\beq
\Gamma_n\sim \frac{1}{N_c} L\Lambda^2 =\frac{B}{N_c}m_n
\label{estgam}
\eeq
where $B$ is a dimensionless coefficient of order one. 
I used here the fact that within the quasiclassical picture the meson 
mass $m_n\sim L\Lambda^2$.
 Thus, the width of the $n$-th excited state
is proportional to its mass which, in turn, 
is proportional to $\sqrt{n}$ for the linear Regge trajectories
\footnote{
Let us note in passing that the $1/N_c^2$ corrections due to 
creation of two quark pairs are of order $L^2/N_c^2$ within this 
picture. Since $L\sim m_n \sim \sqrt{n}$, the expansion parameter is
$\sqrt{n}/N_c$.}. 

This simple estimate was obtained in Ref. \cite{Nussinov} long ago.
 Since the argument bears a very general nature,
the square root dependence of $\Gamma_n$ should take place in all 
models with linear confinement. And it does, indeed!
Recently, the square root formula for $\Gamma_n$
was obtained numerically \cite{BS} in the 't Hooft model
\cite{TM}. 

If Eq. (\ref{estgam}) does indeed take place, it is not difficult to find 
the impact of the resonance widths on the spectral density. 
The infinitely narrow pole is substituted with
$$
\frac{1}{Q^2 + m_n^2} \, \ra\,  \frac{1}{Q^2 + m_n^2- 
im_n\Gamma_n}\,
\ra \, \frac{1}{Q^2 + m_n^2 -i m_n^2B/N_c}\, \ra
$$
\beq
\frac{1}{Q^2 + m_n^2 - \gamma Q^2 \ln Q^2}
\, \ra\, 
 \frac{1}{(Q^2)^{1-\gamma } + m_n^2 }\, , 
\label{resww}
\eeq
where 
\beq
\gamma = \frac{B}{\pi N_c}\, ,
\label{defga}
\eeq
and 
I used the fact that $1/N_c$ is a small parameter.
This explains the last transition in Eq. (\ref{resww}). 
For the same reason, in the third transition
I replaced $m_n^2/N_c$ with $N_c^{-1}\pi^{-1}  Q^2\ln Q^2$.
Near the pole (i.e. at $Q^2 = - m_n^2$) both expressions give the same 
in the leading 
$1/N_c$ approximation, and are equally legitimate
\footnote{The logarithm introduces not only the imaginary part, i.e. 
the width, but a shift in the real part, equivalent to a shift in the
resonance mass. Both effects are of order $1/N_c$.}. 

As a result, all poles, which lied previously at positive
real $q^2$, are now shifted onto unphysical sheets, away from the 
physical cut (Fig. 7).  The $\delta$ functions in the
spectral density are replaced by peaks with finite widths.
Correspondingly, the toy model number two, which replaces Eq. 
(\ref{pipsi})
is
\beq
\Pi (Q^2) = -\frac{N_c}{12\pi^2}\, 
\frac{1}{1-\gamma}\, \psi (z) +\,  \mbox{Const.}\, ,
\label{tm2}
\eeq
where now $z$ is given by the following formula
\beq
z = \frac{(Q^2)^{1-\gamma } + 1}{3}\, .
\label{newz}
\eeq
The additional factor $(1-\gamma)^{-1}$ in the overall normalization 
is chosen to reproduce the free quark result
(\ref{lqpi})
 at large Euclidean $Q^2$. 
Note that the polarization operator defined by Eqs. (\ref{tm2}) and
(\ref{newz}) has the correct  analytical structure: 
it is non-singular in the whole complex $Q^2$ plane, apart from the 
cut at real positive $q^2$. The singularities associated with the poles 
of the $\psi$ function lie on the unphysical sheet (Fig. 7).

\begin{figure}
  \epsfxsize=13cm
  \epsfysize = 7cm
  \centerline{\epsfbox{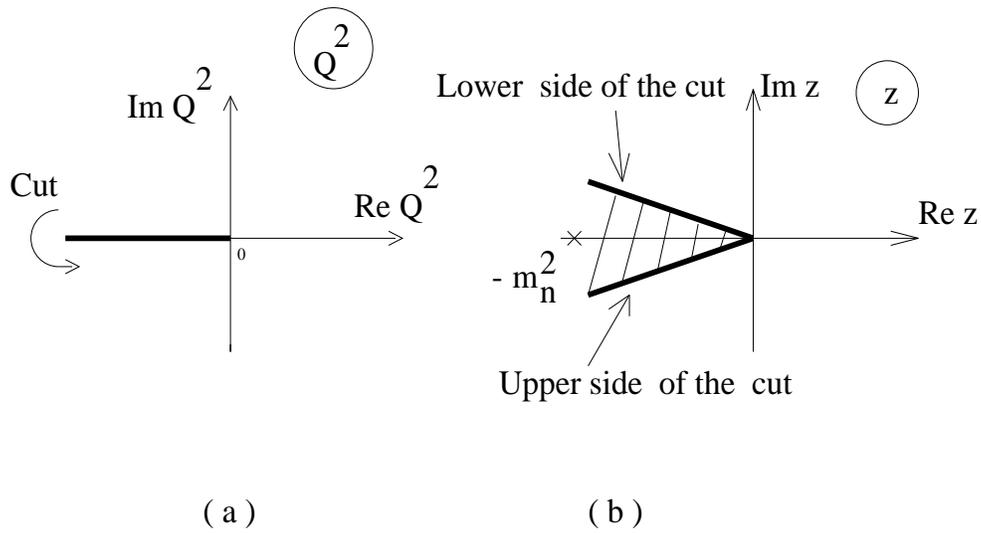}}
  \caption{ Analytical structure of the polarization operator. ($a$)
The polarization operator must be analytic everywhere in the 
complex $Q^2$ plane, except the cut running on the negative real 
semi-axis of $Q^2$ (positive real semi-axis of $q^2$). The imaginary 
part of the polarization operator must be positive at the upper side 
of the $q^2$ cut; ($b$) The mapping of the $Q^2$ plane onto the $z$ 
plane, Eq. (\ref{newz}). The physical sheet on the $Q^2$ plane 
corresponds to the $z$ plane with the shaded sector removed. The 
boundaries of the sector correspond to the lower and upper sides of 
the $q^2$ cut. The angle of the removed sector is proportional to
$\gamma$. }    
\end{figure}

The spectral density (i.e. the imaginary part of
$\Pi$ at $q^2 = s +i\epsilon$) takes the form
of the sum of (modified) Breit-Wigner peaks
\beq
\rho (s) = \frac{3}{\pi (1-\gamma )}\sum_{k=0}^\infty \frac{s^{1-
\gamma}\sin \pi\gamma }{[s^{1-\gamma}\cos \pi\gamma - (1+3k)]^2 
+[s^{1-\gamma}\sin \pi\gamma ]^2}\, .
\label{MBW}
\eeq
It is depicted in Fig. 8 for $B=0.6$ and $N_c=3$.
Although the choice of the 
parameter $B$ is rather arbitrary, it is not unreasonable; it 
corresponds to $\Gamma / m \approx 0.2$. Experimentally
the $\rho$ meson has a close width-to-mass ratio.

 We can see on the picture a pronounced $\rho$-meson peak,
accompanied by a deep dip, and then an almost smooth curve
which approaches the asymptotic value (unity) in an oscillating 
mode. The smearing of the $\delta$ functions into a smooth curve 
occurs because the width of the excited states is proportional to 
$m_n$ while the distance between two neighboring states
$\Delta m_n \sim 1/m_n$. It is assumed, of course, that $N_c$ is 
fixed.
The larger the value of $N_c$, the further we must go, to higher 
excitation numbers, for the resonances to overlap. Figure 8
shows that at $N_c = 3$ a smooth curve starts right after the ground 
state. The first excitation already belongs to the smooth curve.
Remember this pattern of the spectral density --
a conspicuous first peak accompanied by a smooth (and oscillating) 
curve  
approaching its constant asymptotic value. The gross features of this 
picture follow from very general theoretical arguments
\cite{Shif1}.
We will see shortly that experimental data, being different in fine 
details, exhibits the very same type of behavior. 

\begin{figure}
  \epsfxsize=13cm
  \centerline{\epsfbox{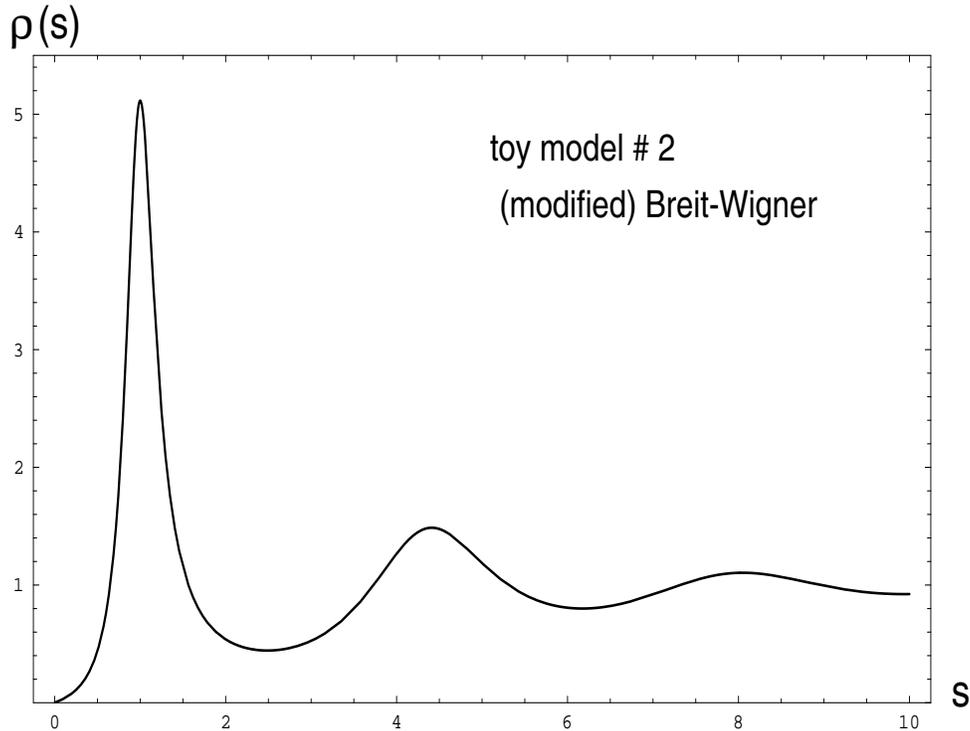}}
  \caption{ The spectral density corresponding to a sum of (modified)
Break-Wigner peaks with one and the same width-to-mass ratio 0.2.
All residues are set equal, and $\gamma \approx 0.06369$, see Eq. 
(\ref{MBW}). The $s$ axis is in the units $m_\rho^2$.}    
\end{figure}

The spectral densities shown in Figs. 4 and 8 at first sight do not  
produce 
an impression of close relatives. Let us integrate them over, with the 
exponential weight function, and compare $I(M^2)$ obtained in this 
way.  The comparison is shown in Fig. 9. The dotted curve 
corresponds to the comb of $\delta$ functions. It was already 
presented in Fig. 6. The solid curve  corresponds to Eq. 
(\ref{MBW}) and Fig. 8. Finally, the dashed curve is obtained with an 
extremely crude model of the spectral density depicted in Fig. 10,
\beq
\rho (s) = 3\delta (s-1) + \theta (s-s_0)\, , \,\,\, s_0 = 2.7\, .
\eeq
It presents an infinitely narrow $\rho$, plus a gap,  plus all higher 
excitations fused into ``continuum" which starts abruptly at $s=s_0$
and coincides (at $s>s_0$) with the asymptotic free-quark expression 
for the spectral density.
This is the original SVZ model \cite{SVZ}. It, obviously, captures only 
gross features, and misses all details: the fact that the gap at 
$s>m_\rho^2$ is not quite empty, the onset of continuum is not so 
abrupt, and the limit $\rho (s) \ra 1$ at $s\ra\infty$ 
is achieved through oscillations. Nevertheless, it is seen that
 $I(M^2)$ in all three models come out very close to each other.
The Breit-Wigner model of Eq. (\ref{MBW}) gives a little bit less 
steep fall 
off to the left of the window, at $M^2\ra 0$. This is quite 
understandable,
since in this model there is a non-vanishing spectral density 
at $s< m_\rho^2$, due to the $\rho$ meson width.  
 
\begin{figure}
  \epsfxsize=13cm
  \centerline{\epsfbox{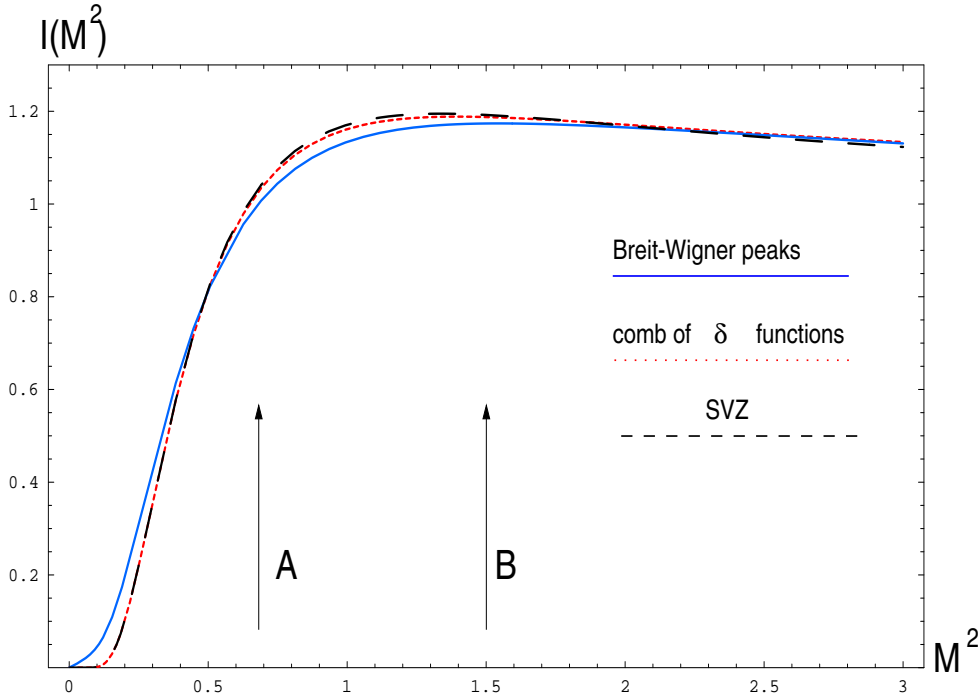}}
  \caption{$I(M^2)$ for three models of the spectral density.}
\end{figure}

\begin{figure}
  \epsfxsize=13cm
  \centerline{\epsfbox{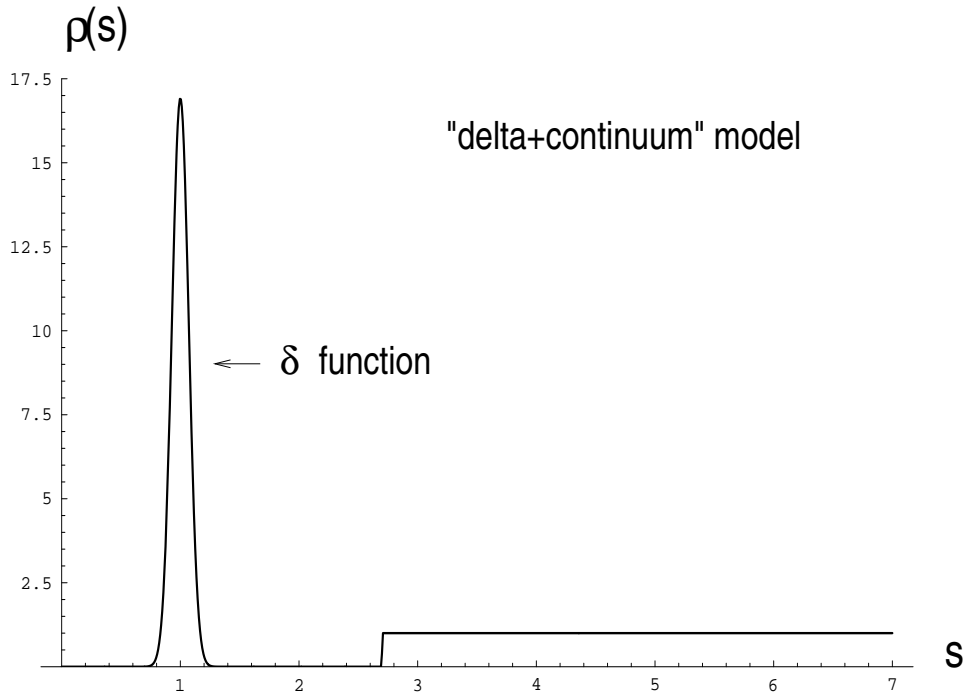}}
  \caption{The original SVZ model of the spectral density.}
\end{figure}

Comparison of three curves in Fig. 9 teaches us a lesson:
by examining $I(M^2)$,  calculated approximately inside the window,
one cannot expect to predict, with any reasonable accuracy, such fine 
structure of the spectral density as the width of the $\rho$ meson,
and the shape of the curve in the continuum (i.e. higher excitations).
As Fig. 9 shows, it is hardly possible to distinguish between
three models under discussion, which, in a sense, represent
extreme situations. 
The best one can hope for is estimating $m_\rho^2$, its residue,
and, to a lesser extent,  an effective value of $s$ where the dip 
following the 
$\rho$-meson
peak ends (i.e. $s_0$). 

In a broader context, this message  is  instructive for
the lattice practitioners too.
Doing calculations in the Euclidean domain -- a common feature of 
the sum
rules and lattice technology -- derives virtually no information about 
the
spectral density beyond the ground state in the given channel.
Indeed, the original SVZ model and the comb of the
$\delta$ functions, representing drastically different
$\rho (s)$ point-by-point,
 lead to $I(M^2)$ coinciding with each other to a high accuracy.
To distinguish between these two scenarios one needs exponential 
precision
going far beyond the level so far achieved in the lattice calculations.

Now we are ready to explore the sensitivity of $I(M^2)$ to the 
$\rho$-meson 
parameters. To this end we will take the SVZ model in the form
\beq
\rho (s) = C_1\delta (s-C_2) + \theta (s-s_0)
\label{origmod}
\eeq
and let the parameters float by, say, 10\% around their ``reference" 
values,
$C_1 = 3, \,\, C_2 = 1$ and $s_0 =2.7$. We then compare $I(M^2)$
obtained in this way with $I(M^2)$ emerging in our (most realistic) 
Breit-Wigner
toy model, see Eq. (\ref{MBW}).

Figure 11 illustrates the sensitivity of the sum rule to the $\rho$ 
meson residue (coupling constant).  The upper and lower dashed 
curves correspond to $C_1 = 3.3$ and 2.7, respectively.
Figure 12 illustrates the sensitivity to the $\rho$ meson mass.
The upper and lower dashed curves correspond to $m_\rho^2 = 0.9$ 
and $1.1$ of the experimental value, respectively.
It is seen that it is quite realistic to expect to get $g_\rho^2$ and 
$M_\rho^2$ with the accuracy  $\sim 10\%$ by inspecting $I(M^2)$ 
in the window and fitting the theoretical prediction by the model 
(\ref{origmod}).  The estimates of $s_0$ are  less accurate. 
One should not be surprised, of course, since the sum rules were 
designed to be most sensitive to the ground-state parameters and
relatively insensitive to the details of the spectral density in the 
continuum. The exponential weight in the definition of $I(M^2)$
takes care of this feature.  Deviations of $s_0$ from its reference 
value at the level of 10\% lead to rather insignificant changes in
$I(M^2)$. Deviations of $s_0$  at the level of 20\% are quite 
noticeable.

I hasten to add, however, that the 10\%  accuracy 
is not a typical outcome of the sum rule analysis.
The $\rho$-meson channel is  most favorable from the point of view 
of applications of the SVZ sum rules. 
In a sense, this is a dream case: the role of continuum with respect to 
$\rho$ is as tempered as it can possibly be,  and higher (unknown) 
condensates in the truncated condensate expansion show up  at 
remarkably low values of $M^2$ so that the working window is 
comfortably wide.  In many other channels, mesonic and baryonic,  
we have to deal with a narrower window. The general rule is: the 
narrower the window
the worse  the accuracy. In some channels, as we will see later,
the window shrinks to zero. Then  the SVZ method fails.
The reasons why it is successful in some cases and fails in others,
when properly understood, give us a unique hint as to the structure 
of the QCD vacuum (see Sect. 8). 

\begin{figure}
  \epsfxsize=12cm
  \centerline{\epsfbox{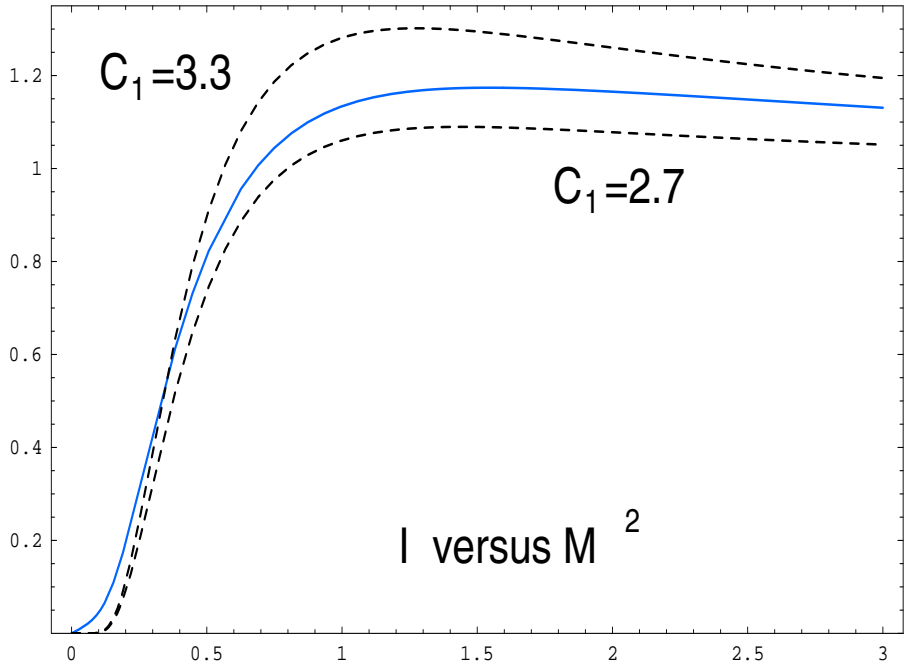}}
  \caption{Changing the $\rho$ meson residue. The solid curve 
corresponds to a ``reference" spectral density presented in Eq. 
(\ref{MBW}). The dashed curves correspond to Eq. (\ref{origmod})
with $C_1 = 3.3$ and 2.7, respectively. The reference value
is $C_1 = 3$.  }
\end{figure}

\begin{figure}
\epsfxsize=12cm
\centerline{\epsfbox{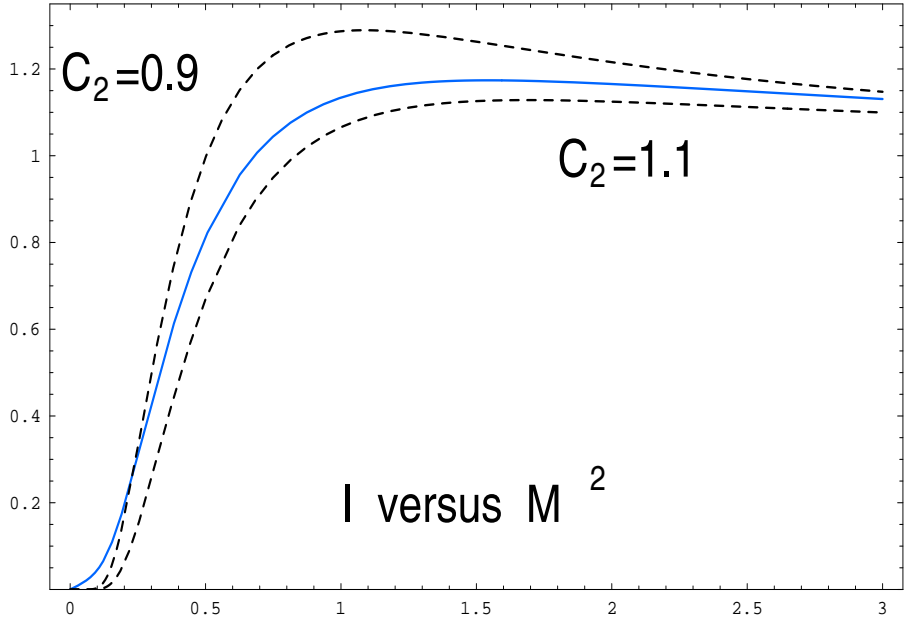}}
\caption{Changing the $\rho$ meson mass. The dashed curves 
correspond to the 10\% variation of $m_\rho^2$, around its reference 
value ($m_\rho^2=1)$. }
\end{figure}

\section{Vacuum Condensates}

It is high time to abandon our baby version of QCD and proceed to 
the real thing. What can be said about $I(M^2)$ now?

Needless to say that no analytic calculation of  $I(M^2)$ at
$M^2 \lsim \Lambda^2$, that would be based entirely on the first 
principles, exists. QCD at large distances is not yet  solved 
analytically.
Under the circumstances it seems reasonable to start at short 
distances and advance to larger distances ``step by step'', staying
on the solid ground -- working with the 
microscopic variables, quarks and gluons, where it is
fully legitimate.
At short distances the quark-gluon interactions are adequately 
described by perturbation theory. Certainly, as we 
have just learned by inspecting the toy models, the (truncated) 
{\em perturbative} series is not going to  yield us  any estimates 
relevant to the $\rho$ meson parameters. To get them we should
add at least some information regarding the large distance dynamics.

In the ideal world this information would be obtained from the 
theory {\em per se}. In the real world we may try to parametrize
the effects caused by the vacuum fields. If the quark-antiquark pair 
injected in the vacuum by the current $J_\mu$ does not propagate 
too far, its impact on the vacuum fields is, hopefully, not drastic.
This means that the polarization operator can be well approximated 
by the interaction of the valence quarks with a few vacuum 
condensates. For the validity of this assumption it is necessary that
the characteristic frequencies of the valence quarks inside the $\rho$ 
meson $\omega$ be  larger than the characteristic scale 
parameter of the vacuum medium $\mu$. Certainly, we can count 
only on an interplay of numbers: since the only dimensional 
parameter of QCD is $\Lambda$,
all quantities of dimension of mass are of order $\Lambda$.
It may well happen, however, that, say $\omega \sim 3\Lambda$
while $\mu\sim \Lambda /2$. In a sense, the success of the SVZ 
sum rules confirms this assumption {\em a posteriori}. 

On general grounds it is  expected that all gauge invariant Lorentz 
singlet local operators built of the quark and/or gluon fields develop
non-vanishing vacuum expectation values (VEV's). It is convenient to 
order all operators (and their VEV's) according to their normal 
dimensions. The higher the dimension, the higher the power of
$1/M$ of the corresponding coefficient. This fact allows one
to control the condensate expansion inside the working window.

The lowest dimension (zero) belongs to the unit operator.
The unit operator is trivial. When one calculates the perturbative
contribution to $I(M^2)$, one actually calculates the
coefficient of the unit operator.
 
The operator of the next lowest dimension (three) is the quark 
density 
operator,
\beq
{\cal O}_q = \bar q q\, .
\label{qdo}
\eeq
No other gauge and Lorentz invariant operators of dimension three
exist.

There is one operator of dimension four, the gluon operator
\beq
{\cal O}_G = \frac{\alpha_s}{\pi} G_{\mu\nu}^aG_{\mu\nu}^a
 \, .
\label{glop}
\eeq
The factor $\alpha_s /\pi $ appears in the definition naturally.
Since both operators, (\ref{qdo}) and (\ref{glop}), play a very
special role we will interrupt
our excursion before passing to higher dimensions, to make several 
important remarks.

In the chiral limit (i.e. when the quark mass term in the Lagrangian 
is put to zero) ${\cal O}_q$ is the order parameter -- its VEV signals 
the spontaneous breaking of the axial $SU(N_f)$ symmetry
and the occurrence of the corresponding  massless pions.
In perturbation theory the chiral
symmetry remains unbroken, of course,
$\langle {\cal O}_q\rangle$ vanishes 
identically. It is tempting to say then that $\langle {\cal 
O}_q\rangle\neq 0$ 
measures deviations from perturbation theory, and the same refers 
to
all other condensates.

In the early days of non-perturbative QCD such an understanding 
was widely spread. 
I hasten to say that this is a wrong understanding.
It is impossible to define the condensates as ``truly
non-perturbative'' residues obtained after subtracting from them a 
``perturbative part'' \cite{David}.
A heated debate took place in the literature in the eighties 
regarding the possibility of defining the condensates,
in a rigorous way,  by
subtracting a ``perturbative part".
 Glimpses 
of the debate
can be seen in Ref. \cite{NSVZ2} where it was clearly emphasized, for 
the first time, that the procedure of isolating
``purely non-perturbative" quantities does not (and must not)
work \footnote{Surprisingly, the
debate was revived recently, over a 
decade after the issue had been seemingly settled, in connection with 
the heavy quark theory. This theory, as it exists today, is also based 
on the operator product expansion and is a close relative \cite{BSUV}
of the SVZ sum rules. One can ask there the very same
questions that are usually raised in connection with the condensate
expansion in the SVZ method. Later on, in Sect. 10.2, we will return 
to 
the issue and discuss some elements of the heavy quark theory.}. 
I will outline  the proper procedure 
below using the gluon condensate as the most instructive example. 

Unlike ${\cal O}_q$, the gluon operator  ${\cal O}_G$
is not an order parameter. There is no known symmetry whose
spontaneous breaking would generate a vacuum expectation value 
 $\langle {\cal O}_G\rangle$. Correspondingly, $\langle {\cal 
O}_G\rangle$
is generated in perturbation theory. 
Physically,  $\langle {\cal O}_G\rangle$ measures the
vacuum energy density $\varepsilon_{\rm vac}$. To see that this is 
indeed the case let us use
the fact that the trace of the energy-momentum tensor
\beq
\theta_\mu^\mu = -\frac{b}{8}\, 
\frac{\alpha_s}{\pi} G_{\mu\nu}^aG_{\mu\nu}^a \, .
\label{aemt}
\eeq
This expression is nothing but  the scale anomaly of QCD 
\cite{Collins}.
Strictly speaking, the coefficient on the right-hand side contains
higher orders in $\alpha_s$; they are not  essential for our
purposes, however,  and  will be  ignored. The vacuum expectation
value
of the left-hand side is obviously $4\varepsilon_{\rm vac}$. Hence
\beq
\varepsilon_{\rm vac} =  -\frac{b}{32}\, \langle  {\cal O}_G\rangle\, .
\label{evac}
\eeq 
Everybody knows that the vacuum energy density badly  diverges
in field theory,  as
the fourth power of the cut-off \footnote{
Let me parenthetically note that in supersymmetric glyodynamics
the vacuum energy density vanishes, and so does the gluon 
condensate.
}. If so, how  can one make sense of
the gluon condensate?

The vacuum fields fluctuate; these fluctuations contribute to the
 vacuum energy density. High frequency modes of the fluctuating 
fields
belong to the weak coupling regime; their contribution in 
the correlation functions is
described by perturbation theory, giving rise to the standard
perturbative expansions.
What we are interested in are the low frequency modes, soft vacuum
fields that are responsible for the peculiar properties
of the vacuum medium. By saturating the condensate by the {\em 
soft modes
only}
we get a consistent definition
of the condensates that automatically solves the problem of 
divergencies. 

Specifically, we must introduce a somewhat artificial boundary,
$\mu$:
all fluctuations with frequencies higher than $\mu$ are supposed to 
be 
hard, those with  frequencies lower  than $\mu$ soft.
The parameter $\mu$ is usually referred to as {\em normalization 
point}.
Only soft modes are to be retained in the condensates.
Thus, the separation principle is ``soft versus hard''
rather than ``perturbative versus non-perturbative''.
This is the basic principle of Wilson's operator product expansion,
which, in turn, is the foundation of the SVZ sum rules. Being defined
in this way the condensates are explicitly $\mu$ dependent. All
physical quantities are certainly
$\mu$ independent; the normalization point dependence of the
condensates is compensated by that of the coefficient functions.

Generalities of
Wilson's approach and its particular implementation in QCD 
will be further discussed in Sects. 5 and 6. 
Here we complete our consideration of the gluon condensate.

It is instructive to trace how the gluon
condensate emerges from a slightly different perspective.
A natural starting point is the 
 diagram of Fig.  13  presenting the one-gluon correction
to
$\Pi (Q^2)$ (two other similar graphs with the one-gluon exchange
are not shown). For convenience we will deal with the so called $D$
function,
\beq
D(Q^2) = - (4\pi^2) Q^2 (d\Pi / dQ^2 )\, ,
\label{dfunct}
\eeq
rather than with the polarization operator $\Pi$ itself.
In terms of $D(Q^2)$ the bare quark loop of Fig. 5 becomes just unity.

\begin{figure}
  \epsfxsize=6cm
  \centerline{\epsfbox{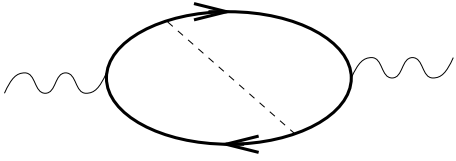}}
  \caption{The one-gluon correction to the $D$ function.}    
\end{figure}

If the gluon momentum is denoted by $k$
the one-gluon  term in $D(Q^2)$ takes the form
\beq
\Delta D(Q^2) =
\alpha_s \int k^2  dk^2\,  \frac{1}{k^2}\,  F(k^2,Q^2 )
\label{begren}
\eeq
where it is implied that the angular integration
over the orientations of the vector $k$ with respect to $q$
is already carried out. The factor $1/k^2$ comes from the gluon
propagator,
while $ F(k^2,Q^2 ) $ represents the rest of the graph.
The function $ F(k^2,Q^2 ) $
was calculated in Ref. \cite{Neubert1} (it can also be extracted from
Appendix B of Ref. \cite{BBB}),
\beq
F(k^2,Q^2 )= \frac{8}{3\pi}\frac{1}{Q^2}\, \left\{ \left(  \frac{7}{4}-
\ln\tau\right)\tau +
(1+\tau ) \left[L_2 (-\tau ) + \ln \tau \ln (1+\tau )   \right]
\right\}\, ,
\label{NBBB}
\eeq
where
$$
\tau = \frac{k^2}{Q^2}\, ,
$$
and
$$
L_2 (x) = - \int_0^x dy y^{-1}\ln (1-y )
$$
is the dilogarithm function.
The analytic continuation of the function
(\ref{NBBB}) to  $\tau > 1$ can be more conveniently rewritten as
$$
F(k^2,Q^2 )= \frac{8}{3\pi}\frac{1}{Q^2}\, \left\{ 1 + \ln\tau  + 
\left(\frac{3}{4} +
\frac{1}{2}\ln \tau   \right)\frac{1}{\tau} \right. +
$$
\beq
\left. 
(1+\tau ) \left[L_2 (-1/\tau ) - \ln \tau \ln (1+\tau^{-1} )   \right]
\right\}\, .
\label{NBBBprim}
\eeq
Substituting Eqs. (\ref{NBBB}) and (\ref{NBBBprim}) in
Eq. (\ref{begren})
and doing the integral over $k^2$ in the most straightforward 
manner
one  gets $\Delta D = \alpha_s /\pi$, the standard well-known 
one-gluon
correction to the $D$ function. 

The plot of the function $F(k^2,Q^2 )$ is shown in Fig. 14. 
Although it has a clear-cut peak at $\tau \sim 0.5$, it presents
a distribution spanning the entire range from $k^2 = 0$ to $k^2 =
\infty$.
The gauge coupling constant $\alpha_s$ in Eq.  (\ref{begren})
is literally a constant provided one limits oneself to the graph of
Fig. 13. In higher orders, however, $\alpha_s$ starts running, and
it is evident that $\alpha_s\ra \alpha_s(k^2)$. Then 
$\alpha_s(k^2)$ should be placed inside the integral in Eq. 
(\ref{begren})  rather than outside. By doing so we effectively 
resum a subset of the multi-gluon graphs.

\begin{figure}
  \epsfxsize=8.5cm
  \centerline{\epsfbox{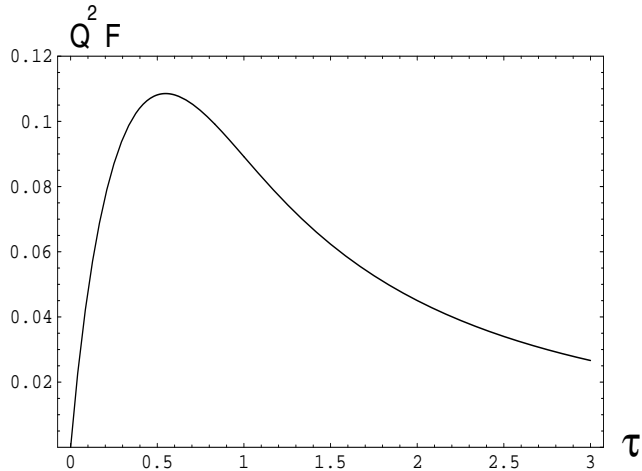}}
  \caption{The distribution function $Q^2F(k^2,Q^2)$ defined in Eq. 
(\ref{begren}) {\em versus} $\tau$.}    
\end{figure}

Certainly, putting 
\beq
\alpha_s(k^2)= \alpha_s(Q^2)\left[ 
1 + \frac{b \alpha_s(Q^2)}{4\pi}\, \ln \frac{k^2}{Q^2}
\right]^{-1}
\label{LLA}
\eeq
 inside the integral in
Eq. (\ref{begren})
we do not account for {\em all} higher-order corrections.
Only those are included that are responsible for the running of
$\alpha_s$ in the leading logarithmic approximation. Usually they 
say
that
the corresponding multigluon graphs are of the bubble chain type
(Fig. 15). Strictly speaking, in the covariant gauges
the gauge coupling renormalization (\ref{LLA}) in QCD (unlike QED) 
does not reduce to the bubble insertions in the 
gluon propagator depicted in Fig. 15. Various graphs of a more
complicated structure are involved too; they combine together to
produce a gauge-invariant expression (\ref{LLA}). The bubble chain
saturation
of Eq. (\ref{LLA}) takes place in the physical gauges, e.g.  in
the Coulomb gauge. In the covariant gauges
one can artificially reduce the problem to  the bubble
chains, 
without explicit identification of the full set of graphs,
by exploiting the so called ``large negative $N_f$ trick''.
We will not dwell on this technical issue, since it is irrelevant for our
purposes. The interested reader is referred to 
Refs. \cite{Neubert1,BBB}.

Even though the bubble chain is a very specific subset of graphs, and 
the majority of the multigluon graphs  are left aside 
in this procedure, it still makes sense to substitute the running 
$\alpha_s(k^2)$
inside the integral in Eq. (\ref{begren}), since in this way we take
into account an essential part of the underlying dynamics:
the fact that the quark-gluon interaction becomes weak if the gluon 
is 
far off-shell, and, on the contrary, becomes stronger as the gluon
virtuality decreases. If we  follow this strategy, then
$$
\Delta D(Q^2) =
 \int_0^\infty \alpha_s( k^2)  dk^2\,   F(k^2,Q^2 )=
$$
\beq
\alpha_s(Q^2)
\int_0^\infty dk^2 \left[ 
1 + \frac{b \alpha_s(Q^2)}{4\pi}\, \ln \frac{k^2}{Q^2}
\right]^{-1}\,  F(k^2,Q^2 )\, .
\label{begrenprim}
\eeq
We immediately run into a problem here. The running $\alpha_s$
has the Landau pole at small $k^2$, the square bracket
in the integrand explodes right inside the integration interval.
The explosion occurs at 
\beq
k^2 = Q^2 \exp \left[-\frac{4\pi}{b\alpha_s (Q^2)} \right]
=\Lambda^2\, .
\eeq
Although the function $F(k^2, Q^2)$ is suppressed at small $k^2$,
i.e. $\tau \ra 0$, 
(see Fig. 14), it by no means vanishes at $\tau = \Lambda^2 /Q^2$, 
rendering the integration
impossible. In order to do the $k^2$ integral
literally, as it is given in Eq. (\ref{begrenprim}),
we must say how the singularity at $k^2 = \Lambda^2$ is to be 
treated.
Some people say: ``let us take the principal value'', others bypass
the singularity by shifting the integration contour
 in the complex $k^2$ plane,
still others do something else. All these prescriptions 
are arbitrary and physically meaningless 
 for obvious reasons. Indeed, let us introduce the normalization point 
$\mu =$
 several units $\times\Lambda$ and split the integration over $k^2$
in two parts: from zero to $\mu^2$ in the first integral and from
$\mu^2$ to $\infty$ in the second.
Then at $k^2 > \mu^2$, in the weak coupling domain,
we do trust Eq. (\ref{begrenprim}) while below $\mu^2$ {\em we do 
not}.
At best, below $\mu^2$ one can use Eq. (\ref{begrenprim}) 
for the purpose of orientation. At small $k^2$
\beq
F(k^2, Q^2) \ra \frac{2}{\pi}\, \frac{k^2}{Q^4}\, ,
\label{flim}
\eeq
implying that
$$
\Delta_{\{k^2<\mu^2\}} D(Q^2) =
 \frac{2\alpha_s(Q^2)}{\pi}\, \frac{1}{Q^4}\, 
\int_0^{\mu^2}k^2  dk^2 \left[ 
1 + \frac{b \alpha_s(Q^2)}{4\pi}\, \ln \frac{k^2}{Q^2}
\right]^{-1} \sim
$$
\beq
 \frac{8}{b}\,\,  \frac{1}{Q^4}\,
\int_0^{\mu^2} k^2 dk^2 \, \frac{ \Lambda^2 }{k^2 - \Lambda^2}\, 
\sim  \, \frac{8\pi}{b}\, \frac{\Lambda^4}{Q^4}\, . 
\label{renprim}
\eeq

This is a very crude   estimate; I substituted the square bracket
by the corresponding pole, and then evaluated  the  integral
by equating it to its imaginary part, i.e. substituting 
$(k^2 - \Lambda^2)^{-1} \ra \pi \delta (k^2 - \Lambda^2)$. 

\begin{figure}
  \epsfxsize=7cm
  \centerline{\epsfbox{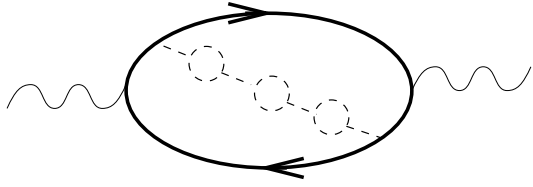}}
  \caption{One of the graphs from the bubble chain set.}    
\end{figure}

From this exercise we learn the following lesson.
The domain $k^2<\mu^2$ gives a contribution to $D(Q^2)$
which scales as $1/Q^4$. The coefficient in front of
  $1/Q^4$ cannot be reliably calculated  and must be  
parametrized by the gluon condensate -- that is the best
we can do. It is not accidental that the $k^2$ expansion of
$F(k^2,Q^2)$ starts from $k^2$, while the term of the zero-th order is 
absent. If it were not the case, the condensate series would 
have to start from ${\cal O}(Q^{-2})$. This is impossible, however, 
since no gauge invariant local operator of dimension two exists in 
QCD. The absence of such operator and the absence of terms
$k^0$ in the distribution function $F(k^2,Q^2)$ are in one-to-one 
correspondence. 

At first sight, one could try to circumvent
the problem of the Landau  pole inside the integration domain by
expanding
the square bracket in Eq. (\ref{begrenprim}) in the series in
$\alpha_s (Q^2)$. This is one of the ``remedies'' sometimes cited in 
this
context. I put the word remedy in the quotation marks since in fact it 
does 
not 
work. True, in every given order of the $\alpha_s$ expansion the 
integral
(\ref{begrenprim}) becomes well-defined. It is easy to see, however, 
that
the expansion is going to be factorially divergent in high orders.
Indeed,
\beq
\int_0^{\mu^2} k^2 dk^2 \left( \ln \frac{Q^2}{k^2}\right)^n
\sim 2^{-n} n!\, , \,\,\, n\gg \ln \frac{Q^2}{\mu^2}\, .
\eeq

It is  seen that perturbation theory {\em per se} carries the
seeds of the gluon condensate. The perturbative series cannot be
consistently defined unless the gluon condensate is introduced.
Once we cut off the $k^2<\mu^2$ tail of the integral from the
perturbative expression obtained from Eq. (\ref{begrenprim}),
the factorial divergence of the $\alpha_s$ series at high $n$ 
disappears,
the series becomes well-defined, and so is the gluon condensate.

The factorial divergence of the perturbative series of the type 
we have just
discussed
is called the {\em infrared renormalon} \cite{Hooft2,renor} for the 
reasons 
which need 
not
concern us here. It is associated  with the
bubble chain graphs of Fig. 15 and is inevitable if one forces the
perturbative integrals to run all the way down to $k^2 = 0$.
Bounding the integration interval in the perturbative part from 
below by $\mu^2$
we get rid of the infrared renormalon altogether.
The domain below $\mu^2$ 
is not lost; it is fully represented by the gluon
condensates.
Higher condensates  appear as higher order terms in the $k^2$ 
expansion of
$F(k^2, Q^2)$.
The intimate relationship between the infrared renormalons
and condensates was first revealed by Mueller \cite{Mueller}.

I hasten to warn that one should not literally equate the infrared 
renormalons to  the condensate expansion.
Even if one sticks to a certain particular definition for summation of 
the bubble chains, that eliminates ambiguities, say the principal 
value prescription in the integral (\ref{begrenprim}), one typically 
gets a numerical value of the $Q^{-4}$ term which is grossly off 
compared to the gluon condensate,
let alone ambiguities associated with various possible choices of the 
regularization prescription. For instance,
in the case at hand,
\beq
D(Q^2) = 1 + \frac{2\pi^2}{3 Q^4}\langle {\cal O}_G \rangle + ...\, ,
\label{addfla}
\eeq
and the ${\cal O}(1/Q^4)$ term is $0.08\,\, \mbox{GeV}^4/Q^4$
under the standard choice of the gluon condensate (see below).
Equation (\ref{renprim})  yields, on the other hand, only 
$0.004\,\, \mbox{GeV}^4/Q^4$ at $\Lambda = 0.2$ GeV
($0.01\,\, \mbox{GeV}^4/Q^4$ at $\Lambda = 0.25$ GeV).
Within the SVZ method, based on Wilson's expansion, the role of the 
infrared renormalons is purely illustrative. They show, in a very 
straightforward manner, that without the condensates the 
perturbative series cannot be defined and, on the contrary,
introducing  the condensates  one eliminates ambiguities
associated with the high-order tails of the perturbative series
\footnote{The infrared renormalon is not the only source of the 
factorial divergence of the $\alpha_s$ series in QCD. The $\alpha_s$ 
expansion of Eq.  (\ref{begrenprim})  
contains
also the so called ultraviolet renormalon -- factorial divergence
of high-order  $\alpha_s$ terms coming from the large $k^2$ domain.
This factorial divergence is readily summable, however, 
as is perfectly clear from the unexpanded expression 
 (\ref{begrenprim}). Indeed, the integrand has no singularities at
large $k^2$ and is well convergent. Below, the  ultraviolet 
renormalons will 
never be mentioned again. Those readers who want to familiarize
themselves with this subject are referred to an excellent review 
\cite{Zakharov}. A factorial divergence of a different nature
showing up in the coefficient functions is briefly discussed in Sect. 
5.}.

On the other hand, in  some (perturbatively infrared stable) 
processes that 
have {\em no} operator product expansion,
most notably in the jet physics, the infrared renormalons,
in spite of  all their obvious  limitations, become poor man's 
substitute for the 
condensate expansion. Although, unlike OPE, the renormalon analysis 
does not provide us with the proper numerical coefficients,
we may infer from the renormalons what powers of $1/E$ (or $1/Q$) 
appear in the nonperturbative parts of such quantities as, say, thrust.
Whether the nonperturbative corrections are
${\cal O}(1/E)$ or   ${\cal O}(1/E^2)$ is clearly a question of 
paramount importance for phenomenology.
Using the infrared renormalons for counting the powers of $1/E$ in 
the processes without OPE is a totally fresh idea which can be 
considered as a distant spin off of the SVZ method, see Ref. 
\cite{RENPHEN} for a review. Previously the nonperturbative 
corrections were just ignored since nobody knew how to approach 
the 
issue scientifically.

Now, when the meaning of the  condensates appearing in Wilson's 
expansion is hopefully clear, we can proceed to a brief discussion of 
their numerical values.  In principle,  two alternative 
approaches are conceivable: (i) calculating the condensates from first 
principles (e.g.
on the 
lattices); (ii) extracting the condensates from the sum rules 
themselves. Attempts of the lattice calculation of the condensates 
were reported in the literature (for a review see \cite{ADG1}).  
The main difficulty, which, to a large extent, devaluates this work is 
the fact that the strategy of ``isolating a non-perturbative residue 
from the perturbative background" was adopted. As we already 
know,
this strategy is doomed to failure.  Instead, one should have 
consistently implemented Wilson's procedure. This has never been 
done, however.
Some initial ideas as to how Wilson's procedure can be implemented 
on the lattices are presented in Ref. \cite{Ji1}. 

Therefore at present, as twenty years ago, one  has  to exploit  the
sum rule themselves in order to determine  the condensates. 
One picks up certain channels where sufficient experimental 
information on the corresponding spectral densities is available.
These channels are sacrificed, i.e. the analysis goes  in the direction 
opposite   to  conventional: the values of the
condensates are extracted from the  Euclidean correlation functions 
obtained through dispersion integrals. The first determination of the 
gluon condensate was carried out exactly in this way,
from the sum rules in the $J/\psi$ channel \cite{VolSVZ,SVZ}. 
The corresponding value will be referred to as standard,
$$
\langle {\cal O}_G\rangle = 0.012 \,\,\mbox{GeV}^4\, .
$$
Since then this estimate was subject to multiple tests, sometimes
with conflicting results. A brief discussion can be found in the 
beginning of Sect. 2 in Ref. \cite{RV}. The overall conclusion is that 
certain deviations 
from the standard value
(say, at the level of $\sim 30\%$)  are  not ruled out \footnote{I 
leave aside extremist
and to my mind unfounded
statements in the literature, that the standard value underestimates 
the gluon 
condensate by a factor of 2 to 5.}.  One of the reasons limiting 
the accuracy is the fact that the gluon condensate was determined
within a simplified (the so called practical) version of the operator 
product expansion, which is strictly speaking somewhat ambiguous.
We will dwell on the issue in Sect. 6. At the present level of 
understanding
it is highly desirable to repeat the analysis within the framework of 
consistent Wilson's procedure. This is not  done so far.
To take into account a certain degree of numerical uncertainty
it is reasonable to accept that
\beq
\langle {\cal O}_G\rangle = 0.012 \, x_G\,\,\,\mbox{GeV}^4\, ,
\label{GCX}
\eeq
where a dimensionless numerical factor $x_G$ is allowed to float
in the vicinity of unity. This parametrization will be used below in 
the $\rho$-meson sum rule.  

Let us pass now to the quark 
 condensate
$\langle {\cal O}_q\rangle$. As was mentioned, this condensate
is the order parameter for the spontaneous breaking of the chiral 
symmetry. Hence, its value can be independently determined from 
the  corresponding phenomenology.
In particular, the celebrated Gell-Mann--Oakes--Renner
formula relates the product of the light quark masses and $\langle 
{\cal O}_q\rangle$ to observable quantities,
\beq
(m_u+m_d) \langle \bar u u + \bar d d\rangle = -M_\pi^2 f_\pi^2\, ,
\label{GMOR}
\eeq
where $f_\pi$ is the pion constant ($f_\pi\approx 133$ MeV)
and $M_\pi$ is the pion mass.  The original estimate
of $\langle {\cal O}_q\rangle$  used  in Ref. \cite{SVZ}
was  obtained  by substituting  the crude estimates of the light quark 
masses that existed at that time \cite{HL1,Wei},
$$
m_u+m_d \approx 11\,\, \mbox{MeV}\, .
$$
In this way one arrives at
\beq
\langle 
{\cal O}_q\rangle = - (250\,\,\mbox{MeV})^3\, . 
\label{svqc}
\eeq
Since then considerable work has been carried out to narrow down
the theoretical uncertainty. First of all,  chiral corrections to 
the Gell-Mann--Oakes--Renner
formula were derived and analyzed. Moreover, 
much effort has been invested in perfecting our knowledge of the 
quark masses.  The progress is summarized in Ref. \cite{HL2}, where
an extensive list of references can be found. 
No dramatic changes occurred. The value of
$m_u+m_d $ extracted from phenomenology of the chiral symmetry 
breaking went up by about 30\%, with an error at the level of 15\%.
Thus, one is tempted to say that the actual value of the quark 
condensate is close to its ``standard"
value quoted in Eq. (\ref{svqc}); perhaps, 
somewhat suppressed \footnote{The quark mass is definition 
dependent.  Usually the
$\overline{\mbox{MS}}$ scheme and a ``reference" normalization 
point $\mu = 1$ GeV are implied. Not to confuse the reader we will 
follow this convention,
in spite of shortcomings of the 
$\overline{\mbox{MS}}$ scheme.  One cannot avoid explicitly 
introducing the normalization point $\mu$ in the case of the quark 
condensate:
 ${\cal O}_q$ has a non-vanishing anomalous (logarithmic) 
dimension. The $\bar q 
q$ 
vertex dressed by gluons is 
logarithmically suppressed. Including the gluons with off-shellness
from $M_0$ down 
to $\mu$ suppresses  the vertex by a factor $\propto
[\alpha_s (\mu )/\alpha_s (M_0)]^{4/b}$. 
This formula  explicitly demonstrates why $\mu$ cannot be put to 
zero in the case at hand:
the logarithm explodes. 
}. 

Events took a dramatic turn in 1996, after  results of the lattice 
calculations of the light quark masses (with two dynamical quarks) 
became known.  The lattice result for $m_u+m_d $ lies significantly  
lower (by a factor of two to three, see e.g. \cite{BG})  than any of the 
reasonable analytic estimates!
Were these results confirmed, this would mean that the quark
condensate is significantly larger than the number given in Eq. 
(\ref{svqc}). 

I would caution against hasty conclusions at this point. 
The discrepancy between the analytic and lattice estimates
may well be an artifact of the lattice procedure. Putting dynamical 
almost massless quarks on the lattice is a notoriously difficult task. 
On the other hand, the discrepancy, if persists, may prove to be a  
signature of a very interesting physical phenomenon --
strong dependence of physical quantities on the number of light
dynamical quarks.  The surprisingly low lattice result
for $m_u+m_d $
was obtained under the assumption that the number of light quarks 
is two, while in fact it is three.  We will return to the issue
of possible strong dependence on $N_f$ in Sect. 12. 

Under the circumstances it seems reasonable, analogously to the 
gluon 
condensate,
to parametrize the quark condensate as
\beq
\langle 
{\cal O}_q\rangle = - (250\,\,\mbox{MeV})^3\, x_q\, ,
\label{svqcprim}
\eeq
and let $x_q$ vary  around unity. 

The full catalog of all relevant operators
up to dimension six, was worked out
in Ref. \cite{SVZ}.
It is not as large as one might think at first sight.
There is only one operator of dimension five,
the mixed quark-gluon operator
\beq
{\cal O}_{qG} = {ig}\, \bar q \sigma_{\mu\nu} G_{\mu\nu} q
\label{qgop}
\eeq
where $ G_{\mu\nu}\equiv  G_{\mu\nu}^a T^a$ ($T^a$ are the color
generators).
This operator plays no role in the $\rho$-meson sum rules,
to be considered below: because of its ``wrong'' chirality it
can enter only being multiplied by the quark mass $m_q$.
It is very important, however, in a wide range of problems
involving baryons \cite{Ioffe}, and mesons built from one light and
one heavy quarks \cite{NovConf}.

At the level of dimension six, there is one operator built
from three gluon field strength tensors (symbolically $GGG$)
and several four-quark operators of the type
\beq
{\cal O}_{4q} = (\bar q_1 \Gamma_1 q_2)( \bar q_3 \Gamma_2 q_4) 
\label{op4q}
\eeq
where $\Gamma_{1,2}$ denote certain combinations of the
Lorentz and color matrices and $q_i$ ($i = 1,...,4$)
are the light quark fields of different flavors, $u,d$ and $s$.
The overall flavor must be conserved of course, but $q_1$ need not
coincide with $q_2$ and so on.

The three-gluon operator is expected to have a significant impact in 
heavy quarkonium
\cite{MiVo}; it  does not appear in the
 $\rho$-meson sum rules, as was first shown in Ref. \cite{DubSm},
 by virtue of a very
elegant theorem. 

As for the four-quark operators,
their vacuum  matrix elements are not known independently.
It became standard to evaluate them in the
factorization approximation, which was first applied in this context
in \cite{SVZ}. Note that using the Fierz identities for the color and 
Lorentz
matrices
 one can always
arrange
the operators ${\cal O}_{4q}$ to  have a natural color flow. By natural
I mean that $\Gamma_{1,2}$ do not contain color matrices, and the
color indices of quarks are contracted inside each bracket in
Eq. (\ref{op4q}).
The color of $q_1$ flows to $q_2$ and that of $q_3$ flows to $q_4$.
If the number of colors $N_c$ were large, then
the expectation value of ${\cal O}_{4q}$ would be totally saturated
by the vacuum intermediate state,
\beq
\langle {\cal O}_{4q}\rangle = \langle \bar q_1 \Gamma_1 
q_2\rangle
\langle \bar q_3 \Gamma_2 q_4 \rangle \, .
\label{vacsat}
\eeq
Corrections to this equation are formally of order of $1/N_c^2$. It is 
clear that in the factorization approximation, after  the natural color
flow is achieved by the Fierz rearrangement, the only surviving
structure is that with $\Gamma_{1,2}=1$.

In the real world $N_c=3$, and we certainly do expect deviations 
from
factorization at a certain level. As far as we can tell
today, factorization works surprisingly well at least for those 
four-quark
operators that appear in the $\rho$-meson sum rule. All attempts
to detect deviations, both on the lattices \cite{mart} and in the sum 
rules
themselves \cite{DFS},
gave results consistent with zero, within errors that typically lie
in the 10\% ballpark. This is quite nontrivial, considering the fact
that
 quite a few examples are known where large numerical coefficients
neutralize formal $1/N_c$ suppression factors. Whatever the reasons
might be, we will accept Eq. (\ref{vacsat}) in what follows.
  
\section{$\rho$ Meson in QCD}

Experience accumulated in Sect. 2 will now be applied 
for explorations  in actual  QCD.
As we learned from the toy models, the polarization operator
$\Pi (Q^2)$ (or its Borel transform $\tilde\Pi (M^2)$) in the deep 
Euclidean domain has an 
expansion  in $\ln Q^2$ (perturbation theory)
plus   power terms $(1/Q^2)^k$ or $(1/M^2)^k$
(the condensates). If in the toy models the power terms
(truncated in certain order) can be made as large or as small
as we want, in QCD they are determined dynamically; through 
interactions of 
the valence quark pair  with the
soft vacuum fields. This interaction gives rise to the condensate 
expansion. The first power term is due to the gluon condensate, while 
the next one is associated with the four-quark condensate.
In principle,  higher condensates were analyzed in the literature 
too, but we will 
not discuss them in this Lecture.

The quark-antiquark pair $q\bar q$ with the total energy $\sqrt{s}$ 
is injected in the QCD vacuum either by the virtual photon ($e^+e^-$ 
annihilation) or $W$ boson ($\tau$ decays). The $q\bar q$ pair, 
being injected, starts 
evolving according to the dynamical laws of QCD. At first the quarks 
do not feel the impact of the vacuum ``medium". As separation 
between them grows, the effects of the medium become more and 
more important, so that eventually it  prevents quarks from 
appearing in the detectors. The injected quarks get dressed and 
materialize themselves in the form of  hadrons. In the sum rule 
approach we control only the beginning of this process. 

In the sum rule framework,  the condensates become important  in 
the 
vicinity of  the 
left edge of the window. To the left of the arrow $A$ they explode, 
and theoretical control is lost. On the other hand, the value of 
$I(M^2)$
near the right edge of the window is determined mostly by ordinary 
perturbation 
theory. All perturbation theory sits in the unit operator.
Technically it is convenient to formulate the corresponding 
calculation in terms 
of the perturbative corrections to the spectral density. We take the 
graph of Fig. 5, and 
attach to it various gluon and quark loops (e.g. Figs. 13 and 15). 
In this way we get the perturbative part of $\rho (s)$; at present in 
the vector isovector channel it  is known  
up to  third order in $\alpha_s (s)$ \cite{RPT}. In the 
$\overline{\mbox{MS}}$
scheme 
the result takes the form 
\begin{equation}
\rho (s) =1 + k_1 a + k_2 a^2 + k_3 a^3
\label{R}
\end{equation}
where 
$ a(s)$ is given  in Eq. (\ref{deflam}). 
 In the low-energy domain the number of active flavors
is $N_f=3$.  For $N_f= N_c =3$ 
the parameter $b=9$, and the values of $k_i$ are
\begin{equation}
k_1= \frac{4}{9}\,,\;\;\; k_2=0.729 k_1\,, \;\;\; k_3=-2.03 k_1\, .
\end{equation}

Equation (\ref{R}) can be immediately translated into the prediction 
for the perturbative 
part of
$I(M^2)$. The perturbative expansions
for $I(M^2)$ and $\rho (s)$ do not coincide. To get the former one 
must
integrate the latter with the exponential weight function. 
The transition is readily carried out with the
aid of the formula \cite{EKSV}
\begin{equation}
\int {ds\over M^2} \frac{\exp(-s/ M^2)} {\left(\ln \frac
{s}{\Lambda^2}\right)^\nu} = 
\frac{1}{\left(\ln \frac {M^2}{\Lambda^2{\rm 
e}^\gamma}\right)^\nu} 
\left[ 1+ \frac{\nu(\nu +1)}{2} \frac{\pi^2/6}{\left(\ln \frac {M^2}
{\Lambda^2{\rm e}^\gamma}\right)^2}
+{\cal O}\left(\frac{1}{\left(\ln \frac {M^2}
{\Lambda^2{\rm e}^\gamma}\right)^4}\right)\right]\, ,
\end{equation}
where  $\gamma$ denotes Euler's constant, $\gamma=0.577$.
Expanding this expression near integer values of $\nu$ one  also gets
necessary expressions 
for the integrals containing logarithm of logarithm ($\ln \ln$).

In this way we arrive at \cite{EKSV}
\begin{equation}
I_{\rm pc} (M^2) = 1+  k_1 a \left( \frac{M^2}{{\rm 
e}^\gamma}\right) + 
k_2 \left[ a \left( \frac{M^2}{{\rm e}^\gamma}\right)\right] ^2 + (k_3 
+ \frac{\pi^2}{6}
k_1)\left[ a \left( \frac{M^2}{{\rm e}^\gamma}\right)\right] ^3\, .
\label{delpc}
\end{equation}
Similar expressions were obtained in Ref. \cite{Kata1}, were
 $a ( {M^2}/{{\rm e}^\gamma}) $ was reexpressed in terms of $ a 
(M^2) $.
It is  more convenient  to work directly
with the expansion (\ref{delpc}). 

The quark mass corrections in $\rho (s)$ and $I(M^2)$ are also 
known,
they are proportional to $m_{u,d}^2/M^2$ and are negligibly small
due to the smallness of the current quark masses. They will be 
discarded 
since the corresponding uncertainty is invisible in the background of 
other 
uncertainties.
 
Now we return to the condensates. 
The modern  perturbative calculations of the 
coefficients $C_n$ are based on the background field technique. 
This is an important aspect of the SVZ approach as it exists today. 
The method is quite 
elegant, to say nothing that it is much more
economic than the original
``brute force" calculations of Ref. \cite{SVZ}. 
We will not discuss the technique {\em per se}, however. 
Conceptually everything is clear 
here. If you need to do a calculation, you just go and learn the 
corresponding technique using, for instance, the review paper
\cite{NSVZBFT}. 

The coefficients $C_G$ and $C_{4q}$, to the leading order, are known 
from the ancient times \cite{SVZ},
$$
I_{\rm npc} (M^2) = 1 + \frac{\pi^2}{3M^4}\, \langle{\cal O}_G\rangle 
-
\frac{8\pi^3}{M^6}\, \left \langle\alpha_s \left( \bar 
u\gamma_\alpha\gamma_5 T^a u - \bar d\gamma_\alpha\gamma_5 
T^a d
 \right)^2 \right\rangle -
$$
\beq
\frac{16\pi^3 }{9 M^6 }\, \left \langle
\alpha_s \left( \bar u\gamma_\alpha T^a u  + \bar d\gamma_\alpha 
T^a d
 \right)\, \sum_q \left(\bar q\gamma_\alpha T^a q\right)  
\right\rangle \, . 
\label{ancient}
\eeq
The four-quark structures appearing in the angle brackets, as well 
as the factor $\alpha_s$ in front of them, 
are normalized at $ Q$. The anomalous (logarithmic) dimension of 
this particular combination nearly vanishes \cite{SVZ}; we will ignore 
it
in evolving the four-quark operator at hand down to $\mu$. 
Applying factorization, as was explained in Sect. 3,  we get for the
four-quark term in $I_{\rm npc} (M^2) $
\beq
C_{4q} \langle {\cal O}_{4q} \rangle = 
-\frac{448\pi^3}{81M^6}\, \alpha_s (\mu ) \langle \bar qq (\mu 
)\rangle^2 \ra -\frac{1}{M^6} \, 0.03\,  x_{4q} \, \, \mbox{GeV}^6\, ,
\eeq
where, as previously, we have introduced a factor $x_{4q} $
to allow for possible theoretical uncertainties. Under the standard 
choice of parameters \cite{SVZ} this factor is equal to unity.
If we are on the right track it is reasonable to expect that
$x_{4q} $ is close to unity. 

A huge amount of  work  was 
carried out  in calculating
two-loop corrections in the coefficients of various  condensates. 
The results can be inferred, say, from Ref. \cite{sursa}
or from the reviews \cite{BGr}. 
 Unfortunately, the existing uncertainty in  the condensates 
themselves precludes us from taking advantage of 
the precision achieved in perturbative corrections. Therefore, we will 
limit ourselves to the leading order results for $C_G$ and $C_{4q}$.

Now, the stage is set and we can finally present the 
sum rule for $I(M^2)$,
$$
I(M^2)=1 + \frac{4}{9} \, a \left( \frac{M^2}{{\rm e}^\gamma}\right)\,
\left\{ 1 + 0.729\, \left[ a \left( \frac{M^2}{{\rm 
e}^\gamma}\right)\right]  -0.386\,
\left[ a \left( \frac{M^2}{{\rm e}^\gamma}\right)\right] ^2  \,\right\} 
\, +
$$
\beq
 0.1 x_G \left( \frac{0.6}{M^2}\right)^2 - 0.14 x_{4q}
\left( \frac{0.6}{M^2}\right)^3\, ,
\label{SRNP}
\end{equation}
where $M$ in the second line of Eq. (\ref{SRNP}) is measured in GeV. 
The term $M^{-4}$ 
is due to the 
gluon 
condensate, while the term $M^{-6}$ is due to the quark condensate.
With the 
``standard" numerical values of the gluon and quark condensates
in the 
the nonperturbative part  $x_G= x_{4q}= 1$; 
 the dimensionless constants $x_G$ and $x_{4q}$ 
allow the gluon and four-quark condensates to ``breathe"
to a certain degree. For instance, $x_G >1$ would imply that
the actual value of the gluon condensate is larger than
the ``standard" value, and so on.  

We are ready to examine 
the situation in the $\rho$-meson channel. Figure 16 shows 
experimental data on the spectral density in the vector isovector 
channel,    measured in the 
$\tau$ decays. The data points are from Ref. \cite{ALEPH}.
Note a remarkably close resemblance with our toy model number 
two.
If we take the beginning of Fig. 8 and expand it
to make the energy scales coincide \footnote{Warning:
the energy scale (the horizontal axis) in Fig. 8 presents $s$  in 
the units of 
$m_\rho^2$ while that in Fig. 13 is in GeV$^2$.} 
the curves will essentially repeat each other. There exist  a few extra 
points in the $\tau$ decays, spanning the interval of $s$ from 2.7 to 
3 GeV$^2$, and some data above 3 GeV$^2$ from  the $e^+e^-$
annihilation, but the corresponding  error bars are so large
that plotting these points would just obscure the picture.

\begin{figure}
\epsfxsize=14cm
\centerline{\epsfbox{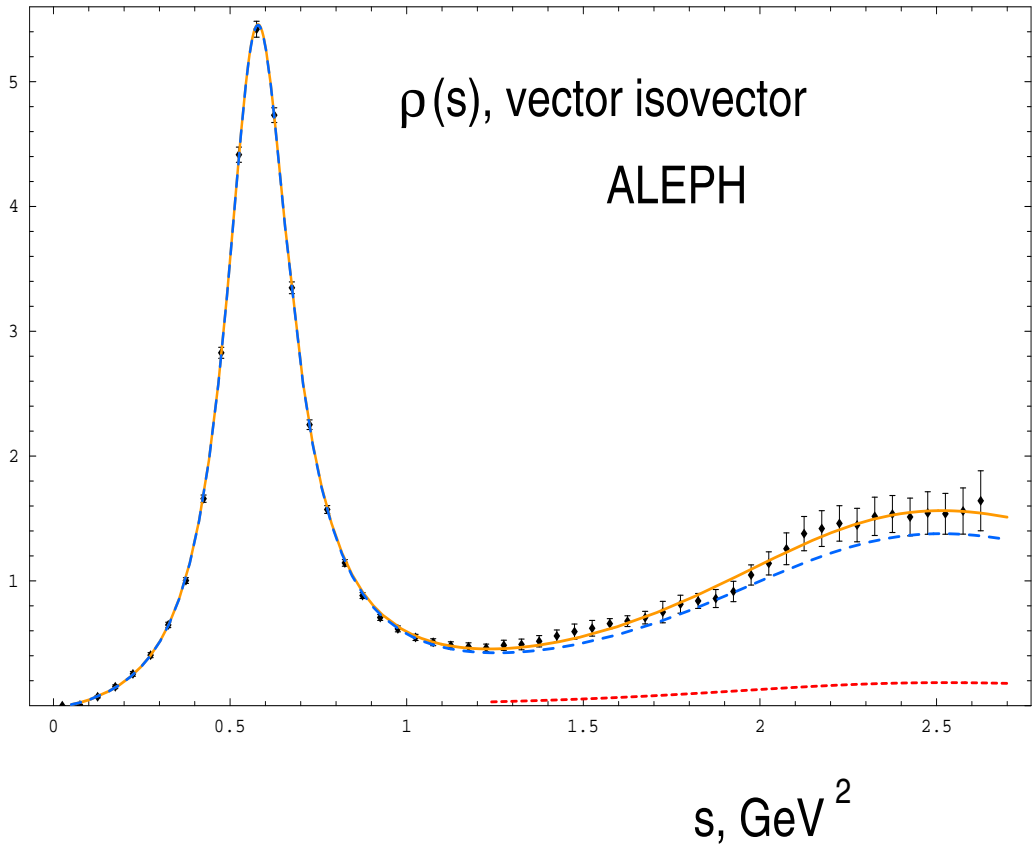}}
\caption{The spectral density in the vector isovector channel 
measured in $\tau$ decays. The data points belong to  
ALEPH. The solid curve is a best fit. The dashed curve is the  best fit 
minus $ K K \pi $   (the latter contribution is 
shown  by the dotted curve) which, I think, should not have  been 
included in the vector spectral density, see text. }
\end{figure}

There is a subtle point which I have to discuss here.
Directly measurable in the hadronic $\tau$ decays is the sum of the 
vector and axial spectral densities. To obtain the spectral densities 
separately one has to sort out all decays by assigning
specific quantum numbers to  each given final hadronic state.
In the majority of cases such an assignment is unambiguous.
For instance, two pions (whose contribution is the largest)
can be produced only by the vector current. Some processes, 
however, can occur in both channels, for instance, the $KK\pi$ 
production. Using certain theoretical arguments it was decided
in Ref. \cite{ALEPH} that around 3/4 of all $KK\pi$ yield
should be ascribed to the vector channel. Other theoretical arguments 
\cite{EKSV},
which seem to be much  more convincing to me,
tell that virtually all $KK\pi$ production should take place in the 
axial 
channel. 
Therefore, in dealing with the data, I will subtract
the $KK\pi$ yield from the ALEPH data points. I hasten to add that  
the subtracted quantity is small (see Fig. 16), 
and  this subtraction is not very essential for a general picture I 
draw here by broad touches. 

The solid curve in Fig. 16 is a best fit by a smooth curve
representing the sum of two Breit-Wigner peaks (the first one is a 
modified Breit-Wigner taking into account threshold effects 
important for the $\rho$ meson). The dashed curve is what I believe
the actual spectral density is. The data points above 1 GeV$^2$
carry a noticeable  uncertainty. Since we are going to use the spectral 
density only in the integrals, where many data points are summed 
over, these individual errors can be neglected, since they will be 
statistically insignificant in the integrals. Systematic uncertainty 
might be important, but since nobody knows how to estimate it, I 
will ignore it for the time being. For our limited
purposes we can consider 
the dashed curve in Fig. 16 as an exact experimental  result for the 
spectral density. 

How does the experimental spectral density might look above 2.7 
GeV$^2$? This question is irrelevant for calculating the 
$\rho$-meson parameters. It is very instructive, however, to have a 
broader perspective.
 I will show you what I call
a ``theoretical experimental" spectral density, and then will make
a short digression explaining where the answer comes from. 

My expectations are summarized in Fig. 17. The  curve to the left of 
the arrow is just the fit to the experimental data, as explained above
(it identically coincides with the dashed curve in Fig. 16.)
To the right of the arrow I have attached a ``tail" which approaches a 
smooth asymptotic prediction for $\rho (s)$ (the dashed line) given 
in Eq. (\ref{R}), in an oscillating manner. The tail 
and the experimental data are smoothly matched at 2.65 GeV$^2$.

\begin{figure}
\epsfxsize=14cm
\centerline{\epsfbox{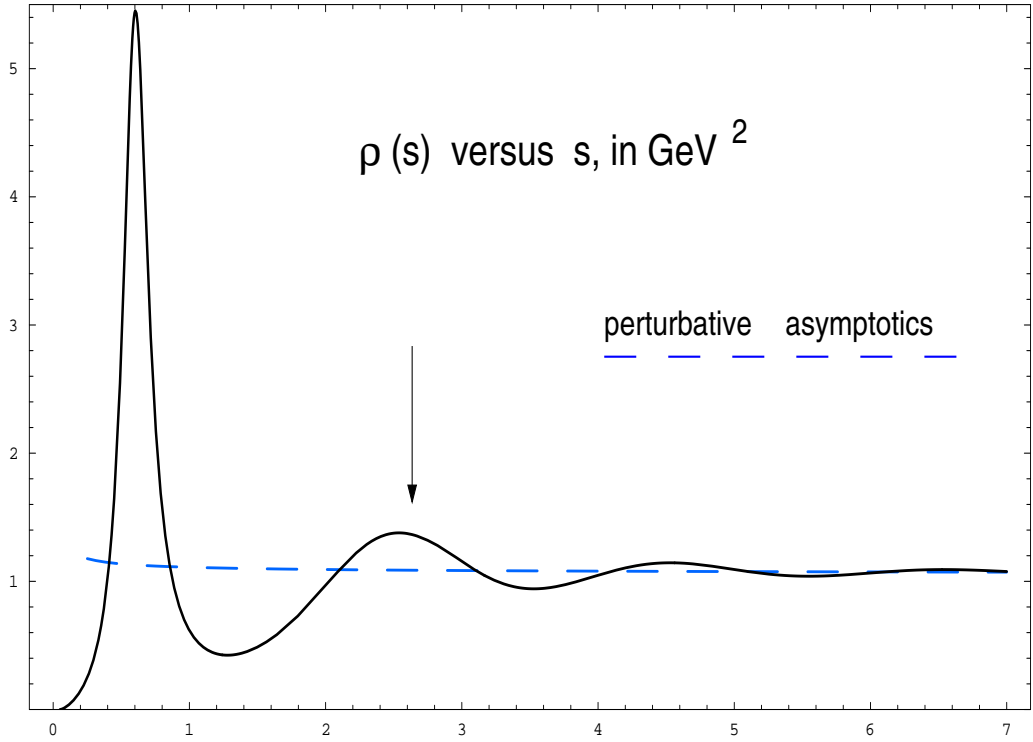}}
\caption{``Theoretical experimental" spectral density in the
$\rho$ meson channel. The solid curve presents experiment (see 
text), while the dashed curve is the perturbative result for $\rho
(s)$, including terms up to $O(\alpha_s^3)$, with $\Lambda = 0.2$ 
GeV. }
\end{figure}

Why I think that the
actual spectral density, when and if it is measured, will approach the   
smooth asymptotic curve with  oscillations rather than 
monotonously?
The reason is of a very general nature \cite{Shif1}. In a nut shell, 
there are exponential terms that are not seen in the truncated OPE 
series for $\Pi (Q^2)$ in the Euclidean domain. When these terms are 
analytically continued to the Minkowski domain, where Im $\Pi (s)$ 
belong, they show up as oscillating terms. The smooth asymptotic 
prediction is obtained from the truncated series and, thus, carries no 
traces of the exponential/oscillating terms. We will return to  the 
issue later on (see Sect. 5). 

Although the very existence of oscillations in the spectral densities
may be considered firmly established, the question of how rapidly 
they die off is highly non-trivial and has no unambiguous answer
in today's theory. In resonance-inspired models 
the damping factor is $\sim \exp (-\mbox{const}\, s)$ (see Sect. 2.2),
in the instanton-based models \cite{Dike} it is $\sim \exp (-
\mbox{const}\, \sqrt{s})$. The tail added in Fig. 17 is taken to be 
proportional to 
$\exp (-\pi\sqrt{s})\cos (\pi s + \mbox{const})$.
This rather eclectic formula is chosen for sophisticated reasons which 
need not concern us here. The task which we address is illustrative, 
anyway. I certainly cannot guarantee that  around 3.5 GeV$^2$ the 
value of $\rho (s)$ is 0.97, as it is shown in Fig. 17,  but I could bet 
that a shortage of the spectral density in this domain will be 
observed in precision measurements, so that  $\rho (s)$ at 3.5 
GeV$^2$ will lie somewhere around unity. I emphasize again that
the precise form of the tail does not affect the sum rule calculation of 
the $\rho$ meson parameters, and is discussed only for completeness 
of the picture. 

With the experimental spectral density in hands, we can calculate the
``experimental" value of the integral $I(M^2)$. The corresponding 
result is presented  in Fig. 18 by the solid curve. 

\begin{figure}
\epsfxsize=14cm
\centerline{\epsfbox{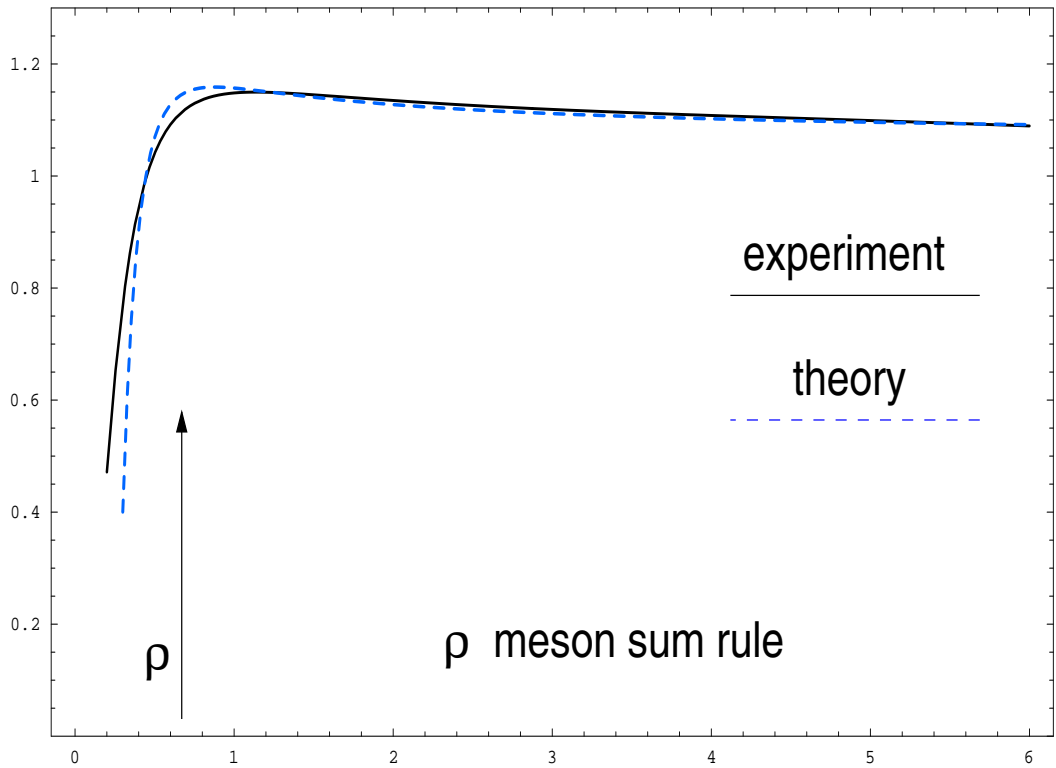}}
\caption{$I$ {\em versus} $M^2$ (in GeV$^2$) in the  $\rho$ meson 
channel: confronting experiment and 
theory. The theoretical curve corresponds to
$x_G = 0.8$, $x_{4q} = 1.3$ and $\Lambda = 0.2$ GeV.}
\end{figure}

The agreement is excellent; it is even better than one could expect 
{\em apriori}. 
We see that the conspicuous  $\rho$ peak and all further  twiddles 
characteristic to $\rho (s)$ are washed out. If we descend from 
larger 
to smaller values of $M$, the  behavior of $I(M)$ is  flat 
down to $M\sim 0.8$ GeV, i.e. down to the $\rho$ meson mass. At 
$M\sim 0.8$ GeV the regime smoothly changes, the curve dives 
down, and at still lower values of $M$, approaches the exponential 
asymptotics (this domain is not shown in the figure). The boundary 
value of $M$ where the regime changes 
is correlated with the mass of the lowest state, the $\rho$ meson in 
the case at hand. This is practically the only characteristic 
dimensional parameter in $I(M)$. 

Let us pretend now that we do not know the spectral density and 
want to use the sum rules to determine the $\rho$ meson mass. 
It is quite clear that the truncated condensate series does not allow 
one to go to the limit $M\rightarrow 0$ where this determination 
would be exact: the series explodes.
We must stay inside the window where the expansion is still under 
control. Correspondingly, our determination of $M_\rho$
is going to be approximate. The strategy  can be formulated as 
follows. Let us assume that somebody shows us a sketch of the 
spectral density presented in Fig. 10 with the numbers along the 
horizontal and vertical axes erased. This sketch is known  to correctly 
reproduce the basic features of the actual spectral density
but leaves us ignorant as to where the $\rho$-meson peak lies and 
what is its height. We insert this sketch in our sum rule,
do the integral 
$$
\frac{1}{\pi M^2} \, \int ds\, \rho (s) e^{-s/M^2}\, ,
$$
fit the numbers in such a way as to be as close to the theoretical
prediction (\ref{SRNP}) inside the window -- click, click --
the scales are restored,  and there come out the $\rho$ mass and the 
coupling constant. Since inside the window the $\rho$ meson 
saturates the
integral at the level $\sim 90 \%$ ,  the fact that  our continuum 
model is a caricature (minor details are lost) is unimportant. 
Even if we are off by a factor of two in this model this will affect
our estimates referring to the $\rho$ meson at the level of $\sim 10 
\%$. 

This example is quite typical. A similar situation takes place for all 
classical low-lying hadrons: those built from the light quarks, heavy, 
and 
light and heavy. I do not define here precisely what the ``classical 
meson" means but  will return to this point later, after considering 
some nonclassical channels. I must admit that the $\rho$ meson is 
the  
example where the SVZ method demonstrates  its best  facets.
The window is sufficiently broad, everything is clean.
As we proceed to higher spins, the mesons become larger in size.
A snapshot of a high-spin meson would  show a string-like picture, 
with a longitudinal size of the ``sausage" much larger than its 
transverse size. Under the circumstances one could hardly expect
that the SVZ method would work. And, indeed, it ceases
to be informative for spins higher than two \cite{MiSh}. 

\section{Basic Theoretical Instrument -- Wilson's OPE}

The theoretical basis of any calculation within the SVZ method is 
the 
operator product expansion (OPE). It allows one to consistently define 
and build the (truncated) condensate series for any amplitude of 
interest in the Euclidean domain. The physical picture lying behind 
OPE was described above: consistent separation of short 
and large  distance contributions. The former are then represented 
by the vacuum condensates while the latter are accounted for in the 
coefficient functions.  
Thus, the operator product expansion in the form engineered by 
Wilson
\cite{Wilson} is nothing but a book-keeping procedure. Wilson's idea
was adapted to the QCD environment in Ref.  \cite{NSVZ2}. 

Although OPE is used in  an innumerable 
amount of works since the mid-1970's, there are no good 
text-books or reviews devoted to this issue. Moreover,  one typically 
encounters a lot of confusion  in the literature. 
Below I will try to  explain  the Wilsonean procedure using,  as an 
example, 
the $T$ product of 
the 
two 
vector currents defined in Eq. (\ref{Tproduct}). 

Technically it is more convenient to deal with the $D$ function
(\ref{dfunct}) rather than $\Pi_{\mu\nu}$,
\begin{equation}
D(q^2)  =\sum_n C_n(q;\mu )\langle {\cal O}_n(\mu )\rangle 
\label{tee}
\end{equation}
where the normalization point $\mu$ is indicated explicitly.  
 The sum in Eq. (\ref{tee}) runs over all possible 
Lorentz and gauge invariant local operators built from the gluon and 
quark fields.  The operator  of  the lowest (zero) dimension is 
the 
unit 
operator {\bf I},
followed by the gluon condensate $G_{\mu\nu}^2$, of dimension 
four.
The four-quark condensate gives an example of dimension-six
operators.

At short distances QCD is well approximated by perturbation theory.
The coefficient functions $C_n$ absorb the short-distance 
contributions. Therefore, as a first approximation, it is reasonable to 
calculate 
them perturbatively, in the form of expansion in $\alpha_s (Q) \sim
(\ln Q)^{-1}$. This certainly does not mean that the coefficients  
$C_n$
are free from non-perturbative, non-logarithmic terms of the type
$\sim Q^{-\gamma}$ where $\gamma$ is a positive number. 

To substantiate the point let us consider the coefficient of the unit 
operator.
The  Feynman graphs for $D(q^2)$ of the lowest order are  depicted 
in Figs. 5 
and  13. Assume that the momentum 
flowing through the graph is Euclidean and large, 
$Q\rightarrow\infty$. If all subgraphs in these and similar graphs
are renormalized at $Q$, the final result, 
being expressed in terms of the running coupling constant 
$\alpha_s(Q)$, 
is finite. The
virtual
momenta saturating the loop integrals scale with $Q$ as the first 
power 
of $Q$. 
At any finite order the perturbative series
is well-defined. 

At the same time, if the number of loops $n$ becomes very large, in 
a random 
graph (Fig. 19)
 each 
loop carries momentum of the same order of magnitude, and the 
total 
momentum 
$Q$ is shared between
many lines, so that the characteristic virtual momentum
is proportional to $Q/n$. The contribution of each graph to $C_I$ is of 
order
$[\alpha_s(Q)]^n$, but the number of distinct diagrams $\nu$ grows 
factorially,
$\nu \sim n!$, so that the expansion in $\alpha_s(Q)$ is asymptotic
(for an exhaustive discussion of this phenomenon see Ref. 
\cite{RVHO}).  

\begin{figure}
  \epsfxsize=7.5cm
  \centerline{\epsfbox{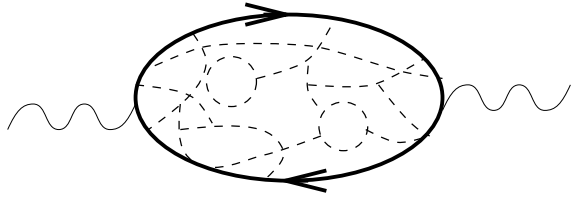}}
  \caption{A typical multiloop graph contributing to the $D$ function. 
The number of such graphs grows factorially with the number of 
loops.}    
\end{figure}

Note that this factorial divergence has nothing to do with the 
renormalon
divergence considered in Sect. 3. There, we had one specific graph of 
the 
$n$-th order whose
contribution  was proportional to $n!$. This factorial was a spurious 
artifact of an  improper calculation where  the domain $k\sim 
\Lambda$
was included. Introducing a lower cut-off at $\mu$ = several units 
times
$\Lambda$ and discarding the contribution coming from 
$k^2<\mu^2$,
as it should be done in the consistent calculation of $C_I(\mu)$, we 
would 
eliminate the renormalon divergence. At the same time, the factor 
$n!$
counting  the  number of graphs of the $n$-th order certainly cannot 
be 
eliminated in this way.

If $n$ is not too large,  so that $Q/n > \mu$, the graphs of Fig. 19 
present a legitimate contribution to $C_I(\mu)$.  It is quite clear 
that the tail of the $\alpha_s$ series with $n \gsim 1/\alpha_s (Q) 
\sim \ln Q$ generates a non-perturbative contribution 
of the type 
\begin{equation}
\Delta C_I \sim \exp {(-2\pi\gamma/\alpha_s (Q))} \sim \left( 
\frac{\kappa \Lambda}{Q}\right)^{b\gamma}\, .
\label{12}
\end{equation}
Here $\gamma$ and $\kappa $ are numerical 
constants.  An explicit 
example of such terms is provided by the so called direct 
instantons \cite{NSVZ}. Integration over the sizes $\rho$ of the 
direct 
instantons is  saturated at $\rho\sim Q^{-1}$. 
The constant $\gamma = 1$ for one instanton, $2$ for two and 
so 
on. 

The correspondence between the 
Feynman graphs with $1/\alpha_s (Q)$ loops and the small-size 
instantons  is not straightforward. I will not go further in this issue, 
referring the interested reader to Ref. \cite{RVHO}. At the qualitative 
level the inevitability of occurrence of the hard non-perturbative 
terms 
(\ref{12})
is evident. At the moment the only semi-quantitative framework we 
have at our disposal  for their evaluation is the  instanton 
mechanism. 

 Equation (\ref{12}) gives  a non-condensate power term.
Generically we will call  such terms 
{\em hard non-perturbative}.
They  may or may not be important numerically depending on
the particular value of $Q$ under consideration.  The  value 
of $b \gamma$ 
  need not be integer, generally speaking; numerically it  is very 
large. 
This means that once we cross the boundary where the argument in 
the brackets in  Eq. (\ref{12}) becomes less than one, $Q > \kappa 
\Lambda$, these terms 
immediately 
become totally unimportant. On the other hand, below this boundary 
they 
are so large that no expansion is possible.  

It is assumed that inside the window the terms (\ref{12}) are 
unimportant;  they 
 are completely 
disregarded
in the {\em practical version}  of OPE (Sect. 6) which, thus, applies to 
the values
of $Q$ above the  critical point. Note that the large value of 
$b\gamma$
results in a large degree of suppression of such terms
after borelization.  

Although the assumption above works very well in many ``classical" 
channels, it seems to fail in exceptional cases of ``non-classical" 
mesons.
This issue will be addressed in more detail in Sect. 8. 

I pause here to make a remark concerning non-dynamical
power terms in $C_I(\mu )$ whose origin is related to the  
introduction of the normalization point $\mu $. Let us return to
the one-loop graph of Fig. 13 and the corresponding expression
(\ref{begren}).  One should not forget that
in doing the loop integrations in $C_I (\mu ) $ we {\em must} discard 
the 
domain
of virtual momenta below $\mu$, by definition. 
Subtracting  
this domain from the perturbative loop integrals we introduce power 
corrections of the type $(\mu^2 /Q^2)^n$ in 
$C_I (\mu )$,
by hand. For instance, in the first order in $\alpha_s$
\beq
D(Q^2) = 1 + 
\left[ \frac{\alpha_s (Q)}{\pi} - \frac{\alpha_s }{\pi}\,  
\frac{\mu^4}{Q^4}
\right] + 
\frac{2\pi^2}{3 Q^4}\langle {\cal O}_G (\mu )\rangle + ...\, ,
\label{crazymu}
\eeq
where I combined Eqs. (\ref{flim}) and (\ref{addfla}). The expression 
in the square brackets is the $\alpha_s$ part of the coefficient $C_I 
(\mu) $. The explicit $\mu$ dependence of this expression
conspires with an implicit $\mu$ dependence residing in 
$\langle {\cal O}_G (\mu )\rangle$ to ensure the  $\mu$ 
independence
of the physical quantity (i.e. the $D$ function). 
If higher-order terms in the $k^2$ expansion of the $F$ function
(\ref{NBBB}) were taken into account, we would get higher powers of 
$\mu /Q$ in the square brackets, which then should have been 
combined with higher condensates, e.g. $GD^2G$ and so on. 

If $\mu$ can be chosen sufficiently low, the explicit and implicit 
$\mu$-dependent terms in Eq. (\ref{crazymu})
 may be numerically
insignificant
 and can be ignored. 

The coefficient functions in front of other operators  generally 
speaking have the same structure as $C_I$. The general situation is 
quite similar. For higher dimension operators it may happen (and, in 
fact,  happens) that some of the loop 
integrations are not saturated at $Q$ or $\mu$ but are, rather, 
logarithmic. Logarithms $[\ln (Q/\mu )]^\gamma$ occurring in this 
way
are associated  with  the anomalous dimensions of the 
operators at hand.

Let us have a closer look at the graph depicted in Fig. 13. As we
already know, some of the lines can be  
in a special regime -- the corresponding virtual  momenta are soft 
and their  off-shellness does not scale with $Q$.  
If the gluon line is soft, this gives rise to the gluon condensate,
if some of the quark lines are soft we deal with the four-quark 
condensate.
Graphically one distinguishes the soft lines by cutting   them;
then the remainder of the graph shrinks to a point while those lines 
that are cut 
form  a  local operator. 
The question is: what happens if we pass to the multiloop graph of 
Fig. 
19 and start cutting more and more lines?

Naturally, in the power series we proceed to the operators of higher 
and higher dimension. When the number of the soft lines becomes 
very large (of order of $Q/\Lambda $) it is conceivable that
there are no hard lines at all:
 the external momentum $Q$ is transferred from
the initial to the final vertex through a very large number of quanta
(growing like a power of $Q$), and none of the quanta carries 
momentum scaling with $Q$. Of course, in this situation one 
can not 
speak of individual quanta, one should rather use the language of 
typical field fluctuations transmitting the momentum $Q$ without 
having any Fourier components with frequencies of order $Q$. 
Contributions of this type are  not seen in OPE truncated at any
finite order.  They correspond to the high-order tail of the 
condensate series which, analogously to the $\alpha_s$ series, is 
factorially divergent. In this way we obtain the so-called {\em 
exponential terms} \cite{Shif1}. 
 
The fact that the condensate series is factorially divergent in
high orders is rather obvious from the analytic structure of 
the polarization operator $\Pi (Q^2)$. In a nut shell, since the cut
in $\Pi (Q^2)$ runs all the way to infinity along the positive real 
semi-axis of $q^2$, the $1/Q^2$ expansion cannot be convergent. 
The actual argument is more subtle than that,
but I would not like to go into details here referring the reader to 
Ref. \cite{Shif1} where a 
careful consideration of the issue is carried out. The final conclusion
is perfectly transparent. It is intuitively clear that the high-order tail
of the (divergent) power series gives rise to 
exponentially small corrections  $\sim\exp (-Q^\sigma )$ where 
$\sigma$ is some critical index. 

The numerical value of  $\sigma$ is correlated with the rate of  
divergence of high orders
in the power series. At the moment very little is known about this 
rate from first principles, if at all. The best we can do is to rely on 
toy 
models. 
The simplest  example is again 
provided by instantons. This time one has to fix the
size of the instanton $\rho$ by hand,  $\rho = \rho_0$. Then the
fixed-size instanton  
contribution is ${\cal O}(\exp (-Q\rho_0 
))$. The exponential factor is the price we pay for
transmitting the large momentum $Q$ through a soft field 
configuration
whose characteristic frequencies are or order $\rho_0^{-1}$. 
 
One can look at this example from a broader perspective. 
Consider the polarization operator $\Pi$ in the coordinate
rather than the momentum space. If the regular terms in OPE
are in one-to-one correspondence with   singularities of 
$\Pi (x)$ at $x=0$, the exponential terms, invisible in the truncated 
OPE,  are related to 
the singularities of $\Pi (x)$ located at a finite distance from the 
origin \cite{DS2}; the distance from the origin plays the same role as 
the 
instanton radius $\rho_0$. For our illustrative purposes it is 
sufficient to consider 
the singularities of the simplest possible structure, namely,
\beq
\frac{1}{x^2+\rho^2_0} ,\,\,\, \ln{(x^2+\rho^2_0)},\,\,\,
({x^2+\rho^2_0}) \ln{(x^2+\rho^2_0)},\,\,\, \mbox{and so on}\, .
\eeq
The Fourier transforms of these expressions have the generic form
\beq
(Q\rho_0)^{-n}K_n(Q\rho_0),\,\,\, n= 1,2,...
\eeq
where $K_n$ is the McDonald function.
At large Euclidean $Q^2$ the corresponding 
contribution dies off exponentially,
\begin{equation}
\Delta\Pi = (Q\rho_0)^{-n}K_n(Q\rho_0)\propto (Q\rho_0)^{-n-1/2} \,  
e^{-
Q\rho_0} \, , 
\label{deltapi}
\end{equation}
in full accord with intuition regarding transmitting a large 
momentum through a soft field fluctuation.

The terms that are exponentially small for Euclidean values
of $Q^2$ become oscillating upon analytic continuation to the 
Minkowski domain, $Q\rightarrow iQ$. 
Analytically continuing $\Delta\Pi $ given in Eq. (\ref{deltapi})
 we arrive at
\begin{equation}
\mbox{Im}\, \Delta\Pi  = (-1)^{n+1}\frac{\pi}{2}
(\sqrt{s}\rho_0 )^{-n} J_n (\sqrt{s}\rho_0 )
\propto  
(-1)^{n+1}\frac{\pi}{2}(\sqrt{s}\rho_0 )^{-n-1/2} 
\cos (\sqrt{s}\rho_0 - \delta_n )\, ,
\label{imdelta}
\end{equation}
where $J_n$ is the Bessel function. 
The oscillating character at $q^2 > 0$ is the most important signature 
of the exponential terms representing the high-order tail of the 
power series.

Having fixed $\rho_0$ is unrealistic. We would do much better 
than that  by 
allowing
$\rho_0$ to vary around a typical value of order of $\Lambda^{-1}$.
In other words, any  reasonable model will  introduce a distribution 
over 
$\rho_0$, with a weight function $w(\rho_0)$ peaked at 
$\Lambda^{-1}$,
\begin{equation}
\langle\mbox{Im}\Delta\Pi\rangle
\propto \int (\sqrt{s}\rho_0 )^{-n} J_n (\sqrt{s}\rho_0 ) w(\rho_0 ) 
\frac{d\rho_0}{\rho_0 }
\label{integral}
\end{equation}
where the angle brackets denote $\rho_0$ smearing. 
The weight function $w(\rho_0 )$ is model-dependent.
Particular details of the 
result will certainly depend on $w(\rho_0 )$, but the general 
pattern -- the exponential fall-off of $\mbox{Im}\Delta\Pi$ 
modulated by 
oscillations --
remains the same under any reasonable choice of $w(\rho_0)$.
This is the reason why the experimental spectral density in Sect.
4  is extrapolated beyond the range of 
the direct measurements  in an oscillating  mode (Fig. 17). 
The exponential suppression factor at large $E$ takes the form
\begin{equation}
\exp(- kE^\sigma)\, ,
\label{expo}
\end{equation}
where the critical index $\sigma$ in various models lies in the 
interval $0<\sigma < 2$. 
 
It remains to be added that the exponential terms are usually 
ignored in the sum rule analyses one encounters with in practical 
applications. 

\section{Practical Version of OPE}

In practical calculations one can hardly go beyond several
lowest-order terms in the $\alpha_s$ expansion of the coefficient 
functions
and in the condensate expansion. For instance, in the $\rho$ meson 
channel, which  is 
 the most advanced, four terms of the perturbative series are 
known in 
the unit operator. 

Thus, the approximation we make at the very first stage of building 
OPE is
truncating both the perturbative and condensate series.
By doing so we expect that those few terms that are kept
represent $D(Q^2)$ or $I(M^2)$ inside the window with sufficient 
accuracy.

Besides truncating the series, one usually ignores the hard 
non-perturbative 
terms in the coefficient functions. The reason why these terms are 
neglected is 
quite obvious: no  framework allowing one to reliably calculate them 
was 
worked out so far. From the instanton studies we know \cite{NSVZ} 
that
such terms must have an extremely steep $M$ dependence,
so that above a critical value of $M$  their  impact is expected to be
totally negligible. It is assumed that the critical value of $M$,
where the hard non-perturbative terms die off, lies to the left of the 
window. 

This is not the end of the story, however.  The third key element 
inherent
to the practical  OPE is a simplified treatment of the $\mu$ 
dependence in the coefficient functions and the condensates.
 The perturbative parts of the coefficient functions are often 
borrowed from 
earlier   calculations of relevant  Feynman graphs
which, sometimes, date back to QED. These calculations
make no distinction between small and large virtual momenta.
Moreover, in analytic QCD it is  technically not always easy to 
explicitly carry 
out the 
Wilsonean separation of virtual momenta (higher than $\mu$ and 
lower than 
$\mu$).  At the very least, such a separation requires dedicated 
analyses.

This is an obvious task for the future. A simplified strategy of today 
is as 
follows. Assume that we do not plan to go beyond the one-loop 
correction
of Fig. 13 in the perturbative part of $C_I$ and beyond the 
leading-order 
result
for $C_G$ (and higher condensates). The corresponding proper 
expression for 
the $D$
function is
\beq
D = 1 +\alpha_s \int_{\mu^2}^\infty \, dk^2 F(k^2, Q^2) + \frac{\pi^3 \, 
F'}{3}
\langle O_G (\mu )\rangle 
\label{pedf}
\eeq
(cf.  Eq. (\ref{crazymu})). Here $F'$  is $dF/dk^2$  at $k^2 = 0$.
The dimension-six condensate $GD^2G$ enters with the coefficient 
proportional 
to ${F'}'$ at $k^2=0$, and so on. 

Now, let us add and subtract to the right-hand side
 the integral
\beq
\alpha_s \, \int^{\mu^2}_0 \, dk^2 F(k^2, Q^2) \,  .
\label{intas}
\eeq
This integral looks exactly as the one-gluon contribution 
naively
continued below $\mu^2$. Of course, below $\mu^2$
the gluon propagator has nothing to do with $1/k^2$, but since the
transformation is identical, we do nothing wrong. Equation 
(\ref{pedf})
then takes the form
\beq
D = 1 +\alpha_s \int_{0}^\infty \, dk^2 F(k^2, Q^2) + \frac{\pi^3 \, 
F'}{3}
\left[ \langle O_G (\mu ) \rangle -\frac{3\alpha_s}{\pi^3} \, 
\int_0^{\mu^2} 
k^2 dk^2 
 \right]
\label{pedfprim}
\eeq
where in the square brackets, I expanded the integrand in $k^2$ 
and kept only the leading term. The next term proportional to $k^4$ 
would be important
at the level of dimension-six condensate, etc. 

The first integral on the right-hand side is the full perturbative one
gluon correction to $C_I$, with no Wilsonean separation (it is equal to
$\alpha_s /\pi$). The term in the square brackets can be called
{\em an effective one-loop} gluon condensate. It is obtained from  
the genuine
condensate by subtracting from it its perturbative one-loop 
expression.
By construction, the $\mu$ dependence of the
effective one-loop gluon condensate cancels provided we do not go 
beyond the 
accuracy specified above. 

Adding the integral (\ref{intas}) to the right-hand side we achieve
the desired goal. Now
the 
coefficient $C_I$, to order ${\cal O}(\alpha_s)$,  is given by full 
perturbative graphs of Fig. 13. To avoid double 
counting  we subtract
the same contribution from the condensates. More exactly,
we subtract {\em almost} the same contribution. Since the sum over 
condensates
is truncated at some finite order $n_0$, the best we can do is to 
subtract from 
the condensate part
the integral
\begin{equation}
\alpha_s\int^{\mu^2}_0 dk^2 \, \sum_{n=1}^{n_0}
\frac{1}{n!} F^{(n)}(k^2=0,Q^2)\, (k^2)^n .
\label{subtr}
\end{equation}
In this way  we subtract from each condensate its ``one-loop 
perturbative 
value".

It is quite clear that this strategy of converting  the genuine Wilson 
OPE
into the practical version 
is admittedly approximate and works only if we limit
ourselves to a given number of loops and given number of
terms in the condensate expansion.  
The effective condensates obtained in this way should be 
supplemented by
a superscript indicating the number of loops 
in the subtracted part. Their numerical value is $\mu$
independent but does depend on the number of loops. 

This approximate procedure (i) leads to a loss of factorization of short 
and 
large 
distance contributions inherent to Wilson's OPE; (ii) can not be 
systematically
generalized to arbitrary  number of loops since there is no way to 
unambiguously 
define the integrand to all orders in the small $k^2$ domain 
\footnote{Within  practical   OPE we automatically 
forbid   to ourselves   
questions concerning any aspects of the high order behavior.};
(iii) only 
approximately avoids
double counting  due to the necessity of truncating the condensate 
series at a 
finite order (the lower the order we truncate the larger the error).
Moreover, the effective condensates are not universal, strictly 
speaking;
they become process-dependent. If this non-universality
were numerically significant, the SVZ method, as it exists now, would 
be 
undermined. 

Therefore, the practical version  is useful in applications
 only provided  $\mu^2$ can be made 
small enough
to ensure that the ``one-loop perturbative" contributions to the 
condensates are
much smaller than their genuine values and, at the same time, 
$\alpha_s 
(\mu^2)/\pi$ is small enough for the expansion to make sense.
The existence of such ``$\mu^2$ window" is not granted {\em a 
priori} and is a 
very fortunate feature of QCD. We do observe this
feature empirically although the roots of the phenomenon
remain unclear.

Although the origin of the $\mu^2$ window, ensuring the 
applicability  of 
practical  OPE  
\footnote{A natural abbreviation for practical OPE
would be POPE, but I do not risk
to put this abbreviation into circulation. }  is not understood, one can 
find 
similar 
situations in  simplified settings. The best example of which I am 
aware  is 
the two-dimensional $O(N)$ sigma model in the limit of large $N$.
In this limit, the model is exactly solvable; therefore any question of 
interest can be exhaustively answered. The model bears remarkable 
parallels to QCD -- it is asymptotically free, and the mass scale is 
generated dynamically. A surprising result was established some 
time ago \cite{NSVZ3}.  In the large $N$ limit all vacuum condensates 
in this model are $\mu$ independent, and the practical version of 
OPE becomes exact; $\mu$ dependence in the condensates and in the 
coefficient functions comes only at the level of $1/N$. If $N$ is large 
these terms are parametrically small. (It would be great to find a 
hidden parameter  which would explain the suppression of the 
$\mu^2 /Q^2$ terms in QCD in the same way $1/N$ explains it in the 
sigma model.)

It is often  asked whether it is legitimate  to retain in the truncated 
expansions 
 logarithmic and power  terms 
simultaneously \footnote{In discussing the issue of
duality violations one deals with the logarithmic, power, and 
exponential
terms simultaneously, see Ref. \cite{Dike}.}. Indeed, formally any 
$1/(\ln 
Q)^n$ term is
parametrically larger that, say, $1/Q^4$  (moreover, any power 
term is parametrically larger than the exponential ones). If so, the 
theoretical uncertainty due to truncation of the logarithmic series is 
formally larger than even the lowest-order condensate correction.

The logarithmic and power terms in practical OPE have distinct
physical nature.
In some instances power terms describe  effects that do not show up 
in 
perturbation theory. This is 
valid  in all cases where the chiral quark condensate is 
involved. In other cases, when  the power terms 
are not 
different in their structure from those occurring in perturbation 
theory, there is a strong numeric enhancement of the power
terms (see Sect.  3). 
This mysterious fact -- the numerical enhancement of the vacuum 
condensates -- explains why
 Wilson's OPE can be 
substituted in QCD by the practical version, at least in the classical 
channels .

In the early days of QCD, in 1970's, it was common to assume that
the coefficient functions in the operator product expansion are
saturated by perturbation theory, with no separation of virtual 
momenta.
Although the difference between  Wilson's OPE and  the practical 
version was realized long ago \cite{NSVZ2},  very little has been done 
to
investigate this difference and perfect the procedure. Only now
we are witnessing attempts in this direction, mostly in connection 
with the 
heavy quark theory (e.g. \cite{BSUV,Dike,Ji}).

\section{Low Energy Theorems}

Although the low-energy theorems are independent of the sum rules, 
at least some of them were derived in connection with the 
development of the SVZ method, and were later combined with
the sum rules giving rise to an intriguing observation of the hadronic 
non-universality. This observation is the topic of the next section.
Here we will dwell on the low-energy theorems related to the trace 
anomaly of QCD.

In the limit of massless quarks (and we will never leave the limit
$m_q =0$) the classical Lagrangian of QCD is free from dimensional 
parameters. The scale parameter appears at the quantum level,
\beq
\Lambda = M_0 \, \exp \left( -\frac{8\pi^2}{bg_0^2}\right)
\label{sparam}
\eeq
where $M_0$ is the ultraviolet cut off, $g_0$ is the bare 
coupling constant, and I ignored the second and higher-order terms 
in the $\beta$ function. (The corresponding modifications are quite 
trivial and are suggested to the reader  as an exercise.)

The occurrence of the scale parameter (\ref{sparam})
is in one-to-one correspondence with the trace anomaly
(\ref{aemt}). Let us denote the trace operator
\beq
\theta_\mu^\mu \equiv \sigma\, .
\label{dto}
\eeq
Then, it is not difficult  to show \cite{NSVZ}
that for arbitrary local operator ${\cal O}$
\beq
\lim_{q\ra 0} \, \left\{  i \,\int\, d^4 x\, e^{iqx} \langle T\{{\cal O}(0),  
\sigma (x)
\rangle_{\rm c}\right\} = - d \, \langle {\cal O}\rangle\, ,
\label{leth}
\eeq
where $d$ is the (normal) dimension of the operator ${\cal O}$,
and the subscript c denotes the connected part
(it will be omitted hereafter). If 
$$
{\cal O} = \sigma\, \propto G^2\, ,
$$
then the dimension $d=4$, 
so that
\beq
 i\int
e^{iqx} d^4 x \, \left\langle T\left\{ \frac{-b\alpha_s}{8\pi}\, 
G^2 (x) ,\,\,\, 
\frac{-b\alpha_s}{8\pi}\, 
G^2 (0) \right\}\right\rangle = (-4) \, \left\langle \frac{-
b\alpha_s}{8\pi}\,  G^2\right\rangle\,\,\,\mbox{at}\,\,\, q^2 = 0\, .
\label{82}
\eeq
For the quark operator $\bar q q $ the 
dimension $d=3$, and so on. 

The derivation is  straightforward. First, we rescale the gluon 
field, ${\cal G}_{\mu\nu}^a = g_0 \, G_{\mu\nu}^a$, so as to factor out
the $g_0$ dependence in the Lagrangian. Instead of Eq. (\ref{qcdlag})
we now  have
$$
{\cal L} = -\frac{1}{4g_0^2}\, {\cal G}_{\mu\nu}^a Ê{\cal 
G}_{\mu\nu}^a + \,\,\,\mbox{quark term}\, .
$$
Then we observe that
$$
i \int\, d^4 x\,  \langle T\{{\cal O}(0),  {\cal G}^2 (x)\}
\rangle =-\frac{\partial}{\partial (1/4g_0^2)} \,\, \langle {\cal O}
\rangle\, .
$$
Since $\langle{\cal O}\rangle = \Lambda^d$, we use Eq. 
(\ref{sparam}) to perform differentiation on the right-hand side.
In this way we arrive at the formula (\ref{leth}).

In fact, the simple derivation outlined above should be 
supplemented by some regularization since the matrix elements 
$\langle {\cal O}
\rangle$, as well as the two-point functions (\ref{leth}), are 
divergent 
unless the cut-off at $\mu$ is introduced. In Ref. \cite{NSVZ} 
(see Appendix B) it is shown that the regularization procedure does 
not affect Eq. (\ref{leth})
provided the cut-off is introduced in a concerted way. 

Denote  the two-point function on the left-hand side of Eq. (\ref{82}) 
as $\Pi_{\rm s} (Q^2)$; the subscript s marks  the scalar glueball 
channel.
If   $\Pi_{\rm s} (0)$ is known, this knowledge can be immediately 
traded 
for a power term in the Borel-transformed sum rule. Indeed,
in this case, instead of borelizing the dispersion relation for
$\Pi_{\rm s} (Q^2)$, one  can deal with the dispersion relation for
$$
\frac{\Pi_{\rm s} (Q^2) - \Pi_{\rm s} (0)}{Q^2}\, .
$$
After borelization one gets a prediction for the integral
$$
\frac {1}{\pi M^2 }\, \int \, \frac{ds}{s}\, \mbox{Im}\Pi_{\rm s} 
(s)\, e^{-s/M^2}\, .
$$
Note the occurrence of an extra factor $s^{-1}$ in the weight function
in the integrand (cf. Eq. (\ref{btdr})). At the very end
 $$
\Pi_s (0) = \frac{b}{2} \, \langle {\cal O}_G\rangle 
$$
 is transferred to the 
right-hand side of the sum rule where it  acts as a coefficient in  the 
corresponding 
power correction.

In this way one arrives at 
\beq
\frac{32\pi^3}{b^2 M^4}\, \int \, \frac{ds}{s}\, \mbox{Im}\Pi_{\rm s} 
(s)\, 
e^{-s/M^2} \, = \, \alpha_s^2 (M) \left[ 1 + \frac{16\pi^4 }{b\alpha_s^2 
(M)} 
\,
\frac{1}{M^4} \langle {\cal O}_G\rangle + ... \right]\, ,
\label{srsch}
\eeq
where the first term in the square brackets corresponds to the 
free gluon loop, the second term is due to $\Pi_{\rm s} (0)$, while 
the 
ellipses stand for all other corrections, perturbative and 
non-perturbative. Comparing the gluon condensate correction to that
in the $\rho$ meson sum rule, Eq. (\ref{ancient}),  we observe, with 
astonishment, a very strong numerical enhancement of the relative 
coefficient, by a factor
$$
\frac{48}{b} \, \left( \frac{\alpha_s}{\pi}\right)^{-2}
\, \sim \,  10^2\, .
$$

The standard condensate expansion, with a few first 
terms kept, does not match this huge low-energy constant.
Attempts to  smoothly match this gigantic power term with other
non-perturbative corrections and to  saturate the sum rule 
in the
scalar glueball channel led us \cite{NSVZ} to the conclusion 
of extremely violent distortions of the spectral density
up to the scale $s_0 \sim 25$ GeV$^2$, to be compared with
$s_0 \sim 1.5$ GeV$^2$ in the $\rho$-meson sum rule, see
Eq. (\ref{origmod}) and Fig. 10.  The scale up to which strong 
violations of asymptotic 
freedom extend in the scalar glueball channel is unconventionally  
large. 

Thus, a new (numerically) 
large scale was discovered, that does not show up in the classical 
channels. It was predicted to have a major impact on the properties 
of the ``non-classical" mesons.  The physical picture lying behind this 
phenomenon is discussed in Sect. 8.  The quantum numbers of the 
mesons affected by the ``superstrong" interaction with the vacuum 
medium were found to be in one-to-one correspondence with the 
presence of the direct instantons. This observation served as an 
initial impetus for the development of the
instanton liquid model of the QCD vacuum \cite{DySh} (for a review 
see \cite{Shuryak}). 

The topic of the QCD low-energy theorems is very vast; it reaches out 
to such profound issues as the subtleties of the $\theta$ dependence,
the $\eta '$ problem, effective dilaton Lagrangians, the 't Hooft 
matching condition, {\em etc.}
It has never been reviewed in full, although it certainly deserves 
attention. This endeavor goes far beyond  the scope of the present 
lecture, and I will limit myself to  fragmentary remarks on the 
literature.  

The low-energy limit of the $n$-point functions generated by 
$G\tilde G$ was studied in \cite{RCr}. A celebrated mass formula for 
$\eta '$  was obtained in \cite{MFE1,MFE2}. An elegant low-energy 
theorem for the coupling of the Goldstone mesons to the gluonic 
sources was derived in Ref. \cite{MBV}. The couplings of photons to 
the gluonic sources
were elaborated in Refs. \cite{NSVZ,LeSh}. 
Low-energy theorems 
instrumental in the sum-rule analyses, additional to those
considered in \cite{NSVZ}, were found in \cite{DIM}. Selected aspects 
of the subject are reviewed in \cite{MiShif}. 

\section{Are All Hadrons Alike?}

The first, superficial, impression of the hadronic family may leave 
one quite bored. There are no obvious small or large parameters,
all quantities are of order one in the appropriate units, and nothing 
special attracts attention. In particular, one might think that the 
characteristic hadronic scale is given in all cases by the slope of 
the Regge trajectory, i.e. universal $\alpha '\sim 1\mbox{GeV}^{-2}$.
The masses of the ``most" typical hadrons,  nucleons and $\rho$ 
mesons, are of this order of magnitude. Roughly at the same 
momentum transfers scaling sets in in deep inelastic scattering
heralding the onset of asymptotic freedom -- the observer sees  
(almost) free quarks with modest corrections due to the gluon 
exchanges.

A careful  reflection shows, however, that the hadronic family
is far from being that monotonous.  Messages of its non-universality
have been coming from various sides, although it was not so easy to 
decipher them.  

Many years ago Witten posed a question \cite{MFE1}:

``There is an obvious troublesome question. If $m_{\eta '}^2 \sim 
1/N_c$ and the $1/N_c$ expansion is usually a good approximation, 
why is $m_{\eta '}$ not much smaller than its actual value of almost 
one GeV?

This is not merely a problem of $1/N_c$ expansion. It is a more 
general phenomenological problem. The $\eta ' -\pi$ mass splitting 
violates Zweig's rule, since it involves $q\bar q$ annihilation.
Since Zweig's rule is usually rather good, why is the $\eta ' -\pi$ 
splitting so large?"

A number of the glueball 
candidates have been experimentally observed in the last decade. 
Still, none was 
unambiguously proved to be a glueball. 
Let us ask ourselves:

Why, unlike the quark mesons, the masses of the glueballs 
are unusually heavy or they  have an unusually  large mixing with 
the quark states which makes their widths large enough
to escape clear  experimental identification, in defiance with the 
$1/N_c$ counting?

The answers to all these questions naturally come with the 
observation \cite{NSVZ} of large scales in the non-classical channels.
Yes, $m_{\eta '}^2 $ and  Zweig's rule violating parameters are 
suppressed by $1/N_c$ and are small, but they are small compared 
to the intrinsic scale $s_0 \sim 25$ GeV$^2$, rather than
1 GeV$^2$.

 The existence of the
diverse scales in the hadronic family is due to the fact that various 
currents injecting the valence quarks/gluons in the vacuum medium 
interact with this medium non-universally. If in the $\rho$ meson 
(and similar classical mesons) 
the impact of the vacuum fields on the valence quark-antiquark pair 
is modest and is reasonably well described by the average densities
of the vacuum fields, in the glueballs the coupling to the vacuum 
fields is much stronger, so that they actually feel fine details (a grain 
structure) of the vacuum medium. 
The strongest coupling to the vacuum fields takes place in the 
channels with the total spin zero, both quark and gluon
(scalar and pseudoscalar quarkonia and glueballs). 
Phenomenologically deviations from Zweig's rule are most drastic 
here.

The occurrence of the
``superstrong" coupling manifests itself 
in an abnormally large critical scale which may or may not 
materialize as the mass of the lowest-lying state with the given 
quantum numbers. By critical scale I mean the energy at which the 
asymptotic  perturbative regime sets in in the spectral density.

A combined analysis of the sum rules and the low-energy theorems
  has led us \cite{NSVZ}  to a number of qualitative (and, in many 
instances, semi-quantitative) 
 predictions. First of all, the hierarchy 
of glueball masses 
was established. The scalar glueball was predicted to be the lightest,
with mass $M(0^+) \sim 1.5$ GeV (see  \cite{MSG}).
Its coupling to the two-pion state was shown to be very large, with 
no trace of the Zweig suppression. The pseudoscalar and tensor $2^+$
glueballs were predicted to lie in the vicinity of 2 GeV. 

In both scalar and pseudoscalar channels  the valence gluons injected 
in the vacuum are affected by  
the vacuum medium in the most drastic way, and the critical scale is 
the largest. In the tensor glue channel the 
critical scale was expected to be  smaller than
in  the scalar/pseudoscalar channels, although larger than for  the 
classical mesons. 

Why the $0^\pm$ channels are so special, with the strongest reaction 
of the medium? It was natural to assume that the phenomenon had 
to do with the structure of the dominant vacuum fluctuations. 
It was noted \cite{NSVZ} that instantons would do the job,
since the direct instantons  show up exactly in the $0^\pm$ channels. 
This was sufficient for a qualitative classification.
Quantitative instanton-based models of the QCD vacuum appeared 
later 
\cite{DySh,Shuryak}; they automatically  incorporate all relevant 
features of 
this 
peculiar picture. Hunting for instantons in the vacuum is the {\em 
vogue} of the day in the lattice community. There are indications that 
they are abundant and almost saturate the string tension. 

It remains to be added that, 
many years after these predictions, we are witnessing how 
other approaches started recovering  the very same pattern
of non-universality. Recent 
calculations 
in  the instanton liquid model produce -- not surprisingly -- a very 
close hierarchy of  
masses and 
mixing parameters \cite{SSh}.  What is interesting, this model 
explains where the large non-perturbative scale in the scalar glueball 
channel is hidden. The violent vacuum fluctuations squeeze
this state to an unusually small radius because of a very strong 
attraction. The radius of
the $0^+$  glueball in this model is only 0.2 fm, to be compared with 
0.6 fm
in the $\rho$ meson case \cite{SSh}. The ratio of the radii squared is 
$\sim 
1/10$. It is clear that no constituent model compatible with quantum 
mechanics can accommodate a state light and narrow simultaneously
(see the end of this section).

This part of the hadronic family is being intensively explored on the 
lattices too.
Of course, the light quarks are usually non-dynamical on today's 
lattices,
and a typical lattice site of 0.2 fm is comparable with the radius of 
the $0^+$  glueball. With all these reservations in mind
it is still instructive to compare what the lattice community learned 
about 
the non-classical hadrons. To this end I borrowed 
relevant numbers from the talks \cite{latrev} and original 
publications to be cited below:

(i)  The lightest glueball is  scalar, and its mass is close to 
     1.6 GeV;

(ii) The tensor glueball is significantly heavier; its mass is slightly 
above 2 GeV; the pseudoscalar glueball is found at approximately the 
same 
place, 
with larger errors; 
 
(iii) The sizes of scalar and tensor glueballs were found to be
     drastically different. This can be inferred from  the
     different magnitude of the  finite size effects (see e.g. 
\cite{VBK}),
     or seen  directly,  in the glueball Bethe-Salpeter amplitudes (or 
``wave functions")
     \cite{Isi, PdF}. The radius of the scalar glueball was found to be 
0.2 fm, while that of the tensor one is much larger, approximately 
0.8 fm. A similar lattice 
measurement of the $\rho$ meson size yields approximately 0.5 fm.

Needless to say that the lattice calculations provide us only with final 
numbers, with no insight as to the mechanisms of the 
non-universality. To the best of my knowledge, no lattice publication 
on this topic
mentions the previous sum rule predictions. True, the lattice
results have a potential to be perfected in the future,
while the accuracy of the sum rule method is admittedly limited.

To make  the highly  nontrivial nature of our findings more contrast 
let us conclude this section by comparing  them to the  expectations 
of conventional 
models, say, the bag model. The quark and (electric) gluon modes in 
the spherical cavity have energies $2.04/R$ and $2.7/R$, 
respectively, where $R$ is the cavity radius. Thus, the glueball states 
are expected to be only marginally heavier than the quark states.
In particular, if the spin-dependent forces are neglected the model 
predicts that $M(2^{++}) \approx 
     M(0^{++})\approx 1$ GeV and $M(0^{-+})\approx 1.3$ GeV.
Including the spin-dependent forces  changes these numbers 
rather
insignificantly \cite{1}. The complex picture of the vacuum medium, 
which most strongly affects the  scalar and pseudoscalar mesons, 
especially glueballs, is completely missed. 

\section{Ecological  Niche}

In the 1980's the original strategy of the SVZ method  was tested, 
with a 
remarkable success, in analyzing practically 
every static property of all established low-lying hadronic states. 
The method 
was developed  in various directions --  three-point functions, 
inclusion of 
external 
electromagnetic and other auxiliary fields, light-cone modifications 
and so on -- which allowed one to expand the range of applicability 
to such advanced problems as magnetic moments \cite{MM}, form 
factors 
\cite{FF} at intermediate momentum transfers, weak 
decays \cite {WD} and structure functions \cite{SF} of deep inelastic  
scattering  at intermediate $x$.  At the initial stage the method was 
essentially 
unchallenged since the lattice calculations were lagging far behind
burdened by multiple  internal problem of lattice QCD.

Now the situation has changed. The machinery of lattice QCD
was bettered, and it became a powerful tool
in many problems where qualitative insight is unimportant, and the 
prime 
emphasis is on numbers. Let us remember that the
SVZ method is admittedly approximate: typically, theoretical 
accuracy 
is at the 
level of  20\%. In some cases it does not work at all.

In order to survive,  the approach based on OPE and the sum rules 
has to find 
a range
of applications in the hadronic physics, where it could successfully 
compete. 
I must say that it  copes with the challenge.
Its strongest advantage is the analytic character of calculations 
which
can be continued to the Minkowski space, term by term. Thus, the 
most 
promising directions of growth are those where the Minkowski 
kinematics
is entangled in this or that way and/or the processes under 
consideration 
develop in more complicated media, rather than in the vacuum 
medium.
(Such problems are very difficult for lattice 
QCD which operates with vacuum correlation functions numerically 
evaluated in the Euclidean space.) 
Let me give three examples. 

\vspace{0.1cm}

1) {\em Particle decays with a large spatial momentum carried away 
by 
produced hadrons, for instance,} $B\rightarrow \rho \ell\nu$. If the 
invariant 
mass
of the lepton pair is small, $q^2 \sim 0$, the spatial momentum of the 
$\rho$
meson is of order $M_B/2 \gg \Lambda$. Treating the problem in 
the 
Euclidean domain we will have to face the necessity of  an extremely 
remote 
analytic continuation. At the same time the light-cone sum rule, to be 
discussed in Sect. 10, are perfectly fit to deal with this problem. 

\vspace{0.1cm}

2). {\em Heavy flavor sum rules}. They allow one to determine
basic parameters of the heavy quark theory -- the masses of the 
heavy 
quarks, form factors at zero recoil, and so on -- in terms of directly 
measurable  quantities. This is a close relative of the SVZ sum rules.
The main distinction is that instead of the vacuum condensates
one deals with the expectation values of various local operators 
over the
heavy hadron state, for instance the $B$ meson. This approach will 
be also 
considered in brief in Sect. 10. 

\vspace{0.1cm}

3) {\em Pre-asymptotic effects in inclusive processes at high 
energies}.
A typical problem from this class is determination of the lifetime 
differences 
in the $b$ quark family. Although the task is somewhat different
from those usually treated in the SVZ sum rules, the method of its 
solution
is exactly the same. The key technical element is the operator 
product 
expansion.  In the given context it is imperative to  address, 
additionally,   an 
issue going  beyond OPE (more exactly,  its practical version)
--  duality violations. The topic of  the deviations from duality
is of paramount practical importance;
it came under  renewed scrutiny 
 recently \cite{Shif1}. 

\vspace{0.1cm}

In all these and other instances
the OPE-based methods remain to be the only theoretical tool
available on the market today. We see that the approach
is
sufficiently rich and flexible to remain 
viable in the future. I am sure it  will continue to play a key role in 
solving  
many applied problems in the hadronic physics  
in the next decade. 
The potential for further growth is there, the active stage is not yet 
over. 

\section{New Developments}

To give you a flavor of what is going on in the 
this field I will pick up two or three subjects which come to  my 
mind 
first. My selection  by no means presents  the full picture.

The heavy quark mass $1/m_Q$ expansions that flourished in the 
1990's
revived interest in the SVZ sum rules for hadrons containing one 
heavy quark 
\footnote{The $1/m_Q$ expansion of the SVZ sum rules {\em per se} 
dates 
back to the work of Shuryak \cite{ShuHQ}.}
(in practice, $b$ quark). It was realized -- and this is a new element 
--
that (i) the hybrid logarithms 
occur in the theoretical part of the sum rules, and (ii) they may 
result in a 
strong enhancement
of the gluon radiative corrections. Hybrid logarithms depend on the
ratio $m_Q/\mu$, they were unknown before 1988. A typical 
example
where summation of the hybrid logarithms significantly shifts the 
answer
is $f_B$. I single out this problem because this parameter is 
fundamentally 
important in the heavy quark 
physics  and because  of an instructive story attached to 
it.
 Early sum rule calculations of $f_B$,   which missed then 
undiscovered
hybrid logarithms, 
yielded a number 
 close to 130 MeV \cite{AE}. The main impact of the hybrid 
logarithms
is a change of the argument of the running coupling constant
in the sum rule: instead of $\alpha_s (m_b)$ the radiative 
corrections, as it 
turns out
\cite{BBBD}, are governed by $\alpha_s (1\,\,\,\mbox{GeV})$. As a 
result, the
sum rule prediction jumped up to  $f_B = 160$ MeV, with the 
theoretical uncertainty 20 to 30 MeV \cite{BBBD}.

Meanwhile, $f_B$ was  the object of the multiple lattice studies.
The first lattice measurements produced an unbelievably
huge value, somewhere around 250 MeV. With elimination of various 
sources
of the systematic uncertainties  the lattice number has been
systematically 
decreasing.
The modern number, quoted at recent conferences,
is close to 180 MeV, with the error in the same ballpark as the sum 
rule
uncertainty. Thus, the initial contradiction between the two methods 
faded 
away giving place to perfect agreement.  

Along with the technical developments of the type I have just 
mentioned
the method experienced ideological developments too. Below we will 
discuss 
some of them.

\subsection{Light-cone sum rules}

This   formalism was  designed to overcome difficulties in the 
traditional sum 
rules for 
three-point functions. Assume we want to calculate a cubic coupling 
constant
corresponding to the amplitude $A\ra B+C$, where $A,B$ and $C$
stand for either hadrons or external currents, say, electromagnetic. 
Typical
examples are
 $D^*\ra D\pi$ or $\gamma^* + \gamma^* \ra \pi$ (here
$\gamma^*$ denotes a virtual photon). In many instances
in order to get the desired constant  it is necessary
to apply borelization with respect to two momenta, $p_A$ and $p_B$, 
independently. 
Then the components of the third momentum, $p_C$, can not be 
small
compared to $p_{A,B}^2$. Indeed, $p_C(p_A+p_B) = p_A^2 - p_B^2 
\sim 
p_{A,B}^2$. The standard condensate expansion
contains a series of operators with derivatives which, eventually, 
give rise to the
 expansion parameter of the type
$p_C(p_A+p_B)/p_{A,B}^2\sim 1$. In other words, all terms in this 
subseries
must be summed over. 

This (partial) summation is carried out automatically if, instead of
the three-point function $\langle j_A, j_B, j_C\rangle$
one considers the correlation function of the currents
$ j_A$ and $ j_B$ sandwiched between the vacuum and the state
$|C\rangle$. The vacuum expectation 
values of local operators (condensates) in the SVZ sum rules are 
substituted 
by the {\em 
light-cone wave functions} ({\em distribution amplitudes}  in the 
American 
literature)
describing the momentum fraction distribution of the components in 
the Fock
wave function of the hadron $C$.
If the SVZ approach is based on the 
short-distance expansion of the current $T$ products in terms of 
local 
operators, and the 
 expansion runs in  {\em dimensions} of the local operators, the 
light-cone 
sum rules exploit  the expansion in {\em nonlocal} ``string''
operators on the light-cone; the expansion runs in  {\em twists},
not dimensions. From the viewpoint of the standard SVZ sum rules, 
in the light-cone approach one performs a partial summation of an 
infinite 
chain 
of operators of arbitrary dimension, but given twist.
 In effect, the light
cone sum rules combine the SVZ approach with the 
 technique
 used in the description
of hard exclusive processes  (see \cite{exclusive}). The price we 
have to pay for the partial summation is rather high:
the large distance dynamics is parametrized not by numbers, as in 
the SVZ 
condensates, but by functions -- the leading twist, next-to-leading
twist, and so on distribution functions. 

A Russian proverb says that it is better to see once than to hear 
hundred 
times. Let us have a closer look at a specific example.
Below I will outline the calculation of the $D^*D\pi$ constant.
This example, as well as the explanations below, are borrowed from 
Ref.
\cite{BBKR}. 

The $D^*D\pi$ coupling constant $g$
is defined as follows:
$$
\langle D^* (p) \pi (q) |D (p+q )\rangle = - g\, q_\mu\epsilon^\mu\, ,
$$
where the hadrons' momenta are indicated in the parentheses, and
$\epsilon^\mu$ is the $D^*$ polarization vector. To determine this 
constant
from the light cone sum rules one considers
the correlation function
\beq
F_\mu (p,q) = i\int \, d^4 x\, e^{ipx}
\langle \pi (q) | T\{\bar d (x) \gamma_\mu c (x), \, \bar 
c(0)i\gamma_5 u 
(0)\}|0\rangle\, .
\label{lcsrcf}
\eeq
The first operator in the $T$ product is the interpolating current for
$D^*$ and the second is the interpolating current for
$D$.

The correlation function (\ref{lcsrcf}) depends on two variables, 
$p^2$ and 
$(p+q)^2$ (note that $q^2 = 0$ since the pion is on mass shell). The 
contribution proportional to the 
$D^*D\pi$ amplitude has two poles, in $p^2$ and $(p+q)^2$, 
corresponding
to the ground state mesons both in the vector and pseudoscalar 
channels.
Namely,
\beq
F_\mu (p,q) = \frac{M_D^2M_{D^*}f_Df_{D^*}}{m_c(p^2-
M_{D^*}^2)((p+q)^2-
M_D^2)}\,\, g\, q_\mu + ...\, .
\label{ppsr}
\eeq
The ellipses on the right-hand side denote  a term proportional to 
$p_\mu$
irrelevant for our analysis, as well as terms with single pole
and non-pole continuum contributions, that will be exponentially 
suppressed 
under double borelization (with respect to $p^2$ and $(p+q)^2$). 
All other notations are self-explanatory.

On the theoretical side we assume that $p^2-M_{D^*}^2$ and $(p+q)^2-
M_D^2$ are Euclidean and 
large 
enough to allow for the operator product expansion of the
correlation function (\ref{lcsrcf}). As usual, the corresponding Borel 
parameters
will lie in the window. The simplest graph pertinent to OPE we have 
to build is 
depicted in Fig. 20. The large distance part of the process is denoted 
by the 
shaded area. 
The light quark lines form the pion, rather than the quark 
condensate,
that was discussed in Sect. 4. More exactly, we deal here with the
leading twist pion wave function whose definition will be given 
shortly.

\begin{figure}
  \epsfxsize=6cm
  \centerline{\epsfbox{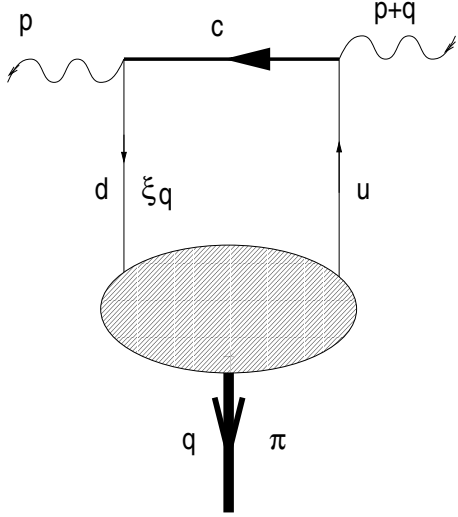}}
  \caption{Light cone sum rule for $D^*D\pi$ in the leading order.
The current labeled by $p+q$ interpolates $D$ while that
labeled by $p$ interpolates $D^*$.}
\end{figure}

If $\xi$ is the fraction of the pion momentum carried by the $d$
quark, the graph of Fig. 20 gives rise to the following expression:
\beq
F_\mu \equiv q_\mu \, F(p^2, (p+q)^2) = m_c f_\pi q_\mu \, \int_0^1 
d\xi\, \frac{\phi_\pi (\xi )}{
m_c^2 - (p+ \xi q)^2} \, + \, ... \, , 
\label{graph20}
\eeq
where $\phi_\pi (\xi )$ is the distribution function, and the ellipses 
denote corrections, both perturbative (the $\alpha_s$ expansion) and 
non-perturbative (due to higher twists). 
The $c$ quark Green's function is substituted, in the leading 
approximation, by that of a free quark. Emission of gluons from this 
line
would result either in $\alpha_s$ corrections or in the distribution 
functions of non-leading twists. 

For very small values of $q$ one could have substituted the
denominator in Eq. (\ref{graph20}) by $m_c^2 - p^2$.
Then the integral $\int d\xi \phi_\pi (\xi ) \ra 1$, and $F_\mu$
would reduce to the matrix element of the operator 
$\bar d\gamma_\mu\gamma_5 u$ sandwiched between the pion 
state and vacuum.  

Expanding the denominator in Eq. (\ref{graph20})  in $q$ we 
readily identify higher dimension operators that are summed
over: the term linear in $q$ originates from
$\bar d \stackrel{\leftarrow}{D}_\alpha\gamma_\mu\gamma_5 u$,
quadratic in $q$ from 
$\bar d \stackrel{\leftarrow}{D}_\alpha 
\stackrel{\leftarrow}{D}_\beta \gamma_\mu\gamma_5 u$,
etc. The actual expansion parameter
is 
$$
\frac{pq}{p^2 - m_c^2} \sim 1\, .
$$
The (leading twist) light-cone wave function $\phi_\pi (\xi )$
is formally defined as
\beq
\langle \pi (q) | \bar d (x) \gamma_\mu\gamma_5 \exp\{ig\int_0^x
A_\alpha (y) dy_\alpha \} u(0) |0\rangle
= - i q_\mu f_\pi \int_0^1 \, d\xi\, e^{i\xi q x}   \phi_\pi (\xi )\, .
\label{dfd}
\eeq
It parametrizes the large-distance dynamics of the
$\bar d u$ pair. 

If we do the double Borel transformation, with
respect to $p^2$ and $(p+q)^2$ the denominator in Eq. 
(\ref{graph20})
becomes
\beq
\frac{M_1^2M_2^2}{M_1^2+ M_2^2}\, \exp\left(-m_c^2\frac{M_1^2+ 
M_2^2}{M_1^2M_2^2} \right) \delta \left(u - \frac{M_1^2 
}{M_1^2+M_2^2} \right)\, ,
\eeq
where $M_{1,2}$ are the corresponding Borel parameters, that are 
expected to lie close to each other inside the window. 
The sum rule is obtained by matching  the double Borel transform of 
Eq. (\ref{ppsr}) with the double Borel transform of Eq. 
(\ref{graph20}).
Thus,  we see that
the theoretical part of the sum rule is  determined by
$\phi_\pi (0.5)$. Of course, there are corrections of higher order in 
$\alpha_s$ and from higher twist distribution functions, but as usual 
inside the window they should be kept at a modest level.
In addition we must check the stability of the result inside the 
window. Provided all this is properly done, one obtains  \cite{BBKR}
$g$ in terms of $\phi_\pi (0.5)$. The latter quantity must be 
extracted
from independent sources. 

(The situation we face here is not quite
generic. {\em A priori} one could expect the coupling $g$
to be expressible in terms of $\phi_\pi (\xi )$, rather than  one 
constant, $\phi_\pi (0.5)$.)

Information on $\phi_\pi (\xi )$ is quite abundant; there is some 
limited information on the higher-twist wave functions too.
The issue of $\phi_\pi (\xi )$ is almost as old as the SVZ sum rules 
themselves.  The key element of the corresponding theory, which 
was studied in great detail in connection 
with hard exclusive processes, is
 the (approximate) conformal 
symmetry of QCD.  Other elements are provided by the sum rule and 
lattice calculations and models. The issue is not completely settled,
although it seems that a heated debate of the 1980's gradually 
wanes. Out of two competing  scenarios --
a double hump distribution function of Chernyak and Zhitnitsky
\cite{exclusive} and a narrow distribution of Radyushkin and 
collaborators \cite{ARad} that is close to the asymptotic form of 
$\phi_\pi 
(\xi )$ --
the second is gaining more recognition now. Experts lean towards
a narrow almost asymptotic  distribution. As usual, the best strategy
is sacrificing  one of the sum rules
in order to extract  $\phi_\pi (\xi )$ from data. A recent attempt
based on the $\gamma^* + \gamma\ra \pi^0$ data has been
undertaken 
in Ref. \cite{RecKh}.  
I will not go into further details here, referring the reader to a very 
rich original  literature
devoted to this topic.

It remains to be added that (i) the term  ``light-cone sum rules'' first 
appears in  \cite{BBD}; (ii) the idea of the approach dates back to 
Refs. \cite{CS}; and (iii) a review of this topic,
with a representative list of references was published  recently 
\cite{VolBr}. 

If the partial summation of the higher-dimension operators turns out 
to be so successful in the light-cone sum rules,
it is natural to ask \cite{MRRad} why it is not applied in
the vacuum correlation functions appearing in the  SVZ sum rules.
In this case, in order to partially sum, say, the quark operators
we must substitute the quark condensate $\langle \bar q q\rangle$
by a `` string expectation value" in the vacuum,
\beq
 \langle \bar q (x) \,\exp\{ig\int_0^x
A_\alpha (y) dy_\alpha \} q(0) \rangle\,  ,
\label{nlcr}
\eeq
which is a function of $x$. 

In the light cone sum rules one can ascribe certain twist to each 
given
distribution function. Each distribution function
sums up a tower of the local operators, with all dimensions but given 
twist.
Therefore, the  distribution functions can be systematically ordered 
in twist.
Twist is obviously irrelevant in the vacuum correlation functions.
Therefore, the partial summation implemented by the 
``string expectation values" (\ref{nlcr}) is hard to justify
theoretically. A principle that would allow us to
order distinct ``string expectation values"  has to be elaborated.

Another major difference 
with the light cone sume rules is that the pion distribution function
 is known at least at a certain level, while
next to nothing is known about the vacuum function (\ref{nlcr}).
Until recently even the large $x$ asymptotic behavior of this
function was erroneously assumed to be Gaussian.
We are at the initial stage, when first observations are being 
made regarding properties of the function (\ref{nlcr})
following from the general structure of QCD \cite{Shif1}. 
I do not rule out that,  as our understanding progresses,
the corresponding modification of the sum rule approach 
 will be elaborated. 

\subsection{Heavy flavor sum rules}

Although the character of problems in which this
formalism is instrumental is somewhat different from the classical 
applications of the SVZ sum rules, the basic techniques --
OPE, matching of the theoretical expansions with the 
phenomenological expressions represented by dispersion integrals
over the observable spectral densities, etc. -- are the same.
The vacuum condensates are replaced by the expectation values of 
local operators over the heavy flavor states, say, $B$ mesons.
All questions that can be asked in connection with the vacuum 
condensates exist
 here too; the answers are more transparent, however.
We will consider below a sample application. Our main goal is 
discussing subtle nuances associated with the condensate expansion
(Wilsonean OPE versus practical version, see Sect. 6).

 The heavy flavor sum rules were engineered \cite{BSUV}
to treat the inclusive heavy flavor decays, for instance
the semileptonic decays 
$B\ra X_c \ell\nu$ where $X_c$ is arbitrary hadronic state 
containing $c$ quark. The lowest-lying state is
the $D$ meson, then comes $D^*$ and then higher 
excitations/continuum. By varying the momentum of the lepton pair
one can measure various spectra
in  the transition $b\ra c$ induced by the weak
current. The object of the theoretical study is the {\em transition 
operator}
\beq
{\hat T}_{ab}(q) = i\int d^4 x \;{\rm e}^{iqx} \, T\{ j_a^\dagger (x) 
j_b (0)\} \, ,
\label{tpro}
\end{equation}
that will be eventually sandwiched between the states
$|B\rangle$ and $\langle B|$ (see Fig. 21). 
Here $j_a$ denotes a current of the type 
$\bar{c}\Gamma_{a}b$ with an arbitrary Dirac matrix 
$\Gamma_a$; $q$ is the momentum carried away by 
the lepton pair. 

\begin{figure}
  \epsfxsize=8cm
  \centerline{\epsfbox{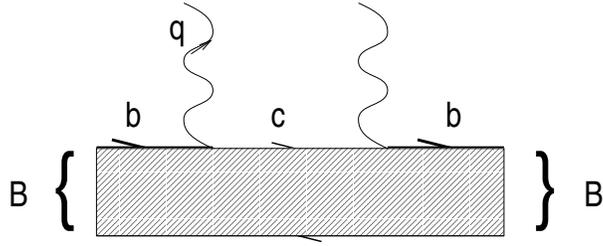}}
  \caption{The operator product expansion for $h_{ab}$.
For large negative $\epsilon$ the $c$ quark is far off shell, and the 
points of the current emission and absorption are close. The $B$ 
meson in the initial and final states is assumed to be at rest.}
\end{figure}

The average of ${\hat T}_{ab}$ over 
the heavy hadron  state $B$ with the
momentum $p_{B}$  represents a forward scattering 
amplitude (the so-called hadronic tensor),
\begin{equation}
h_{ab}(p_{B},q) = \frac{1}{2M_{B}} 
\langle B|\hat{T}_{ab} | B\rangle\;.
\label{htsr}
\end{equation}
The
structure functions $w_{ab}$ are obtained from the
hadronic tensor by taking its imaginary part, 
$$
w_{ab} = (1/i)\, {\rm disc}\,\, h_{ab}\, .
$$

All  observable spectra  are represented as certain integrals
over $w_{ab}$.
The hadronic tensor $h_{ab}$  and the structure functions $w_{ab}$
can be expanded in terms of various kinematic  structures.
For each set of the currents generically we  may have  up to five
distinct kinematic  structures; these details need not bother us here.
For illustrational purposes it is sufficient to limit ourselves to  the 
axial currents, so that $\Gamma_{a,b} \ra \gamma_\mu\gamma_5$,
and to the first function,
$$
h_{\mu\nu} = -h_1 g_{\mu\nu} + ...\, , \,\,\,  w_{1} = (1/i)\, {\rm 
disc}\,\, h_{1}\, .
$$
The lowest-lying state produced by the axial current from $B$ is 
$D^*$.
To end up  with the kinematical aspects we must specify the
lepton pair momentum $q$. It will be assumed that
the spatial momentum $\vec q$ is small but non-vanishing,
$\Lambda \ll |\vec  q 
|\ll M_D$. 
 The terms
${\cal O}({\vec q}^2)$ will be kept while those of higher order in
$|\vec q |$ will be neglected. Since $\vec q$ is fixed, 
$h_1^{AA}$ becomes a function of one (complex) variable,
$q_0$. Instead of $q_0$ it is more convenient to work with
$\epsilon$ defined as
\begin{equation}
\epsilon = q_{0max}-q_0
\end{equation}
where 
\begin{equation}
q_{0max} = M_B- E_{D^*},\,\,\, E_{D^*}= M_{D^*}+\frac{{\vec 
q}^2}{2M_{D^*}}.
\label{q0max}
\end{equation}
When $\epsilon$ is real and positive we are on the
physical  cut where the imaginary part of $h_1^{AA}$ is measurable. 
For negative $\epsilon$ we are below the cut, in the Euclidean 
domain, where the $T$ product (\ref{tpro}) 
can be computed as an expansion in $1/m_{c,b}$. 
The leading operator in this expansion is $\bar bb$. It has dimension 
three. No operators of dimension four exist. At the next
level, of dimension five, there are two operators,
$\bar b (i\vec D )^2b$ and $(ig/2)\bar b \gamma^\mu\gamma^\nu
G_{\mu\nu} b$. The first operator (sometimes called the
{\em kinetic energy operator}) is not a Lorentz scalar. In the problem 
considered the operators to be retained in the expansion
need not necessarily be Lorentz scalars since, unlike  the vacuum 
case, the very presence of the $B$ meson in the initial/final state 
singles out a reference frame.

The whole procedure is
perfectly analogous to the $1/Q^2$
expansion of the polarization operator discussed in Sect. 4. 
The operator $\bar bb$ occupies the same position in the hierarchy 
as the unit operator in the expansion of the vacuum correlators.
The operators of dimension five and higher generate $1/m_{c,b}$
corrections analogous to the condensate corrections. In the problem 
at hand all
 local operators appearing in the  expansion of the $T$ product
(\ref{tpro}) are 
averaged
over the $B$ meson state. In this way we get a theoretical expression 
for $h_1^{AA}$  off the cut. 

A bridge between $h_1^{AA}$ at negative 
$\epsilon$ and $w_1^{AA}$ at positive $\epsilon$
is provided by the dispersion relation. Expanding the 
denominator in the dispersion relation in $1/\epsilon$ we obtain an 
infinite set of the sum rules. The third of them  is as follows
\cite{Grozin}:
\begin{equation}
\frac{1}{2\pi}\int d\epsilon \, \epsilon^2\, w_1^{AA}(\epsilon ) 
 = \frac{1}{3}\mu_\pi^2\, {\vec v}^2 + ...\, .
\label{srgr}
\end{equation}
Here $\vec v$ is the velocity of the produced  heavy hadron (in the 
$B$ rest frame), and 
$\mu_\pi^2$ is defined as the expectation value
\beq
\mu_\pi^2 =\frac{1}{2M_B}\langle B|\bar b (i\vec D )^2b|B\rangle 
\, .
\eeq

This parameter, $\mu_\pi^2 $,  has the meaning of the average 
spatial momentum squared of the 
heavy quark inside $B$ meson, and is one of the fundamental
parameters of the heavy quark theory. One encounters
$\mu_\pi^2 $ in a large number of applied problems. It is fair to say 
that in the heavy quark physics $\mu_\pi^2 $ plays the same role as
the gluon condensate  in classical applications of the SVZ method. 

The ellipses in Eq. (\ref{srgr})
denote corrections suppressed by powers of $1/m_{c,b}$. 
Since the weight function in Eq. (\ref{srgr}) is proportional to
$\epsilon^2$, the ``elastic" $B\ra D^*$ transition drops out, and the 
integral on the left-hand side is saturated by ``inelastic" transitions,
i.e. decays of the $B$ meson in the excited $X_c$ states.
For such transitions $w_1^{AA}(\epsilon ) $ is proportional to
${\vec v}^2$ at $|{\vec v}|\ll 1$.
It is convenient then to introduce
a spectral function $\sigma (\epsilon )$ which does not vanish
in the limit $|{\vec v}|\ra 0$,
\beq
\sigma (\epsilon ) = |{\vec v}|^{-2}w_1^{AA}(\epsilon ) \, , \,\,\, 
\epsilon > 0\, .
\label{defsde}
\eeq

Our brief excursion in the heavy flavor sum rules is almost over.
In terms of the new spectral density
the sum rule we are interested in takes the form
\begin{equation}
\frac{1}{2\pi}\int d\epsilon \epsilon^2\, \sigma (\epsilon ) 
 = \frac{1}{3}\mu_\pi^2 \, .
\label{srgroz}
\end{equation}
As with any sum rule it can be read in both directions:
if $\mu_\pi^2$ is known this is a prediction for the
spectral density. On the other hand, if  the
spectral density is measured we can take  the integral and get 
$\mu_\pi^2$. Below we will pursue the latter strategy assuming that
$\sigma (\epsilon ) $ is measured. In actuality this is not the case,
but we have a pretty good idea of how $\sigma (\epsilon )$
will look like when it is measured.  First of all, it is 
positive-definite.
At small values of $\epsilon$, of order of a few units time 
$\Lambda$, there is one, perhaps, two resonances;
then at higher values of $\epsilon$ the spectral density 
must approach its perturbative asymptotics which is known too:
$\sigma (\epsilon ) \ra C\alpha_s/\epsilon$, where
$C$ is a constant which was calculated in Ref. \cite{BSUV}.  
The numerical value of $C$ is unimportant for our purposes. 
The asymptotic regime is achieved through oscillations --
 the  reasons are the same as  in Sects. 2.2 and 4. 
A schematic plot of  $\epsilon^2 \sigma (\epsilon )$ is depicted in Fig. 
22.

\begin{figure}
  \epsfxsize=9cm
  \centerline{\epsfbox{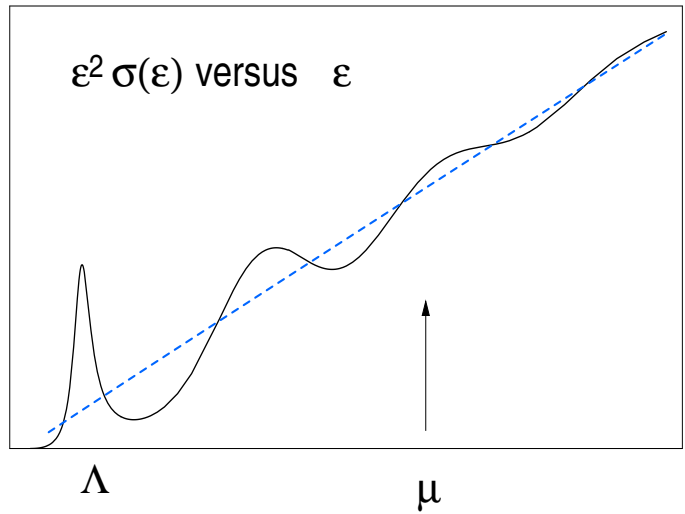}}
  \caption{A sketch of $\epsilon^2 \sigma (\epsilon )$ 
in the $b\ra c$ transition at small $|\vec v|$ (solid curve). The 
spectral density $\sigma (\epsilon )$ is defined in Eq. (\ref{defsde}).
The dashed curve represents $\epsilon^2 \sigma (\epsilon )$  in 
perturbation 
theory at one loop.}
\end{figure}

If you think about Eq. (\ref{srgroz}) you will, probably, find it 
remarkable: it gives a condensate-like non-perturbative parameter
in terms of an integral over the observable spectral density.
Further examination raises a question. Indeed, if the integral extends 
to infinity, it badly diverges. 

One need not be concerned, however. 
This infinity reflects the fact that the proper definition of 
$\mu_\pi^2$ requires a normalization point, just in the same way as 
the for the gluon condensate. In the case at hand this normalization 
point has a very transparent physical meaning --
$\mu$ is the maximal excitation energy of the hadronic state 
belonging to the class of ``soft excitations" (Fig. 22).  Higher 
excitations are 
considered to be hard, and are well approximated by perturbation 
theory. I remind that the hard contributions do not belong to 
$\mu_\pi^2$; rather, they determine the coefficient functions in OPE. 
In the perturbative 
calculation of the coefficient functions one must remove
the part with $\epsilon < \mu$.
In fact,  the cut off in the excitation energy
as a normalization point 
was suggested in the given context long ago in Ref.  \cite{NIMW}.  
Thus, in the 
Wilsonean approach
$\mu_\pi^2 (\mu)$  is determined as
\begin{equation}
\mu_\pi^2 (\mu ) = \frac{3}{2\pi}\int_0^\mu  d\epsilon \epsilon^2\, 
\sigma (\epsilon ) 
\, .
\label{muwils}
\end{equation}

In  practical OPE we deal, instead, with subtracted ``surrogates".
Say, if one  limits oneself to one-loop accuracy,
$$
``Ê\mu_\pi^2 " = \frac{3}{2\pi}\int_0^\mu  d\epsilon \epsilon^2\, 
\left[ \sigma (\epsilon ) -\sigma_{\rm 1-loop} (\epsilon )  \right]\, .
$$
The right-hand 
side in this expression is  $\mu$ independent at one loop. Its $\mu$
dependence shows up only at two loops. If we agree to discard all
effects ${\cal O}(\alpha_s^2)$ and higher, {\em just totally ignore 
them}, then 
the integration in $``Ê\mu_\pi^2 " $ can be extended to infinity. 
Correspondingly, the coefficient functions 
must be  calculated at one-loop, following the standard Feynman 
rules, with no 
removal of the $\epsilon < \mu$ part. 

\section{Sum Rules and Lattices}

In this Lecture it has been already  noted, more than once, 
that certain aspects of the condensate approach, as well as a rich 
experience in the analysis of various spectral densities, might be 
useful for lattice practitioners. Remarks to this effect are scattered 
here and there. Of particular importance are qualitative observations, 
such as an enhanced sensitivity to the vacuum fermion loops in 
certain channels (this effect is neglected in the quenched 
approximation), the existence of new, abnormally large scales
in the $0^{\pm}$ glueball channels, and so on. Surprising though it is, 
all these
 hints, which might have produced a strong impact on the lattice 
theory,
are so far almost totally ignored
\footnote{I am aware of two dedicated works where the sum rule 
results where analyzed in parallel  with the lattice calculations 
\cite{SRLC}. 
This seems to be a rare exception.}. The reaction of rejection of any 
ideas coming from outside the lattice theory {\em per se} plagues the 
lattice community. This happens even in those cases 
where the overlap is quite 
obvious.
  
Let me give an almost anecdotal example. As was mentioned in 
Sect. 7, in 1981 a wide class
of ``QCD scale anomaly" low-energy theorems was obtained
\cite{NSVZ} in connection with the SVZ sum rules. These theorems 
relate to each other $n$-point functions with arbitrary number of 
insertions of the operator
$
\sigma (x) = \theta^\mu_\mu (x) \, \propto G^2$ (see Eq. 
(\ref{aemt}))
at vanishing momentum. One of this theorems can be cast in the  
form
of a prediction for the expectation value of
$G_{\mu\nu}^2$ in the presence of the Wilson loop operator $W$,
\beq
\left\langle W\right\rangle^{-1}
\left\langle \int d^3 x \sigma (x) W\right\rangle_{\rm E,c} =
V(R) + R\frac{\partial V (R)}{\partial R}
\label{let}
\eeq
where $V(R)$ is the static potential between the heavy quarks
separated by distance $R$, 
$$
W =\oint_C \,\mbox{Tr}\, \exp (igA_\mu (x) dx_\mu )\, ,
$$
the contour $C$ in the definition of the 
Wilson loop is depicted on Fig.  23, and the subscript E reminds us 
that all definitions and calculations refer to the Euclidean space. 
Moreover, the integration in Eq. (\ref{let})  is performed
in three directions perpendicular to the Euclidean time;
the subscript c refers to the connected part, i.e. the disconnected
part of the correlation function (\ref{let}) must be discarded.
One can interprete the operator $\sigma(x)$ in Eq. (\ref{let})
as a probe of the energy density of the vacuum medium at the point 
$x$
in the presence of a pair of heavy static quark sources 
separated by distance $R$ \cite{NSVZ}. 
In the case of purely linear potential the right-hand side
reduces to $2V(R)$. In this form the theorem (\ref{let}) was derived 
in Ref. \cite{NSVZ} (see Appendix B). The generalization to the case of 
arbitrary 
potentials is due to Dosch {\em et al.} \cite{Dosch}.

\begin{figure}
  \epsfxsize=7cm
  \centerline{\epsfbox{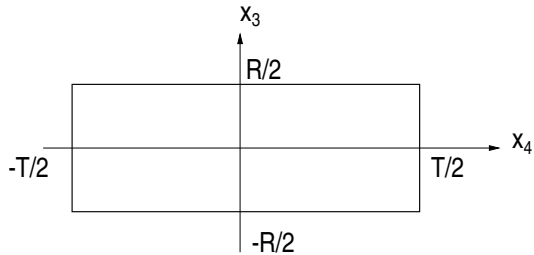}}
  \caption{Rectangular contour in the Wilson loop operator. One can 
consider it as a trajectory of infinitely heavy quarks (in the 
imaginary time). $R$ is the distance between the quark and 
antiquark, $T\ra\infty$. }
\end{figure}

Derivation of the theorem  (\ref{let}) is very similar to other 
theorems discussed in Sect. 7.  The functional-integral representation 
for $\langle W \rangle$ is differentiated with respect to
$1/g_0^2$. We then use the fact that 
\beq
\frac{\partial V (R)}{\partial (1/4g_0^2)} =-\frac{32\pi^2}{b}
\left(ŽÊV(R) + R\frac{\partial V (R)}{\partial R} \right) \, .
\label{valphad}
\eeq
The latter equality, in turn, follows from consideration similar to that 
presented in  Sect. 7. Thus, the low-energy  theorem  
(\ref{let}) is a reflection of the scale anomaly of QCD, and as such,
represents a fundamental aspect of the theory. 

Now I come to a culmination of the story. In 1987 the theorem 
(\ref{let}) was rediscovered in the lattice community
\cite{Michael}, where it goes under the name of the ``action Michael 
sum rule".  The original derivation was erroneous, the second term 
on the right-hand side of Eq. (\ref{let}) was omitted. The error was 
corrected in Ref.  \cite{Dosch} which was specifically devoted to the 
issue. Nevertheless, the theorem continues to circulate in the lattice 
community as the ``Michael sum rule" , see e.g. \cite{Rothe}. 

This example of a ``lattice xenophobia"   is by no means unique. 
Needless to say that I consider the situation as absolutely unhealthy. 
The problems we face in the hadronic world are too complicated  to 
afford neglecting the work done in related areas.
The analytic and numerical approaches should complement each 
other. I sincerely hope that the attitude of the lattice community will 
start changing after recent breakthrough discoveries in 
supersymmetric gauge theories. Supersymmetry is a powerful tool
which allows us to reveal features of strong coupling gauge dynamics 
we never suspected of. One of the lessons which seems undeniable is
the proof that massless (or light) quarks play  much more significant 
a role than is universally believed among the lattice practitioners. 
In the quenched approximation, the most wide spread today,
the dynamical quarks are neglected altogether. 
Even the most advanced lattice investigations, which do not 
rely on the quenched approximation, routinely use extrapolations in 
the number of the massless quarks. Supersymmetric gauge
theories teach us that increasing the number of massless quarks 
from two to three
or changing the color representation of the fermion fields from
fundamental to adjoint is not necessarily a smooth process.
Dynamics may change drastically!

Penetration of ideas and insights obtained analytically,
in the practice of lattice calculations, where they can be used, at the 
very least, to test  reliability of various approximations, is 
unavoidable. First steps in this direction have been already reported 
\cite{EHS}. 

\section{Vacuum Fluctuations Are Subtle Creatures}

The rich SVZ phenomenology confirms the wide-spread belief that
the structure of the hadronic family, in all its gross features and  
intricacies, derives from a complicated organization of the vacuum 
medium. The vacuum fluctuations are violent, but  fine-tuned in a 
very specific way. Within the  method, as it exists today, we do not 
calculate the vacuum condensates. This does not mean, of course,  
that the origin of the condensates, and their relative values, are of no 
interest. I merely wanted to say that the problem  
of the vacuum structure in analytic QCD is not solved in full. 
There are indirect methods, however, which show that this structure
is very subtle -- much more subtle than one might think of {\em a 
priori} -- and depends on seemingly insignificant details of the 
theory. For instance, changing the number of colors from three to two
dramatically changes the pattern of the spontaneous chiral 
symmetry breaking resulting in a drastic rearrangement in the 
Goldstone sector of the hadron family. Instead of $N_f^2 -1$  QCD 
``pions"  ($N_f$ stands for the number of the light quark flavors) 
we get $2N_f^2 -N_f -1$ massless Goldstone hadrons
\footnote{It is curious to note
that the existence of the extra massless states was discovered
in the lattice  $SU(2)$ theory, which is being used   as a 
theoretical laboratory for decades, only recently. The result is known 
in analytic QCD since the mid-eighties. This is another manifestation 
of the ``lattice xenophobia", see Sect. 11.}. This conclusion can be 
achieved 
\cite{VKS} from consideration of the 't Hooft matching conditions
\cite{Hooft3} (for earlier analyses based on less rigorous  arguments 
see 
Ref. \cite{Peskin}).

Let us dwell on this issue. 
Assume that we have $N_f$ massless (light) quarks but the 
color gauge group is $SU(2)$. The quarks belong to the fundamental 
representation of $SU(2)$. Unlike QCD, in which the color triplets and 
antitriplets are independent representations, the  $SU(2)$ doublets 
are essentially the same as antidoublets. In other words, all 
representations of $SU(2)$ are (quasi)real.

This simple observation has far reaching consequences. Indeed, the 
flavor symmetry of the Lagrangian is $SU(2N_f)$ now, rather than
$SU(N_f)\times SU(N_f)$ we deal with in QCD. Out of $4N_f^2 -1$ 
currents $2N_f^2 -N_f -1$ are axial; all  correspond to the 
spontaneously broken symmetries since there is no way one can 
achieve the matching of the triangular AVV anomalies otherwise
\cite{VKS}.
This means that the pattern of the chiral symmetry breaking 
is $SU(2N_f) \rightarrow Sp(2N_f)$ rather than the $SU(N_f)\times 
SU(N_f)\rightarrow SU(N_f)$ we got used to in QCD. The very gross 
features of the vacuum structure do depend on the number of colors!

Not only do they depend on the number of colors, they are also 
sensitive 
to the fermion contents of the theory.

Indeed,  let us substitute the standard quarks of QCD with
one Majorana fermion $\lambda$ in the adjoint representation of the 
color group. The theory thus obtained is nothing but  
supersymmetric gluodynamics. Superficially it looks very similar to 
conventional QCD. At first sight, one can hardly expect any 
conspicuous deviations from the conventional pattern of behavior.

It has been known for many years  that the supersymmetric gauge 
theories posses some miraculous properties. In particular, in
supersymmetric gluodynamics the Gell-Mann--Low function is 
known {\it exactly} \cite{NSVZbeta}
\begin{equation}
\beta (\alpha_s ) = -\frac{\alpha^2_s}{2\pi}\,\,  \frac{3N}{1-
(N\alpha_s
/2\pi )}\, ,
\label{beta}
\end{equation}
where the $SU(N)$ gauge group is assumed. 

As QCD,  supersymmetric gluodynamics is believed to be
confining in the infrared domain.  Unlike QCD, however, the 
condensate $\langle \lambda^2\rangle$ may or may not develop.
It was argued recently \cite{KovS} that two distinct phases coexist in 
this theory -- the conventional chirally asymmetric phase
with $\langle \lambda^2\rangle \neq 0$, that reminds QCD,
and a very unusual  chirally symmetric phase where $\langle 
\lambda^2\rangle = 0$. The arguments leading to this conclusion are 
too technical to be reproduced here. They would lead us far astray.
The interested reader is referred to Ref. \cite{KovS}.

Instead, let us dwell on another example which is pretty close to 
actual QCD. In fact, we will consider just conventional QCD with more 
than three massless flavors. 

The Gell-Mann-Low function in QCD has the form \cite{3loop}
$$
\beta (\alpha_s ) = -b  \frac{\alpha_s ^2}{2\pi}
- b_1 \frac{\alpha_s ^3}{4\pi^2} - ...\, ,
$$
\beq
b = 11 -\frac{2}{3} N_f\, , \,\,\, b_1 = 51 -\frac{19}{3} 
N_f\,  .
\label{betaQCD}
\eeq
At small $\alpha_s$ it is negative (asymptotic freedom!) since the 
first term always 
dominates. With the scale 
$\mu$ decreasing the running gauge coupling constant grows,
and the second term becomes important. Generically the second term 
takes over the first one at $\alpha_s/\pi \sim 1$, when all terms in 
the $\alpha_s$ expansion are equally important, i.e. in the strong 
coupling regime. We do not know what exactly happens at 
$\alpha_s/\pi \sim 1$.
Assume, however, that for some reasons the first 
coefficient $b$ is abnormally small, and this smallness does 
not 
propagate to higher orders. Then the second term catches up with 
the first one when $\alpha_s/\pi \ll 1$, we are in the weak coupling 
regime, and higher order terms are inessential. Inspection of Eq. 
(\ref{betaQCD}) shows that this happens when $N_f$ is close
to $33/2$, say 16 or 15 ($N_f$ has to be less than $33/2$ to ensure 
asymptotic freedom). For these values of $N_f$ the second coefficient
$b_1$ turns out to be negative. This means that the $\beta$ 
function develops a zero in the weak coupling regime, at
\beq
\frac{\alpha_{s}^{*}}{2\pi} = \frac{b}{-b_1} \ll 1\, .
\eeq 
(Say, if $N_f = 15$ the critical value is at 1/44 and is indeed small.)
This zero is nothing but the infrared fixed point of the theory. 
At large distances $\alpha_s \ra \alpha_{s}^{*}$,
and $\beta (\alpha_{s}^{*}) = 0$, implying that the trace of the 
energy-momentum vanishes. Then the theory is in the conformal 
regime. There are no localized particle-like states in the spectrum of 
this theory; rather we deal with massless unconfined interacting 
quarks and gluons; all correlation functions at large distances exhibit 
a power-like behavior. In particular, the potential between two 
heavy static quarks at large distances $R$ will behave as
$\sim \alpha_{s}^{*}/R$, i.e. we have a pure Coulomb behavior. The 
situation is not drastically different 
from 
conventional QED.  As long as $\alpha_{s}^{*}$
is small, the interaction of the massless quarks and gluons
in the theory is weak at all distances, short and large, and is 
amenable to the standard perturbative treatment (renormalization 
group, etc.). 
 The chiral symmetry 
is not broken spontaneously, the quark condensates do not develop. 
QCD becomes a fully calculable theory. 

The fact that at $N_f$ close to 16 QCD becomes
conformal and weakly coupled in the infrared limit is known
for about 
 20 years 
\cite{MB,BZ}. If the conformal regime takes place at $N_f= 16 $
and 15,  let us ask ourselves how far can we descend in $N_f$
without ruining the conformal nature of the theory in the infrared?
Certainly, when $N_f$ is not close to 16 the gauge coupling 
$\alpha_{s}^{*}$ is not small. The theory is {\em strongly coupled}, as 
conventional QCD,  and, simultaneously, 
conformal in the infrared. The chiral symmetry presumably 
is unzbroken,  $\langle \bar qq\rangle = 0$. The vacuum structure is 
totally different, and so are all properties of the theory. 

The range of $N_f$ where this phenomenon occurs is called
the {\em conformal window}. The right edge of the conformal 
window is at $N_f= 16 $. There are indications that
the left edge of the conformal window $N_{f}^*$ lies not too far from 
actual
QCD (i.e. not too far from $N_f=3$). 

First, in the instanton liquid model it was found
\cite{CW1} that at $N_f=5$ the chiral condensate disappears,
$\langle \bar qq\rangle = 0$. Although nothing is said about the 
onset of the conformal regime in this work, the vanishing
of $\langle \bar qq\rangle $ may be a signal.  More definite is the 
conclusion of the lattice investigation \cite{CW2}. It is claimed 
that at $N_f=7$ not only the quark condensate disappears,
$\langle \bar qq\rangle = 0$, but the decay law of the correlation 
functions at large distances changes from the exponential to 
power-like, i.e.
the theory switches from the confining regime to the conformal one.
One could speculate that at $N_f = 5$ and 6
we are in the intermediate phase, when confinement is still operative 
at large distances but the chiral symmetry is unbroken. 
Previously it was believed that 
that confinement of color in QCD implies the chiral symmetry 
breaking \cite{Casher}. Today's wisdom tells us that this need not be
necessarily  the case. At least in some supersymmetric gauge 
theories we have both: confinement and unbroken chiral symmetry
(see e.g. \cite{KovS}).  In any case, we see that
ascending  from $N_f=3$ to $N_f=5$ or 7 --  quite a modest 
variation-- 
we 
drastically change the vacuum structure of the theory,
with the corresponding abrupt change of the 
picture of the hadron world.

An interesting question is what happens slightly below the left edge 
of the conformal window $N_{f}^*$. If $N_f$ were a continuous 
parameter, and the phase transition in $N_f$ were of the second 
order,
slightly below $N_{f}^*$ the string tension $\sigma$ would be 
parametrically small
 in its natural scale given by $\Lambda^2$. In actuality
$N_f$ changes discretely. Still it may well happen that
at $N_f = N_{f}^*-1$ the ratio $\sigma/\Lambda^2$
is numerically small. This would mean that the string and its 
excitations, whose scale is set by $\sigma$, are abnormally light.
Under the circumstances one might hope to build
effective low-energy approaches analogous
to the  chiral Lagrangians of actual QCD. In the limit 
$\sigma/\Lambda^2
\ra 0$ the string  becomes, in a sense, classic; the quantum 
corrections 
are unimportant. I do not rule out that the string representation of 
QCD -- the holy grail of two generations of theorists --
is easier to construct in this limit. One could even dream of an 
expansion in $N_{f}^*-3$. At this moment, this is a pure speculation, 
however, and I have to wind up.

\section{Instead of Conclusions}

As time passes the hopes that the full analytic solution of QCD will be 
found fade away. After all, the problem is with us for over a quarter 
of the century, and none of numerous theoretical attacks reached the 
goal. 
Many theorists whose philosophy is ``all or nothing"
abandoned the field \footnote{The ``all or nothing" philosophy is 
wide-spread and, unfortunately, not only in theoretical physics.
This is a favorite child of the so called revolutionaries
in all times and in all countries. The misfortunes it brought to our
world are innumerable. Needless to say, I personally think this is a 
rotten philosophy.}. Does it mean that the theory  of hadrons
is an aging science at the verge of submerging into a 
permanently dormant state? 

I do not think so. Those theorists who stayed in the field can claim 
many partial successes. The field is messy and down-to-earth but -- 
what can we do? -- this is the only world God gave to us.
The theory of hadrons is alive.  
Our understanding of the vacuum structure of quantum 
chromodynamics and various dynamical regimes it can support
continues to grow. Especially fruitful were the last years,
with many breakthrough discoveries in supersymmetric
gauge theories. These discoveries are to be used as hints and insights 
in actual QCD.

OPE-based methods -- and the SVZ sum rules is one of them --
continue to grow too. They are admittedly approximate,
but their analytic nature and a graphic physical interpretation 
make them indispensable in many practically important problems.
In some issues they share their role with other methods,
which were developed later, for instance, lattice QCD.
Since the theory of hadrons is so difficult it is in our best interests
to combine all sources of information. We are in no position to 
neglect one of them in favor of the other.
Many ideas and theoretical devices surfaced in connection with the 
SVZ method (e.g. the low-energy theorems, Sect.
7); the method  also produced a number of variations and spin-offs.
This is a healthy process which will hopefully continue.

\section*{Acknowledgments}

 It is my pleasure to thank 
Prof. 
T. Suzuki and Prof. T. Kunihiro who were my hosts during my
1997 visit to Japan, for their kind hospitality. Prof. T. Suzuki kindly 
arranged this lecture in the framework of the International
Workshop {\em Non-Perturbative QCD -- Structure of the QCD 
Vacuum}, 
Yukawa Institute for Theoretical Physics, Kyoto, December 2--12,
1997.

Useful communications with V. Braun, H.G. Dosch, B.L. Ioffe, A. 
Kaidalov, A. 
Khodjamirian,  A. Radyushkin   and E. Shuryak are 
acknowledged. I would like to thank A. Vainshtein for numerous
illuminating discussions. 

This work was supported in part by DOE under the grant number 
DE-FG02-94ER40823. 

\newpage

\section{Appendix. Some Useful Definitions and Formulae.}

\renewcommand{\theequation}{A.\arabic{equation}}
\setcounter{equation}{0}

The Gell-Mann--Low function is defined as
\beq
\frac{\partial\alpha_s}{\partial\ln \mu} \equiv
\beta (\alpha_s) =  -b \frac{\alpha_s^2}{2\pi} 
-b_1\frac{\alpha_s^3}{4\pi^2}
+ ...
\label{aone}
\eeq
This definition is standard except that in some sources the first
coefficient is called $\beta_0$, the second $\beta_1$, and so on
(cf. Ref. \cite{3loop}). 
At two loops the $\beta$ function is scheme independent.

The coefficients are
\beq
b = 11 -\frac{2}{3}\, N_f\, , \,\,\, b_1 = 51 -Ê\frac{19}{3}\, N_f\, .
\label{atwo}
\eeq
The running coupling $\alpha_s (\mu)$ is parametrized
as follows
\begin{equation}
\alpha_s (\mu) 
=\frac{4\pi}{b\, \ln \frac{\mu^2}{\Lambda^2}}\,
\left( 1-\frac{2b_1}{b^2}\, \frac{\ln \ln 
\frac{\mu^2}{\Lambda^2}}{\ln 
\frac{\mu^2}{\Lambda^2}} + ...
\right)\, .
\label{athree}
\end{equation}
A more convenient form of the very same expression
is
\beq
\frac{2\pi}{\alpha_s (\mu) } = b\, \ln \frac{\mu}{\Lambda}
+\frac{b_1}{b}\ln \ln \frac{\mu^2}{\Lambda^2}+ ...\, .
\label{afour}
\eeq
Being expanded, at two loops, the latter  expression is  equivalent
to the former; Eq. (\ref{afour})
effectively sums up some higher order terms.

The scale parameter $\Lambda$ following from Eq. (\ref{athree})
is
\beq
\Lambda^b =\left(\frac{2}{b} \right)^{b_1/b}
\mu^b \exp\left\{ -\left[ \frac{2\pi}{\alpha_s (\mu)} -\frac{b_1}{b} 
\,
\ln  \frac{2\pi}{\alpha_s (\mu)} \right] \right\}\, .
\label{afive}
\eeq
The first factor on the right-hand side is an (inconvenient) artifact 
of the definition. It could have been easily avoided.
Since the definition is standard, we will keep it not to cause 
confusion. 

Perturbative calculations in QCD, as a rule, are carried out in the so 
called
modified minimal subtraction ($\overline{\rm MS}$) scheme
\cite{Bardeen}, using dimensional regularization. For massless quarks 
and 
gluons it works nicely; the treatment of the heavy quark mass 
thresholds
in this approach is ugly. The most common is the step-function 
approximation
\cite{Barnett}: immediately above the threshold the quark
is declared massless, while below the threshold it is treated as
infinitely heavy, so that it is frozen out and does not participate in 
loops.
This is equivalent to treating $N_f$ as a function of
$\mu$ constructed from several step functions,
$N_f= 3$ below $m_c$, then it jumps to $N_f=4$
between $m_c$ and $m_b$, etc. Matching conditions at thresholds
require equivalence of one effective theory with $N_f$ massless 
quarks
to another effective theory with $N_f-1$ massless quarks.

If the step-function approximation is applied,
at one loop  the coupling constant is obviously continuous,
while the first $\mu$ derivative experiences a jump.
At two and three loops the coupling itself becomes discontinuous
if the matching is done at the quark masses \cite{Wetzel}.
Several suggestions as to how one can smooth out the running 
coupling 
constant
using various ``physical" definitions were presented  in the
literature \cite{BGR}. I propose a somewhat different approach 
inspired by 
supersymmetry. In supersymmetric theories the quark mass 
thresholds
can be  accounted for {\em exactly}, to all orders
\cite{ShiSUSY}. For instance, consider supersymmetric QCD with one 
flavor,
with the mass term $m$.  Assume that we start our evolution at a 
high
normalization point $M_0$ (the corresponding coupling constant is 
$\alpha_{s0}$), pass the threshold and descend down to $\mu \ll m$.
A continuously running $\alpha_s$ can be obtained from the formula
\beq
\frac{2\pi}{\alpha_s(\mu )}= \frac{2\pi}{\alpha_{s0}} - 3N\ln
\frac{M_0}{\mu (\alpha_{s0}/\alpha_s(\mu ))^{1/3}}
+\ln \frac{M_0}{\mu  Z(\mu )}\, ,
\label{asix}
\eeq
for SU(N) gauge group. Here $Z(\mu ) $ is the $Z$ factor of the matter 
fields and $\mu$ is arbitrary: larger or smaller than $m$.
At  $\mu = m$ and below  the second logarithm freezes at 
$\ln M_0/m_0$ where $m_0= mZ$ is the mass parameter
normalized at $M_0$. 
The one loop $Z$ factor implies  $\alpha_s(\mu )$
 at two loops. Needless to say that $Z(\mu ) $
varies continuously at one loop. 
At the two-loop level the high- and low-energy
scale parameters are related as follows:
\beq
\Lambda_{\rm LOW}^{3N}
= \Lambda_{\rm HIGH}^{3N-1}
\left( \frac{2}{3N}\right)^N \left( \frac{3N-1}{2}\right)^{N-
\frac{2C_2}{3N-1}}\, 
\left[ m_0 \left( \frac{2\pi}{\alpha_{s0} } \right)^{2C_2/(3N-
1)}\right]\, ,
\label{aseven}
\eeq
where
$$
C_2 =\frac{N^2-1}{2N}\, .
$$
At this level of accuracy $m_0$ and $\alpha_{s0} $
in the square brackets can be treated  in the leading logarithmic 
approximation. In this approximation the expression in the
 square brackets is renormalization-group invariant. In principle, the 
procedure can be extended to all loops, but we will not pursue this 
goal here.

In non-supersymmetric QCD the exact treatment of the mass 
thresholds seems impossible. However, numerically the situation is 
quite close to what we have in supersymmetric QCD. One can work 
out compact formulae applicable at two loops. Say, for the charm
threshold
\beq
\frac{2\pi}{\alpha_s(\mu )}= \frac{2\pi}{\alpha_{s0}} - 9\ln
\frac{M_0}{\mu (\alpha_{s0}/\alpha_s(\mu ))^{32/81}}
+\frac{2}{3}\ln \frac{M_0}{\mu Y(\mu )}\, ,
\label{aeight}
\eeq
where 
$$
Y= \left[ \frac{\alpha_{s0}}{\alpha_s(\mu )}\right]^{3y/25}
$$
and
$$
y = \frac{107}{18}\, .
$$
Correspondingly,
\beq
\Lambda_{\rm LOW}^{9}
= \Lambda_{\rm HIGH}^{25/3}
\left( \frac{2}{9}\right)^{32/9} \left( \frac{25}{6}\right)^{77/25}\, 
\left[ m_0\left( \frac{2\pi}{\alpha_{s0}  } 
\right)^{12/25}\right]^{2/3}\, 
\left[ \frac{2\pi}{\alpha_s (m
)} \right]^{7/45}\, 
.
\label{anine}
\eeq
If it were not for the last factor (which is quite close to unity) the 
relation between the low- and high-energy scales would be similar to 
that in supersymmetric QCD. 

Matching at two thresholds, charm and beauty, yields
\beq
\Lambda_{\rm LOW}^{9}
= \Lambda_{\rm HIGH}^{23/3}
\left( \frac{2}{9}\right)^{32/9} \left( \frac{23}{6}\right)^{58/23}\, 
\left[ m_1^*m_2^*\right]^{2/3}\, 
\left[ \frac{2\pi}{\alpha_s (m_2 
)} \right]^{21/115}\, \left[ \frac{2\pi}{\alpha_s (m_1 
)} \right]^{7/45}\, 
,
\label{aten}
\eeq
where
$$
m_i^* = m_{i0}\, \left[ \frac{2\pi}{\alpha_{s0}}\right]^{12/23}\, .
$$

\newpage

\begin{center}
{\bf Recommended Literature}
\end{center}

Two review papers

\vspace{0.1cm}

B.L. Ioffe, {\it Acta Phys. Polon.}
{\bf B16} (1985) 543;

L.J. Reinders, H. Rubinstein, and S. Yazaki, {\it Phys. Reports} {\bf 
127} (1985) 1

\vspace{0.1cm}

\noindent were published in the mid-1980's. They may serve for an 
initial 
exposure but they do not provide an overview of modern 
developments. Neither do they reflect modern understanding
of conceptual issues. A dedicated  comprehensive review on this 
subject is long  overdue.

\vspace{0.1cm}

\noindent A brief survey of the light-cone sum rules is given in

\vspace{0.1cm}

V. Braun, hep-ph/9801222.

\vspace{0.1cm}

\noindent SVZ sum rules in the heavy quark theory are discussed in

\vspace{0.1cm}

M. Neubert, {\it Phys. Reports} {\bf 245} (1994) 259.

\vspace{0.1cm}

\noindent For a review of the sum rule applications in weak decays
see

\vspace{0.1cm}

A.  Khodjamirian and R. R\"{u}ckl, hep-ph/9801443. 

\vspace{0.1cm}

\vspace{0.1cm}

\noindent Some topics related to the SVZ sum rules are covered in

\vspace{0.1cm}

M. Shifman, {\it Phys. Reports} {\bf 209} (1991) 341;

T. Sch\"{a}fer and E. Shuryak, hep-ph/9610451 (Rev. Mod. Phys., to 
appear).

\vspace{0.1cm}

\noindent A collection of the original papers with an extended 
commentary 
reflecting the state of the art in the late 1980's can be found in

\vspace{0.1cm}
{\it Vacuum Structure and QCD Sum Rules}, Ed. M. Shifman
(North-Holland, 

Amsterdam, 1992).

\vspace{0.1cm}

\noindent Some useful formulae are compiled in

\vspace{0.1cm}

  S. Narison, 
   {\em QCD Spectral Sum Rules},   (World
   Scientific,  Singapore, 1989).

\vspace{0.1cm}

\noindent I do not agree with the treatment of many conceptual 
issues and 
technical details in this book. 

\end{document}